\documentclass[12pt]{iopart}
\usepackage{graphics}
\usepackage{epsfig}
\usepackage[usenames]{color}

\newcommand{\Sch}{Schr\"odinger}
\newcommand{\CPL}{{{\it Chem.~Phys.~Lett.}}~}
\newcommand{\PRA}{{{\it Phys.~Rev.}~A~}}

\newcommand{\CMP}{{\it Commun.~Math.~Phys.~}}
\newcommand{\CPC}{{\it Comput.~Phys.~Commun.~}}

\newcommand{\beq}{\begin{equation}}
\newcommand{\eeq}{\end{equation}}
\newcommand{\beqa}{\begin{eqnarray}}
\newcommand{\eeqa}{\end{eqnarray}}
\newcommand{\cY}{\mathcal{Y}}
\newcommand{\cA}{\mathcal{A}}
\newcommand{\cB}{\mathcal{B}}
\newcommand{\cU}{\mathcal{U}}
\newcommand{\cF}{\mathcal{F}}
\newcommand{\cI}{\mathcal{I}}
\newcommand{\cD}{\mathcal{D}}
\newcommand{\cG}{\mathcal{G}}
\newcommand{\cM}{\mathcal{M}}

\newcommand{\cO}{\mathcal{O}}

\newcommand{\tblue}[1]{\textcolor{Blue}{#1}}

\newcommand{\fff}{$\left(f_1,f_2,f_3\right)$}
\newcommand{\rxb}{$\left(\rho,\xi,\beta\right)$}
\newcommand{\vrj}{$\mathbf{r}_j$}

\newcommand{\bfr}{{\bf r}}
\newcommand{\bfx}{{\bf x}}
\newcommand{\bfy}{{\bf y}}
\newcommand{\bfL}{{\bf L}}
\newcommand{\bfxy}{{\bf x, y}}

\newcommand{\hlqsxy}{$\left\{h_{l-q+\lambda,q}^{(\lambda)}(\mathbf{x,y}),\,\lambda \leq q
\leq l\right\}$}
\newcommand{\psilq}{$\left\{\psi_q^{(\lambda)},\,\lambda \leq q \leq l\right\}$}

\newcommand{\jab}{$\left\{ J_{\pm
m,n}^{(l)}(\xi,\beta),\,m,n\geq 0 \right\}$}

\newcommand{\lpx}{$\left\{L_{p}^{(2l+4)}(x),\,p=0,\,1,\,2,\,\cdots \right\}$}
\newcommand{\lpxlam}{$\left\{L_{p}^{(2l+2\lambda+4)}(x),\,p=0,\,1,\,2,\,\cdots \right\}$}
\newcommand{\rmM}{{\rm M}}

\newcommand{\rmL}{{\rm L}}

\begin{document}

\title[Quantum three body problem]{Kinetic energy operator approach to the quantum
three-body problem with Coulomb interactions}
\author{Xuguang Chi$^1$, Wuyi Hsiang$^2$
and Ping Sheng$^1$\footnote{Corresponding author email: sheng@ust.hk}}
\address{$^1$Department of Physics and the Institute of Nano Science and
Technology,  \\Hong Kong University of Science and Technology,
\\Clear Water Bay, Kowloon, Hong Kong, China }
\address{$^2$Department of Mathematics,  Hong Kong University of Science
and Technology, \\Clear Water Bay, Kowloon, Hong Kong, China }

\begin{abstract}
We present a review of the quantum three-body problem, with emphasis
on the different methodologies, different three-body atomic systems
and their historical interest.  With the review as the background, a
more recently proposed non-variational, kinetic energy operator
approach to the solution of quantum three-body problem is presented,
based on the utilization of symmetries intrinsic to the kinetic
energy operator, i.e., the three-body Laplacian operator with the
respective masses.  Through a four-step reduction process, the nine
dimensional problem is reduced to a one dimensional coupled system
of ordinary differential equations, amenable to accurate numerical
solution as an infinite-dimensional algebraic eigenvalue problem.  A
key observation in this reduction process is that in the functional
subspace of the kinetic energy operator where all the rotational
degrees of freedom have been projected out, there is an intrinsic
symmetry which can be made explicit through the introduction of
Jacobi-spherical coordinates. A numerical scheme is presented
whereby the Coulomb matrix elements are calculated to a high degree
of accuracy with minimal effort, and the truncation of the linear
equations is carried out through a systematic procedure. The
resulting matrix equations are solved through an iteration process.
Numerical results are presented for (1) the negative hydrogen ion
$\textrm{H}^-$, (2) the helium and helium-like ions ($Z=3 \sim 6$),
(3) the hydrogen molecular ion $\textrm{H}_2^+$, and (4) the
positronium negative ion $\textrm{Ps}^-$. Up to
thirteen-significant-figure accuracy is achieved for the ground
state eigenvalues when double precision programming is used.
Comparison with the variational and other approaches shows our
ground state eigenvalues to be comparable, generally with less
decimal digits than the variational results, but can yield highly
accurate wavefunctions as by-products. Results on low-lying excited
states and their wavefunctions are obtained simultaneously with the
ground state properties, some at accuracies not achieved by other
methods. In particular, for the doubly excited state $^3P^e$ of
$\textrm{H}^-$ and the $^{1,3}P^e$ states of helium, some results
are obtained for the first time. Also, we have calculated fourteen
$\textrm{H}_2^+$ excited states, up to its dissociation level.
Analysis of the wavefunction characteristics, especially in relation
to the electron-electron correlation effects, are presented. A
significant advantage of the kinetic energy operator approach is its
general applicability to different three-body systems, with only the
charges, masses, and the symmetry of the desired state as the
required inputs.  Potential applications of the present approach to
scattering and other problems are noted.
\end{abstract}

\maketitle
\tableofcontents

\section{Introduction}
\label{sec:intro}

The dynamics of three interacting bodies constitutes one of the
oldest challenges in physics. Studies in this field can be traced
back to the work of Euler in the 18th century. In the beginning of
the 20th century, the failure of the Bohr-Sommerfeld quantization to
correctly describe the ground state of helium has stimulated the
development of the ``new" quantum theory, formulated by \Sch~and
Heisenberg. Almost 100 years after the founding of modern quantum
mechanics, there is still a continuing effort to improve the
solution methods or to invent new approaches for the seemingly
simple three-body Coulomb system.

In the early calculations of the two-electron systems, the focus was
usually on the bound-state spectra, with the helium and helium-like
ions as the proving ground.  The spectra could be calculated
efficiently with the help of the Hartree-Fock self-consistent-field
method based on the variational principle. Very high accuracy can be
achieved for eigenvalues such that they may be compared with high
precision measurements.  Bether and Salpeter~\cite{Bethe:1977} have
summarized the early works in this area.  The more recent
calculations of the bound states have extended such routines with
new numerical schemes and judicious choices of basis functions.

In a seminal experiment by Madden and Codding~\cite{Madden:1963},
the discovery of strong electron-electron correlation effects in
doubly excited resonance states of helium has triggered the
development of group-theoretical and adiabatic quantum
approximations to understand these effects. The two-electron
dynamics were again at the forefront of a revival.  As doubly
excited resonant states could not be tracked by an effective
single-particle method, over the past four decades the effort to
understand doubly excited resonances has stimulated much of the
theoretical research on two-electron atoms. The role of electron
correlation has become more important with increasing order of
double excitations, and the correlated three-body Coulomb dynamics
have been found to cause an extremely rich and complicated resonance
spectrum. Hence two-electron atoms have come to represent a
prototype of the few-body systems strongly affected by electronic
correlation.

A more difficult problem is the three-body scattering, such as
hydrogen-electron scattering. This problem has attracted much
attention recently. By using the finite element method, Levin and
Shertzer~\cite{Levin:1988} analyzed the S-wave phase shifts for
low-energy positron-hydrogen scattering. Botero and
Shertzer~\cite{Botero:1992} directly solved the \Sch~equation for
electron-hydrogen scattering, and Rescigno~\etal\cite{Rescigno:1999}
and Baertschy~\etal\cite{Baertschy:2001} used supercomputers to
obtain a complete numerical solution of the hydrogen atom ionization
through electron collision.

Three-body systems, especially the two-electron systems, remain an
active field of research today.  This persistent interest can be
traced to the fact that the three-body problem is just complex
enough for rather sophisticated theoretical concepts, yet simple
enough to provide accurate numerical and experimental tests.

More recently, Hsiang and
Hsiang~\cite{HsiangHsiang:1996,Hsiang:1997,HsiangHsiang:1997a,HsiangHsiang:1997b}
have outlined a new approach to the quantum three-body problem which
was based on the systematic exploitation of all the intrinsic
dynamic symmetries of the three-body kinetic energy operator~(the
three-body Laplacian with the respective masses), some of which  not
fully recognized previously. The purposes of this work are to
implement this new mathematical formulation and to compare the
present approach with the conventional variational approach in terms
of numerical results for both eigenvalues and eigenfunctions of a
number of quantum three-body systems.

The main conclusions of this work are that the present approach
offers not only systematic computability, requiring minimal ad hoc
inputs in the computational process, but also achieves numerical
accuracy for both the eigenvalues and eigenfunctions. The latter
advantage is particularly significant for excited states. In
particular, the wave function characteristics of excited states with
strong electron-electron correlation can be accurately delineated.

In what follows, we first review the various approaches to the
quantum three-body problem in \sref{sec:review}, followed by a
detailed presentation of the kinetic energy operator approach in
\sref{sec:formulation}. The formulation essentially consists of a
four-step reduction process, in which the dynamic symmetries
intrinsic to the kinetic energy operator are fully utilized.  The
end result of the reduction is a one-dimensional coupled system of
ordinary differential equations, solvable as an infinite-dimensional
algebraic eigenvalue problem. In \sref{sec:scheme}, a scheme is
presented for the numerical implementation of our approach.   Due to
the high precision required for the Coulomb matrix elements, a
special integration technique is used to evaluate both the matrix
elements as well as the product of the potential energy matrix with
a vector. A sparse matrix solver is then applied to solve the linear
equations iteratively. Truncation of the infinite linear system is
carried out by following a rule implied by the asymptotic behavior
of the eigenvalues, leading to sequences of numerical data from
which one can apply a systematic extrapolation procedure. The
results for some typical Coulomb systems are presented in
\sref{sec:results}, with comparisons to those obtained via other
approaches.  We also explore the properties of three-body wave
functions, and discuss some of their physical significances. It
should be noted that all the results were obtained by using the same
program. Inputs are the symmetry of the state~(i.e., $S$, $P$, $D$,
or $F$), the mass ratios, and the sign of the charges.  The article
concludes by noting some potential applications of the present
approach.

\section{Review of the various approaches}
\label{sec:review}

\subsection{Variational method}

Quantum variational method is most suitable for obtaining accurate
results for the ground state or low-lying states, and in this regard
it is superior to the perturbation methods.  The basic idea of the
variational method, sometime also denoted as the Ritz variational
method, named after the pioneer of the approach,  is to write the
trial wave function $\Psi_{tr}$  in some arbitrarily chosen
mathematical form with variational parameters,
\begin{equation}
\Psi_{tr}=\Psi_{tr}(\alpha,\beta,\gamma,\cdots),
\end{equation}
and then adjust the parameters to obtain the minimum energy
\begin{equation}
E_{tr}(\alpha,\beta,\gamma,\cdots)=\frac{\int
\Psi_{tr}^*H\Psi_{tr}\rmd\tau}{\int |\Psi_{tr}|^2\rmd\tau }
\end{equation}
through the solution of a system of  coupled equations:
\begin{eqnarray}\label{eq:e_tr}
\frac{\partial E_{tr}(\alpha,\beta,\gamma,\cdots)}{\partial
\alpha}&&=0
\nonumber \\
\frac{\partial E_{tr}(\alpha,\beta,\gamma,\cdots)}{\partial
\beta}&&=0
\nonumber \\
\frac{\partial E_{tr}(\alpha,\beta,\gamma,\cdots)}{\partial
\gamma}&&=0
\nonumber \\
&&\vdots
\end{eqnarray}
It should be noted that the trial wave function(s) must satisfy the
symmetry condition imposed by the Fermi-Dirac statistics.  The
solution to \eref{eq:e_tr} yields minima of energy in the
multidimensional parameter space. The lowest energy minimum is
treated as the approximate ground state eigenvalue.  As an example,
a simple trial wave function for the ground state of helium-like
atoms is $\psi(r_1,r_2,r_{12})=\exp(-(Z-\sigma)(r_1+r_2))$, where
$\sigma$ represents the screening effect in an approximate way.
Minimizing the energy functional gives $\rm{E}=-(Z-5/16)^2$, i.e.,
E=-2.85~a.u. for the helium ground state~\cite{Bethe:1977}. However,
for $\textrm{H}^-$ this trial wave function is noted to fail in
obtaining a bounded ground state.  For the excited states, the
Hyllerass-Undheim theorem states that the remaining energy minima of
\eref{eq:e_tr}, $\lambda_2,\,\,\lambda_3,\,\,\cdots$, are also upper
bounds to the exact eigenvalues $E_2,\,\,E_3,\,\,\cdots$, provided
that the spectrum is bounded from below.  However, the calculation
of precise excited state energies is more difficult compared to that
of the ground state, due to the appearance of the subsidiary
condition that the eigenfunction of every excited state must be
orthogonal to the eigenfunctions of all the lower-order states. This
condition reduces considerably the number of available trial
functions which may be chosen to approximate the eigenfunction~(to
be inserted in the variation integral) of the particular excited
state.  As a result, convergence is not nearly as good.

Historically, Kellner~\cite{Kellner:1927} was the first to use the
variational principle to obtain a rather precise ground state
energy, E=-2.895~a.u.. His results were improved upon by the
variational calculations of Hylleraas~\cite{Hylleraas:1928,
Hylleraas:1929}, who obtained E=-2.9037~a.u.~using a trial wave
function with 38 variational parameters.  This method was later used
by Kinoshita~\cite{Kinoshita:1959} in large-scale variational
calculations. In a bold move, Frankowski and
Pekeris~\cite{Frankowski:1966} used more than 200 parameters in
trial functions to obtain the energies of helium-like systems that
were not surpassed for almost two decades. With the appearance of
powerful modern computers, however, people can now include more than
one thousand parameters in trial wave functions in order to obtain
high accuracy results for the ground state of helium, hydrogen-like
ions and some muonic molecular
ions~\cite{Frolov:1992,Frolov:1993,Bishop:1994,Burgers:1995,Frolov:1998,
Drake:2002}.

As a modified version of the variational method, the complex
rotation method, based on the dilatation analytic
continuation~\cite{Aguilar:1971,Balslev:1971,Simmon:1972}, was
extensively used to calculate the doubly excited states of the
two-electron
systems~\cite{Ho:1979,Reinhardt:1982,Ho:1983,Ho:1986,Ho:1994}. The
basic idea is that after a complex rotation, $r\rightarrow r
\rme^{\rmi\theta}$, is applied to the radial coordinate, the
resulting Hamiltonian becomes complex, i.e., $H(\theta)=\rme^{-2\rmi
\theta}T+\rme^{-\rmi\theta}V$, where $T$ is the kinetic energy
operator and $V$ the Coulomb potential.  The complex Hamiltonian
will yield complex eigenenergies as a result of applying the Ritz
variation, in which the real part would correspond to the position
of the doubly excited state, and the imaginary  part its life-time.

\subsection{Hyperspherical coordinates method}

Hyperspherical coordinates were first introduced into atomic physics
by Gronwall~\cite{Gronwall:1937} to study the analytic structure of
the \Sch~equation for the ground state of helium atom.  The basic
idea of the hyperspherical approach to the three-body systems is to
express the two relative~(to center of mass) coordinates as a single
six-dimensional vector, and the nonrelativistic \Sch~equation in the
six-dimensional space is to be solved without reference to the
``wave functions" associated with individual particles. The
condition of particle exchange symmetry then becomes a boundary
condition on the three-body wave function on the hypersurface.

The hyperspherical approach has been applied to solve bound states
and scattering problems in many different fields of physics and
chemistry. Many of the earlier works dealt with the basic structure
of the mathematical functions encountered in hyperspherical
coordinates. Here we introduce the hyperspherical coordinates for
two-electron atomic systems such as helium and hydrogen negative
ions where the mass of the nucleus is treated as infinite. The
hyperspherical coordinates are then obtained by defining
\begin{equation}
\rho=\sqrt{r_1^2+r_2^2},\quad \alpha=\arctan(r_2/r_1),
\end{equation}
where  $\rho$ is the hyperradius which measures the size of the
system, and  $\alpha$   is the hyperangle.  Thus the two vectors
$\bfr_1 $ and  $\bfr_2$  are replaced by six coordinates
$(\rho,\Omega)$, where
$\Omega=(\alpha,\theta_1,\phi_1,\theta_2,\phi_2)$ denotes
collectively the five angles, with  $(\theta_i,\psi_i)$  being the
spherical angles of electron $i$. In hyperspherical coordinates the
two-electron equation is given by
\begin{equation}
\left(-\frac{1}{2}\Delta_1-\frac{1}{2}\Delta_2-\frac{Z}{r_1}
-\frac{Z}{r_2}+\frac{1}{r_{12}}-E\right)\Psi(\bfr_1,\bfr_2)=0.
\end{equation}
After eliminating the first-order derivatives in the differential
operators by expressing
\begin{equation}
\Psi(\bfr_1,\bfr_2)
=\psi(\rho,\Omega)/(\rho^{5/2}\sin\alpha\cos\alpha),
\end{equation}
an equation in terms of  $\psi(\rho,\Omega)$   is obtained:
\begin{equation}
\left( -\frac{\partial^2}{\partial
\rho^2}+\frac{\Lambda^2}{\rho^2}+\frac{2C}{\rho}+2E
\right)\psi(\rho,\Omega)=0,
\end{equation}
where
\begin{equation}
\Lambda^2= \left( -\frac{\partial^2}{\partial \alpha^2}+\frac{{\bf
l}_1^2}{\cos^2\alpha} +\frac{{\bf l}_2^2}{\sin^2\alpha}
\right)-\frac{1}{4}
\end{equation}
is the grand angular momentum operator, ${\bf l}_i$  being the
angular momentum operator  for electron $i$ and  $C/\rho$  is the
total Coulomb interaction potential among the three charged
particles, with $C$ given by
\begin{equation}
C(\alpha,\theta)=-\frac{Z}{\cos\alpha}-\frac{Z}{\sin\alpha}+\frac{1}
{\sqrt{1-\sin 2\alpha\cos \theta_{12}}}.
\end{equation}
Here $\theta_{12}$  is the angle between the two electrons with
respect to the nucleus (as the origin).

The straightforward solution approach is to expand
$\psi(\rho,\Omega)$  by the eigenfunctions of $\Lambda^2$, called
hyperspherical  harmonics.  This method has been applied by a number
of authors to $\textrm{H}^-$  and $\textrm{He}$
systems~\cite{Knirk:1974,Haftel:1983,Frey:1987,Jerjian:1987}, but
the rate of convergence is rather slow.  To improve the rate of
convergence, Haftel and Mandelzweig~\cite{Haftel:1987,
Haftel:1988a,Haftel:1988b, Haftel:1990, Haftel:1993} introduced an
exponential factor in the expansion, $\psi = \chi \Phi$, where
$\chi$  is chosen to be of the form
$\chi=\exp\{-a(r_1+r_2)+br_{12}\}$, with $a$ and $b$  to be obtained
variationally or by some ansatz.  If $a$ and $b$ are appropriately
chosen, the singularity in the Coulomb potential can be removed, and
it is then possible to expand $\Phi$  in terms of hyperspherical
harmonics with rapid convergence.  Another common approach, the
adiabatic expansion, was introduced by Fano and first applied by
Macek~\cite{Macek:1968}. Details of this method can be found in the
review article by Fano~\cite{Fano:1983} and a relevant
book~\cite{Fano:1986}.

\subsection{Perturbation, Hartree-Fock, and the finite element methods}

In the perturbation method, the Hamiltonian is split into two parts,
$H=H_0+\lambda H_1$, where the perturbation term $\lambda H_1$  is
small in a relative sense.  For the Hamiltonian equation
\begin{equation}
(H_0+\lambda H_1-E)\Psi=0,
\end{equation}
the eigenfunction $\Psi$  and eigenvalue $E$  are expanded in powers
of the small parameter $\lambda$, namely
\begin{equation}
E=\sum_{n=0}^{\infty}\lambda^nE_n,\quad
\Psi=\sum_{n=0}^{\infty}\lambda^n \Psi_n.
\end{equation}
Substitution of these expansions into the \Sch~equation and equating
the coefficients for each power of $\lambda$ to zero lead to an
infinite set of coupled linear equations:
\begin{eqnarray}
H_0\Psi_0-E_0\Psi_0&=&0\\
H_0\Psi_1+H_1\Psi_0-E_0\Psi_1-E_1\Psi_0&=&0\\
.............................................\nonumber \\
H_0\Psi_n+H_1\Psi_{n-1}-\sum_{m=0}^{n}E_m\Psi_{n-m}&=&0.
\end{eqnarray}
If we consider  $\Psi_0$ and $E_0$ as known, we can obtains the
first order perturbation energy
\begin{equation}
E_1=\int\Psi_0^*H_1\Psi_0 \rmd\tau.
\end{equation}
For the calculation of $E_2$, $E_3$, $\cdots$, more efforts are
needed. There is no theorem to clearly tell one how to choose the
perturbation $\lambda H_1$, and there can be different choices for
the same problem. For large $Z$  in helium-like ions, the
interaction $1/r_{12}$  between the electrons can be treated as the
perturbation; for the excited state, there is the unsymmetrical
choice~\cite{Bethe:1977}, $H_0=T+V_1+V_2$ and $\lambda H_1=W$, where
\begin{equation}
V_1(r_1)=-\frac{Z}{r_1},\quad V_2(r_2)=-\frac{Z-1}{r_2},\quad
W=\frac{1}{r_{12}}-\frac{1}{r_2}.
\end{equation}

In the Hartree-Fock method, the wave functions for the two-electron
atom must obey overall antisymmetry. That means
\begin{equation}
\Psi=\frac{1}{\sqrt{2}}[ \psi_1(\bfr_1) \psi_2(\bfr_2) \pm
\psi_1(\bfr_2) \psi_2(\bfr_1)],
\end{equation}
where the $+$ sign in the above equation is used when the spin state
has antisymmetry.  The form of the trial wave functions
$\psi_1(\bfr)$ or $\psi_2(\bfr)$  is unknown, but the aim is to find
the most accurate form possible for the two functions $\psi_1(\bfr)$
or $\psi_2(\bfr)$  that would minimize the expectation value of the
Hamiltonian, which is regarded as a functional of the two trial wave
functions.  The application of the general variational principle
leads to two coupled differential-integral equations~(the
Euler-Lagrange equations). In essence, the Hartree-Fock method is a
mean-field theory. It differs from the Ritz variational approach in
that the variation in the Ritz method is carried out by using
parameters, whereas in the Hartree-Fock method the variation is
carried out by solving coupled integral-differential equations.

Recently, the finite element method (FEM)~\cite{Ram-Mohan:2002} has
been used to directly solve the \Sch~equation, especially for the
$S$ state of some systems. The domain of the wave function is
segmented into tetrahedrons, each serving as the domain of a local
polynomial basis set. The approximation of the wave function as a
linear combination of these local polynomials is called a
finite-element description. FEM treatment for the helium atom in the
infinite nuclear mass
approximation~\cite{Levin:1985,Braun:1993,Bottcher:1994,Scrinzi:1995,Ackermann:1996},
and the hydrogen molecular ion $\textrm{H}_2^+$ in the
Born-Oppenheimer approximation, have been presented by several
authors~\cite{Laaksonen:1983,Ford:1984,Schulze:1985,Yang:1991,Ackermann:1993,Ackermann:1994}.
An adaptable FEM approach was used by
Ackermann~\cite{Ackermann:1995} and co-workers to obtain the energy
values to a precision of $10^{-11}$ with moderate computational
effort. The advantage of FEM lies in its flexibility because of the
local basis, but the drawback is that one often has to face huge
sparse matrices.

\section{Formulation}
\label{sec:formulation}

In a three-body Coulomb system, the position vectors and the masses
of the three particles are denoted by \vrj~and  $\rmM_j$,
$j=1,\,2,\,3$.  The relative masses are defined as
$m_j=\rmM_j/\rmM$, where $\rmM$ is the total mass,
$\rmM=\sum{\rmM_j}$, and $\sum{m_j}=1$. \Sch~equation is given by
\begin{eqnarray}\label{eq:schr}
 &-\frac{1}{2\rmM}\Delta\Psi+V \Psi=E\Psi,
 \nonumber \\
&\qquad \Delta=\sum_{j=1}^{3}\frac{\Delta_j}{m_j},
\end{eqnarray}
where  $\Delta_j$ is the Laplacian operator with respect to the j-th
particle, $\Delta$ is defined as the kinetic energy operator, and
$V$ is the Coulomb potential of the three interacting particles:
\begin{equation}\label{eq:potential0}
V=\frac{Z_2Z_3}{|\bfr_2-\bfr_3|}+\frac{Z_3Z_1}{|\bfr_3-\bfr_1|}
+\frac{Z_1Z_2}{|\bfr_1-\bfr_2|}.
\end{equation}
Here  $Z_j$  is the electric charge of the $j$-th particle. In the
above and in what follows, the length unit is the Bohr radius,
$\hbar^2/m_e e^2$, $m_e$ is the electron mass, also the mass unit,
and $m_e e^2/\hbar^2$ is the energy unit. The above system is
uniquely determined by the six parameters,
$\{m_j,\,Z_j,\,\,j=1,\,2,\,3\}$. The Coulomb potential depends only
on the distance between each pair of particles, and the system is
invariant under spatial translation, rotation, and inversion.
Furthermore, if the system consists of identical particles,
invariance with respect to the permutation of identical particles
must be imposed.

\subsection{Overview of the approach}

As implied by the name, the focus of the present approach is on the
kinetic energy operator $\Delta$.   Before delving into the details,
it would be helpful to give an overview of the formulation, which in
essence consists of four reduction steps.  In the first
step~(\sref{sec:Jacobi}), we reduce the nine dimensional problem to
a six dimensional problem by simply using coordinates relative to
the center of mass.  Jacobi vectors will be introduced at this stage
to facilitate later developments.  In the second
step~(\sref{sec:angular}), angular momentum eigenfunctions will be
presented which are in the null space of the kinetic energy
operator. That is, when the angular momentum eigenfunctions are
operated on by the kinetic energy operator, the result is zero. Thus
when the total wave function is expanded in terms of the angular
momentum eigenfunctions, the coefficients of the expansion, which
are rotationally invariant, satisfy a reduced \Sch~equation which is
three-dimensional. The set of functions that satisfy this reduced
\Sch~equation constitutes a subspace in which all the rotational
degrees of freedom have been projected out.  In the third step of
the reduction~(\sref{sec:Spherical}), this 3D reduced \Sch~equation
is expressed in terms of the Jacobi-spherical coordinates, leading
to a form with an angular operator whose eigenfunctions are the
Jacobi polynomials~(hence the denotation of Jacobi-spherical
coordinate).  It should be emphasized that the Jacobi-spherical
symmetry is particular only to the kinetic energy operator and not
to the Coulomb potential, in contrast to the rotational symmetry.
Hence the total Hamiltonian does not have this symmetry.  However,
the Coulomb potential becomes separable in the Jacobi-spherical
coordinates.  That is, the Coulomb potential can be expressed as a
product of two terms, one of which depends only on the radial
coordinate of the Jacobi-spherical coordinates.  The wave functions
of this reduced \Sch~equation can therefore be expanded in terms of
the Jacobi polynomials with coefficients depending only on the
radial coordinate of the Jacobi-spherical coordinates.  In the
fourth and last reduction step~(\sref{sec:ODEs}), the substitution
of the this expansion into the reduced \Sch~equation leads to a set
of coupled ordinary differential equations~(ODEs) which is now only
one dimensional, i.e., the solution depends only on the radial
coordinate. The solution to the set of ODEs can be expanded in terms
of the Laguerre polynomials, and the coefficients of this expansion
satisfy an infinite linear system which is amenable to numerical
solution~(\sref{sec:Linear}).

\subsection{Jacobi vectors and coordinate transformations}
\label{sec:Jacobi}

The configuration space of a given three-body system is a
nine-dimensional space consisting of a triplet of position vectors,
\begin{equation}
\Re^9=\left\{\bfr_1,\,\,\bfr_2,\,\,\bfr_3\right\},
\end{equation}
where \vrj~is the position vector of $j$-th particle. Without the
loss of generality, one can assume that the center of gravity is
fixed at the origin, thus reducing the configuration space to the
following six-dimensional reduced configuration space:
\begin{equation}
\Re^6=\left\{(\bfr_1,\,\,\bfr_2,\,\,\bfr_3); \,\,\sum m_j \bfr_j=0
\right\}.
\end{equation}
Following Jacobi, we introduce the kinematic metric on the reduced
configuration space by defining the inner product as
\begin{equation}\label{eq:jac_metric}
\left< \{\mathbf{a}_j\},\{\mathbf{b}_j\} \right>:=\sum_{j=1}^{3}m_j
\mathbf{a}_j\cdot \mathbf{b}_j.
\end{equation}
This definition is related to the Lagrange's least-action principle,
which was reformulated by Jacobi, leading to the geometric
explanation of mechanics.  The metric  $\rmd s^2$ is defined in
terms of the kinetic energy  $T$ by setting
\begin{equation}
\rmd s^2=\frac{2T}{\rm M}\rmd t^2.
\end{equation}
For example, in the case of an n-body system
\begin{equation}
 \rmd s^2=\sum_{i=1}^{n}m_i
\left(\rmd x_i^2+\rmd y_i^2+\rmd z_i^2\right),
\end{equation}
where $\rm{M}$  is the total mass, and $m_i$  is the percentage of
mass for the $i$-th particle. Lagrange's least-action principle of
classical mechanics has a simple geometric explanation: trajectories
of a given mechanical system are exactly those geodesic curves in
the metric space of $\rmd s^2$, defined in the configuration space.

To each given m-triangle, i.e., triplet  $\{\bfr_j\}$ with $\sum m_j
\bfr_j=0$, we define a pair of Jacobi vectors $\bfx=(x_1,x_2,x_3)$
and $\bfy=(y_1,y_2,y_3)$  in the reduced space $\Re^6$, given by
\begin{equation}
\fl \qquad {\bfx}=\sqrt{\frac{m_1}{1-m_1}}
\left(m_2({\bfr_1-\bfr_2})+m_3({\bfr_1-\bfr_3})\right),\quad
{\bfy}=\sqrt{\frac{m_2m_3}{1-m_1}} \left({\bfr_2-\bfr_3}\right).
\end{equation}
It should be noted here that in this work, indices 2, 3 are used to
label identical particles in our three-body system~(e.g., the two
electrons in the two-electron systems).  Hence the exchange
antisymmetry of the Fermi-Dirac statistics is manifest in letting
$\bfy \rightarrow\, -\bfy$.  The reduced configuration space can be
represented by $\Re^3\oplus \Re^3=\left\{(\bfx,\,\bfy);\bfx,\,\bfy
\in \Re^3\right\}$, and the metric $\rmd s^2$,  becomes
\begin{equation}
\rmd s^2=\rmd\bfx \cdot \rmd\bfx+\rmd\bfy \cdot \rmd\bfy,
\end{equation}
with the kinetic energy operator given by
\begin{equation}
\Delta=\Delta_{\bfx}+\Delta_{\bfy}.
\end{equation}
Also, the total angular momentum operator is given by
\begin{eqnarray}
\bf{L} &=&\bf{L_{\bfx}}+\bf{L_{\bfy}} \nonumber
\\
&=& -\rmi\left(\bfx\times\nabla_{\bfx}
           +\bfy\times\nabla_{\bfy} \right).
\end{eqnarray}
By using the Jacobi vectors $\bfx$ and $\bfy$, the motion of the
center-of-mass is removed from \eref{eq:schr}, leading to a
six-dimensional equation:
\begin{equation}
-\frac{1}{\rm {2M}}(\Delta_\bfx+\Delta_\bfy)\Psi+V\Psi=E\Psi.
\end{equation}

\subsubsection{Rotationally invariant variables}

For given two vectors $\bfx$ and $\bfy$, we can construct three
rotationally invariant polynomials,
\begin{equation}
f_1=\bfx\cdot\bfx,\quad f_2=\bfy \cdot \bfy,\quad f_3=\bfx \cdot
\bfy.
\end{equation}
Any rotationally invariant quantity with respect to $\bfx$ and
$\bfy$ should be a function of \fff. Since the metric $\rmd s^2$ is
rotationally invariant, thus it can be expressed as
\begin{equation}
\rmd s^2=\sum_{i,j}^{3}g_{ij}\,\rmd f_i\, \rmd f_j,
\end{equation}
where $(g_{ij})$ is the inverse matrix of the following $(g^{ij})$:
\begin{equation}
(g^{ij})=(\bigtriangledown f_i \cdot \bigtriangledown f_j).
\end{equation}
The three variables \fff~will be used to express the reduced
\Sch~equation below.

\subsubsection{Jacobi spherical coordinates}
The Hamiltonian of the three-body problem~(Laplacian plus Coulomb
potential) has space-rotation symmetry, but the symmetry of the
kinetic energy operator is higher than that of the potential
energy~(Coulomb interaction).  Moreover, there is a direct relation
between the kinetic energy operator and the kinetic metric $\rmd
s^2$. It is obvious that \fff~is not orthogonal; hence terms
$\{g_{ij}\,\rmd f_i\,\rmd f_j,i \neq j \}$   do not vanish in  $\rmd
s^2$. By making use of the following coordinate
transformation~\cite{HsiangHsiang:1997a,HsiangHsiang:1997b}, namely
\begin{eqnarray}\label{eq:sph_trans}
&\rho= \sqrt{f_1+f_2},\qquad \cos\theta =
{2(f_1f_2-f_{3}^{2})^{1/2}}{(f_1+f_2)^{-1}},
\nonumber \\
&\tan\beta= {2f_3}{(f_2-f_1)^{-1}}, \quad
\sin\beta=2f_3\left[(f_2-f_1)^2+4f_{3}^{2}\right]^{-1/2},
\nonumber \\
&(0\leq \rho<\infty,\quad 0\leq \theta \leq \pi/2,\quad -\pi \leq
\beta \leq \pi),
\end{eqnarray}
the metric $\rmd s^2$  becomes
\begin{equation}
\rmd s^2=\rmd\rho^2+\frac{1}{4}\rho^2(\rmd\theta^2+\sin^2\theta
\rmd\beta^2).
\end{equation}
It is thus obvious that there is a semi-spherical structure in $\rmd
s^2$, which implies some form of symmetry of the kinetic energy
operator, beyond those associated with rotational symmetries.  We
will soon learn the importance of this symmetry.  Moreover, it is
easy to get the inversion of the transform, namely
\begin{equation}
f_1 =\frac{\rho^2}{2}(1-\xi \cos\beta),\quad f_2
=\frac{\rho^2}{2}(1+\xi \cos\beta),\quad f_3 =\frac{\rho^2}{2}\xi
\sin\beta,
\end{equation}
where $\xi=\sin\theta$.  For convenience, we shall use \rxb~instead
of $\left(\rho,\theta,\beta\right)$  throughout this work, and call
them the Jacobi spherical coordinates for reason that will become
obvious later (see  \eref{eq:V_in_sph} and \eref{eq:V_xibeta}).

\subsubsection{Interparticle distance  ${r}_{ij}$ expressed in Jacobi
spherical coordinates} From the cyclic permutations of three sets of
Jacobi coordinate vectors, i.e.,
\begin{eqnarray}\label{eq:sets}
\fl \qquad &\bfx^{(1)}=\sqrt{\frac{m_1}{1-m_1}}\left(
m_2(\bfr_1-\bfr_2)+m_3(\bfr_1-\bfr_3)\right),\quad
\bfy^{(1)}=\sqrt{\frac{m_2m_3}{1-m_1}}(\bfr_2-\bfr_3),\nonumber
\\
\fl & \bfx^{(2)}=\sqrt{\frac{m_2}{1-m_2}}\left(
m_3(\bfr_2-\bfr_3)+m_1(\bfr_2-\bfr_1)\right),\quad
\bfy^{(2)}=\sqrt{\frac{m_3m_1}{1-m_2}}(\bfr_3-\bfr_1), \nonumber
\\
\fl & \bfx^{(3)}=\sqrt{\frac{m_3}{1-m_3}}\left(
m_1(\bfr_3-\bfr_1)+m_2(\bfr_3-\bfr_2)\right),\quad
\bfy^{(3)}=\sqrt{\frac{m_1m_2}{1-m_3}}(\bfr_1-\bfr_2),
\end{eqnarray}
and two simple relations
\begin{eqnarray}\label{eq:setr}
\left(
  \begin{array}{c}
  \bfx^{(2)}\\
  \bfy^{(2)}
  \end{array}
\right) = \left(
    \begin{array}{cc}
    \cos\beta_{2} &\sin\beta_{2}\\
    -\sin\beta_{2} & \cos\beta_{2}
    \end{array}
\right) \left(
  \begin{array}{c}
  \bfx^{(1)}\\
  \bfy^{(1)}
  \end{array}
\right) ,\nonumber
\\ \nonumber
\\
\left(
  \begin{array}{c}
  \bfx^{(3)}\\
  \bfy^{(3)}
  \end{array}
\right) = \left(
    \begin{array}{cc}
    \cos\beta_{3} &\sin\beta_{3}\\
    -\sin\beta_{3} & \cos\beta_{3}
    \end{array}
\right) \left(
  \begin{array}{c}
  \bfx^{(1)}\\
  \bfy^{(1)}
  \end{array}
\right),
\end{eqnarray}
where
\begin{eqnarray}
&&\cos\beta_{2}=\sqrt{\frac{m_1m_2}{m_3+m_1m_2}},\quad
\sin\beta_{2}=-\sqrt{\frac{m_3}{m_3+m_1m_2}},   \nonumber \\
&&\cos\beta_{3}=\sqrt{\frac{m_1m_3}{m_2+m_1m_3}},\quad
\sin\beta_{3}=\sqrt{\frac{m_2}{m_2+m_1m_3}},
\end{eqnarray}
we can write  $\left|\mathbf{r}_{ij}\right|^2$  in the
Jacobi-spherical coordinates $\left(\rho,\xi,\beta\right)$:
\begin{eqnarray}\label{eq:r123}
\left|\bfr_{23}\right|^2 &=&k_1\rho^2(1+\xi\cos\beta),
\nonumber \\
\left|\bfr_{31}\right|^2 &=&k_2\rho^2(1+\xi\cos(\beta+2\beta_2)),
\nonumber \\
\left|\bfr_{12}\right|^2 &=&k_3\rho^2(1+\xi\cos(\beta+2\beta_3)),
\end{eqnarray}
with \beq k_1=\frac{1-m_1}{2m_2m_3},\quad
k_2=\frac{1-m_2}{2m_3m_1},\quad k_3=\frac{1-m_3}{2m_1m_2}. \eeq For
the special case of helium-like ions with infinite nuclear mass, we
take \beq \label{eq:hexy} \bfx=\frac{\bfr_1+\bfr_2}{\sqrt{2}},\quad
\bfy=\frac{\bfr_2-\bfr_1}{\sqrt{2}}, \eeq as the Jacobi coordinate
vectors, where $\bfr_1$ and $\bfr_2$ are the position vectors of two
electrons, and the distances become \beqa \label{eq:r123_inf}
|\bfr_1|^2&=&\frac{1}{2}\rho^2(1-\xi \sin\beta),
\nonumber \\
|\bfr_2|^2&=&\frac{1}{2}\rho^2(1+\xi \sin\beta),
\nonumber \\
|\bfr_{12}|^2&=&\rho^2(1+\xi \cos\beta). \eeqa From \eref{eq:r123}
and \eref{eq:r123_inf},   it is straightforward to express the
Coulomb potential in terms of \rxb. It will be seen that the Coulomb
potential becomes separable in the Jacobi-spherical coordinates.

\subsection{Symmetries and angular momentum eigenfunctions}
\label{sec:angular}

\Sch~equation is invariant under the spatial rotation and coordinate
inversion.  That means both the total angular momentum operator
$\bfL^2(=\rmL_1^2+\rmL_2^2+\rmL_3^2)$, its z-direction component
$\rmL_3$, and the parity operator commute with the Hamiltonian,
therefore one has the angular momentum quantum numbers $(l,m)$,
where $m$ characterizes the azimuthal component of the angular
momentum, plus the parity quantum number $\lambda=0,\,1$ for even
and odd parities, respectively.  In addition, the wavefunctions must
also satisfy the Fermi-Dirac statistics, manifest as antisymmetry
under the exchange of two electrons.  It should be noted that since
the antisymmetry applies to the product of spin state with the
wavefunction, hence the wavefunction can be either symmetric~(for
antisymmetric spin state) or antisymmetric~(for symmetric spin
state).

In this section, we use a set of  bi-harmonic functions as the
eigenfunctions of $\bfL^2$ and $\rmL_3$, and expand  the wave
functions in these basis functions  so as to separate out the
rotational degrees of freedom from the \Sch~equation.

The system is independent of the choice of z-direction, hence it is
enough to consider the special case, $m=l$.  As is well known,
$\mathcal{Y}_m^l(\bfr)=r^lY_m^l(\theta,\phi)$,  where the set
$\left(r,\theta,\phi\right)$ denotes the spherical coordinates of a
vector,  is a homogeneous polynomial of degree $l$ with respect to
the components of $\bfr$,  and satisfies the Laplace equation. It is
the eigenfunction for the angular momentum operator.  The
polynomials
$\mathcal{Y}_m^q(\bfx)\mathcal{Y}_{m^\prime}^{l-q}(\bfy)$ can be
combined to form the eigenfunctions of the total angular momentum
$\bfL^2$  ($\bfL=\mathbf{L_x}+\mathbf{L_y}$) by the Clebsch-Gordan
coefficients~\cite{Zare:1998}.    For the special case, $m=l$, one
has two types of bi-homogeneous functions
~\cite{Hsiang:1997,HsiangHsiang:1997a}  with respect to $\bfx$ and
$\bfy$:
\begin{eqnarray}
h_{a,b}^{(0)}(\bfx,\bfy)&=&\frac{1}{a!b!}u^{a}v^{b} , \quad a,b \geq
0,
\nonumber \\
h_{a,b}^{(1)}(\bfx,\bfy)&=&\frac{1}{(a-1)!(b-1)!}u^{a-1} v^{b-1}w ,
\quad a,b \geq 1,
\end{eqnarray}
where
\begin{eqnarray}
u&=&x_1+\rmi x_2,\quad v=y_1+\rmi y_2,\nonumber \\
w&=&(x_2y_3-x_3y_2)+\rmi (x_3y_1-x_1y_3).
\end{eqnarray}
 This specific family of basis functions,
$h_{a,b}^{(\lambda)}(\bfx,\bfy)$ ($\lambda=0$ or 1), has the
following properties.
\begin{enumerate}
\item They are in the null space of the kinetic energy operator. In
particular, they have zero eigenvalues for the operators
$\Delta_{\bfx}$, $\Delta_{\bfy}$ and $\Delta_{\bfx\bfy}$, i.e., they
exhibit bi-harmonicity: \beq
\Delta_{\bfx}h_{a,b}^{(\lambda)}(\bfxy)=\Delta_{\bfy}
h_{a,b}^{(\lambda)}(\bfx,\bfy)
=\Delta_{\bfx\bfy}h_{a,b}^{(\lambda)}(\bfxy)=0, \eeq where
$$
\Delta_{\mathbf{xy}}=\sum_{j=1}^{3}\frac{\partial^2}{\partial x_j
\partial y_j}.
$$
\item They are bi-homogeneity in $x_j$ and $y_j$.
\item They are the common eigenfunctions of $\bfL^2$ and $\rmL_3$,
i.e., \beqa
\bfL^2h_{a,b}^{(\lambda)}&=&(a+b-\lambda)(a+b-\lambda+1)h_{a,b}^{(
\lambda)},
\nonumber \\
{\rmL}_3h_{a,b}^{(\lambda)}&=&(a+b-\lambda)h_{a,b}^{(\lambda)}.
\eeqa

\item They exhibit spatial inversion symmetry under
$(\bfxy)\longrightarrow(\mathbf{-x,-y})$: \beq
h_{a,b}^{(\lambda)}(\mathbf{-x,-y})
=(-1)^{a+b}h_{a,b}^{(\lambda)}(\bfxy). \eeq

\item They exhibit symmetry under the permutation of particle 2 and
particle 3, $\bfy\rightarrow-\bfy$: \beq
h_{a,b}^{(\lambda)}(\mathbf{x,-y})=(-1)^{b}h_{a,b}^{(\lambda)}
(\bfxy). \eeq
\end{enumerate}
By using the ladder operator $\rmL_{\_}$$(=\rmL_1-\rmi\, \rmL_2)$,
it is easy to construct the common eigenfunctons of $\bfL^2$,
$\rmL_3$, and the parity operator from $h_{a,b}^{(\lambda)}$:
\begin{equation}
h_{a,b;k}^{(\lambda)}=\rmL_{\_}^{k}h_{a,b}^{(\lambda)}.
\end{equation}
It is simple to verify that the  set of functions
\begin{equation}
\left \{ h_{l-q+\lambda,q;l-m}^{(\lambda)},\, \lambda \leq q\leq l
\right \}
\end{equation}
are the common eigenfunctions for $\bfL^2$, $\rmL_3$, and the parity
operator with the eigenvalues  $l(l+1)$, $m$, and
$(-1)^{l+\lambda}$,  respectively.

\subsubsection{Wave function expansion}
For the special case of  $m=l$,  one can expand the  wave function
with parity  $(-1)^{l+\lambda}$ in the following form:
\begin{equation}\label{eq:wf0}
\Psi^{(\lambda)}=\sum_{q=\lambda}^{l}\,\psi_{q}^{(\lambda)}\,h_{l-q+\lambda,
q}^{(\lambda)} (\bfxy),
\end{equation}
where $\lambda=0$ or 1, and $\left\{\psi_{q}^{(\lambda)},\,\lambda
\leq q \leq l\right\}$  are ($l+1-\lambda$)-tuples of functions of
the rotationally invariant variables  \fff. They are the
coefficients of the vector space \hlqsxy. This expansion is unique
because of the orthogonality of \hlqsxy.

\subsubsection{Reduced \Sch~ equation}
These operators of $\Delta_{\bfx}$, $\Delta_{\bfy}$, $\nabla_{\bfx}$
and $\nabla_{\bfy}$, on $\psi(f_1,f_2,f_3)$ may be expressed in
terms of $f_i$  derivatives as {
\begin{eqnarray}\label{eq:deq}
\fl &\Delta_{\bfx}\psi=\left(4f_1\frac{\partial^2}{\partial
f_{1}^{2}}
              +4f_3\frac{\partial^2}{\partial f_1\partial f_3}
              +f_2\frac{\partial^2}{\partial f_3^{2}}
              +6\frac{\partial}{\partial f_1}\right)\psi,
\nonumber \\
\fl &\Delta_{\bfy}\psi=\left(4f_2\frac{\partial^2}{\partial
f_{2}^{2}}
              +4f_3\frac{\partial^2}{\partial f_2\partial f_3}
              +f_1\frac{\partial^2}{\partial f_3^{2}}
              +6\frac{\partial}{\partial f_2}\right)\psi,
\nonumber \\
\fl &\Delta\,\psi=\left(\Delta_{\bfx}+\Delta_{\bfy}\right)\psi
\nonumber \\
      \fl &\qquad=\left(
             4\left(f_1\frac{\partial^2}{\partial f_{1}^{2}}
              +f_2\frac{\partial^2}{\partial f_{2}^{2}}
              \right)
             +4f_3\left(\frac{\partial^2}{\partial f_1\partial f_3}
                   +\frac{\partial^2}{\partial f_2\partial f_3}\right)
             +(f_1+f_2)\frac{\partial^2}{\partial f_3^{2}}
             +6\left(\frac{\partial}{\partial f_1}
                +\frac{\partial}{\partial f_2}\right)
             \right)\psi,
\nonumber \\
\fl &\nabla_{\bfx}\psi=\left(
               2x_1\frac{\partial}{\partial
f_1}+y_1\frac{\partial}{\partial f_3},
               2x_2\frac{\partial}{\partial
f_1}+y_2\frac{\partial}{\partial f_3},
               2x_3\frac{\partial}{\partial
f_1}+y_3\frac{\partial}{\partial f_3}
               \right)\psi,
\nonumber \\
\fl &\nabla_{\bfy}\psi=\left(
               2y_1 \frac{\partial}{\partial
f_2}+x_1\frac{\partial}{\partial f_3},
               2y_2 \frac{\partial}{\partial
f_2}+x_2\frac{\partial}{\partial f_3},
               2y_3 \frac{\partial}{\partial
f_2}+x_3\frac{\partial}{\partial f_3}
               \right)\psi.
\end{eqnarray}}
From \eref{eq:wf0} and~\eref{eq:deq},  one has {\small
\begin{eqnarray}\label{eq:laplace}
\fl&\Delta\Psi^{(\lambda)} =\sum_{q=\lambda}^{l} \left(
\Delta\psi_{q}^{(\lambda)} h_{l-q+\lambda,q}^{(\lambda)}
            +2\nabla\psi_{q}^{(\lambda)}\cdot\nabla
h_{l-q+\lambda,q}^{(\lambda)} \right)
          \nonumber \\
          \fl &\quad=
            \sum_{q=\lambda}^{l}
            \left[
            \left(\Delta+4(l-q+\lambda)\frac{\partial}{\partial f_1}
            +4q\frac{\partial}{\partial f_2}\right)\psi_{q}^{(\lambda)}
            +2(q-\lambda)\frac{\partial\psi_{q-1}^{(\lambda)}}{\partial
f_3}
            +2(l-q)\frac{\partial\psi_{q+1}^{(\lambda)}}{\partial f_3}
            \right]
            h_{l-q+\lambda,q}^{(\lambda)},\nonumber \\
            \fl
\end{eqnarray}}where $\lambda=0$ or $1$.  Substituting \eref{eq:laplace} into
the \Sch~equation, one obtains  a system of coupled partial
differential equations~(PDEs) for the $(-1)^{l+\lambda}$  parity
states, in terms of the expansion coefficients:
\begin{equation} \label{eq:reduced_sch0}
\fl \qquad-\frac{1}{\rm 2M} \left\{
\begin{array}{l}
        \Delta\psi_{q}^{(\lambda)}
        +4(l-q+\lambda)\frac{\partial\psi_{q}^{(\lambda)}}{\partial f_1}
        +4q\frac{\partial\psi_{q}^{(\lambda)}}{\partial f_2}
\\
        +2(q-\lambda)\frac{\partial\psi_{q-1}^{(\lambda)}}{\partial f_3}
        +2(l-q)\frac{\partial\psi_{q+1}^{(\lambda)}}{\partial f_3}
\end{array}
\right\} +V\psi_{q}^{(\lambda)}=E\psi_{q}^{(\lambda)},
\end{equation}where $\lambda=0$ or 1, and $\lambda \leq q \leq l$. For a given
angular momentum $l$ ,  one has $l+1$ coupled PDEs for the $(-1)^l$
parity state, and $l$  coupled PDEs for the $(-1)^{l+1}$ parity
state.  In terms of atomic spectrum, these $(-1)^{l+1}$  parity
states are called doubly excited states (DES), such as the spectra
$2p^2 \,\,{}^{3}P^e$ and $2p3d \,\,{}^{1,3}D^o$. Below we note three
special cases of interest.
\begin{enumerate}
\item The case of $l=0$ \\
In this case, the wave function is a rotationally invariant
function, and the \Sch~equation becomes \beq
-\frac{1}{2\rmM}\Delta\psi+V\psi=E\psi. \eeq
\item The even parity case of $l=1$ \\
In this case, the wave function \beq \Psi=\psi h_{1,1}^{(1)}, \eeq
and the coupled PDEs reduce to one single PDE: \beq -\frac{1}{2\rmM}
\left\{ \Delta\psi
        +4\frac{\partial\psi}{\partial f_1}
        +4\frac{\partial\psi}{\partial f_2}
        \right\}+V\psi=E\psi.
\eeq
\item The odd parity case of $l=1$ \\
In this case, the wave function \beq
\Psi=\psi_0h_{1,0}^{(0)}+\psi_1h_{0,1}^{(0)}, \eeq and the the
coupled PDEs become
\begin{eqnarray}
&-\frac{1}{2\rmM} \left\{ \Delta\psi_0
        +4\frac{\partial\psi_0}{\partial f_1}
        +2\frac{\partial\psi_1}{\partial f_3}
        \right\}+V\psi_0=E\psi_0,
\nonumber \\[0.5em]
&-\frac{1}{2\rmM} \left\{ \Delta\psi_1
        +4\frac{\partial\psi_1}{\partial f_2}
        +2\frac{\partial\psi_0}{\partial f_3}
        \right\}+V\psi_1=E\psi_1.
\end{eqnarray}
\end{enumerate}

In the above form of \eref{eq:reduced_sch0}, the rotational degrees
of freedom have been completely projected out  from the
\Sch~equation. This is achieved by expanding the wave function in
the bi-harmonic basis, \hlqsxy, leading to a set of coupled PDEs for
the expansion coefficients functions \psilq, expressed with respect
to the three rotationally invariant variables \fff.  The problem is
hence three dimensional. The number of the coupled PDE system is
noted to be finite, $l+1$ or $l$, and there are no singularities,
which is noted to contrast with the case if the Euler angles were
used.

\subsection{Expansion of reduced \Sch~equation in Jacobi-spherical
coordinates} \label{sec:Spherical}

We introduce the Jacobi-spherical coordinates \rxb~to rewrite the
reduced \Sch~equation. It will be seen that the present step
naturally leads to the introduction of the Jacobi polynomials as the
eigenfunctions of the angular differential operator, hence the
denotation of Jacobi-spherical coordinates.  It should be noted that
the Jacobi-spherical coordinate has been used by
Simonov~\cite{Simonov:1969} and Whitten~\cite{Whitten:1969}, the
latter used it to study the expression of pair potential.
Mandelzweig and
co-workers~\cite{Haftel:1989,Haftel:1991,Barnea:1990} have also used
this coordinate system to calculate some properties of the Coulomb
system.  In the present case, the Jacobi-spherical coordinates are
applied to the reduced \Sch~equation for the three-body system. As
such, it offers a natural coordinate system to delineate the
intrinsic symmetry of the reduced three-body kinetic energy
operator.  We first note some physical interpretation and properties
of this coordinate system.
\begin{enumerate}
\item The variable $\rho=\sqrt{I}$, where
$I=\sum_{j=1}^{3}m_jr_j^2$  is the moment of inertia in the
center-of-mass frame, thus $\rho$ provides a natural measurement of
the ``size" of the system.
\item For the three sets of Jacobi coordinate vectors of \eref{eq:sets},
there are three simple relations among them:
\begin{eqnarray}
   \rho^{(1)}=\rho^{(2)}=\rho^{(3)}=\rho, \nonumber
    \\
    \xi^{(1)}=\xi^{(2)}=\xi^{(3)}=\xi, \nonumber
    \\
    \beta^{(1)}=\beta^{(2)}-2\beta_{2}=\beta^{(3)}-2\beta_{3}=\beta,
\end{eqnarray}
where $\rho$ and $\xi$ are unchanged, and  $\beta$  is shifted. This
property has been used  to express ${r}_{ij}$.
\item For the permutation of particle 2 and particle 3,
$\bfy \rightarrow -\bfy$, $(\rho,\xi,\beta) \rightarrow
(\rho,\xi,-\beta)$.  This property will be  useful for us to deal
with the case  in which  there are identical particles in the
system.
\end{enumerate}
In terms of the coordinate system \rxb, one has the following
differential  relations  for  the function $\psi(f_1,f_2,f_3)$:
\begin{eqnarray}\label{eq:der}
\frac{\partial \psi}{\partial f_1}&=& \frac{1}{2\rho}\frac{\partial
\psi}{\partial \rho} -\frac{1}{\rho^2} \left[ \xi\frac{\partial
\psi}{\partial\xi} +\left( \cos\beta\frac{\partial
\psi}{\partial\xi} -\frac{\sin\beta}{\xi}\frac{\partial
\psi}{\partial\beta} \right) \right],
\nonumber \\
\frac{\partial \psi}{\partial f_2}&=& \frac{1}{2\rho}\frac{\partial
\psi}{\partial \rho} -\frac{1}{\rho^2} \left[ \xi\frac{\partial
\psi}{\partial\xi} -\left( \cos\beta\frac{\partial
\psi}{\partial\xi} -\frac{\sin\beta}{\xi}\frac{\partial
\psi}{\partial\beta} \right) \right],
\nonumber \\
\frac{\partial \psi}{\partial f_3}&=&\frac{2}{\rho^2} \left(
\sin\beta\frac{\partial \psi}{\partial\xi}
+\frac{\cos\beta}{\xi}\frac{\partial \psi}{\partial\beta} \right),
\end{eqnarray}
 and
\begin{equation}\label{eq:der_lap}
\Delta \psi=\frac{\partial^2
\psi}{\partial\rho^2}+\frac{5}{\rho}\frac{\partial
\psi}{\partial\rho} +\frac{4}{\rho^2}A^{(0)}\psi,
\end{equation}
where
\begin{equation}\label{eq:al}
A^{(0)}\psi=(1-\xi^2)\frac{\partial^2\psi}{\partial\xi^2}
+\frac{1-3\xi^2}{\xi}\frac{\partial\psi}{\partial\xi}
+\frac1{\xi^2}\frac{\partial^2\psi}{\partial\beta^2}.
\end{equation}

\subsubsection{Reduced \Sch~equations in terms of \rxb}
In general, there are two quantum states of different parities for
the special case of  $m=l$, whose wave functions are given by
$\Psi^{(\lambda)}=\sum_{q=\lambda}^{l}\psi_{q}^{(\lambda)}
h_{l-q+\lambda,q}^{(\lambda)} (\bfxy)$, where $\lambda=0$ or $1$.
Here \psilq~are $(l+1-\lambda)$-tuples of rotationally invariant
functions satisfying the systems of coupled PDEs. Substitution of
the derivative relations, \eref{eq:der} and \eref{eq:der_lap}, into
the reduced \Sch~equation, the coupled PDEs of
\eref{eq:reduced_sch0}, leads to the following coupled PDEs in the
Jacobi- spherical coordinates \rxb:
\begin{eqnarray}\label{eq:pde2}
\fl &-\frac{1}{\rm 2M} \left \{\left[ \frac{\partial^2
\psi_q^{(\lambda)}}{\partial\rho^2}+\frac{5+2l+2\lambda}
{\rho}\frac{\partial \psi_q^{(\lambda)}}{\partial\rho}\right]
+\frac{4}{\rho^2}
\left[(1-\xi^2)\frac{\partial^2\psi_q^{(\lambda)}}{\partial\xi^2}
+\frac{1-(l+\lambda+3)\xi^2}{\xi}\frac{\partial\psi_q^{(\lambda)}}
{\partial\xi}
 \right. \right.
\nonumber \\
\fl &\left. \left .
+\frac1{\xi^2}\frac{\partial^2\psi_q^{(\lambda)}}{\partial\beta^2}+
(l-2q+\lambda)B_1 \psi_q^{(\lambda)}+(q-\lambda)B_2
\psi_{q-1}^{(\lambda)}+(l-q)B_2\psi_{q+1}^{(\lambda)} \right
]\right\} +V\psi_q^{(\lambda)}=E\psi_q^{(\lambda)}.
\nonumber \\
\fl &\quad (\lambda=0,1;\lambda \leq q \leq l)
\end{eqnarray}
The above system of PDEs has remarkable simplicity and uniformly.
Notice that the differential operators involving partial derivatives
in $\rho$ are the same for each component function
 $\psi_q^{(\lambda)}$, i.e., $\frac{\partial^2
\psi_q^{(\lambda)}}{\partial\rho^2}+\frac{5+2l+2\lambda}{\rho}
\frac{\partial \psi_q^{(\lambda)}}{\partial\rho}$; while the partial
derivatives in   $\xi$ and $\beta$ can be organized into the
following two parts.
\begin{enumerate}
\item A uniform part for each component function $\psi_q^{(\lambda)}$,
denoted  $A^{(l+\lambda)}$, which comes from the angular part of the
kinetic energy operator~(see \eref{eq:al}):
\begin{equation}
A^{(l+\lambda)}=(1-\xi^2)\frac{\partial^2}{\partial\xi^2}
+\frac{1-(l+\lambda+3)\xi^2}{\xi}\frac{\partial}
{\partial\xi}+\frac1{\xi^2}\frac{\partial^2} {\partial\beta^2}.
\end{equation}
\item A second part consisting of $(l-2q+\lambda)B_1
\psi_q^{(\lambda)}+(q-\lambda)B_2
\psi_{q-1}^{(\lambda)}+(l-q)B_2\psi_{q+1}^{(\lambda)}$, where
\begin{equation}
B_1=\cos\beta \frac{\partial }{\partial
\xi}-\frac{\sin\beta}{\xi}\frac{\partial}{\partial \beta} , \quad
B_2=\sin\beta \frac{\partial }{\partial
\xi}+\frac{\cos\beta}{\xi}\frac{\partial}{\partial \beta}.
\end{equation}
This part comes from the first order derivative terms (with respect
to $f_i$ ) in \eref{eq:reduced_sch0}.
\end{enumerate}
Below we write out the reduced \Sch~equations explicitly for those
cases of particular interest.
\begin{enumerate}
\item{The case of $l=0$:}
\begin{equation}
-\frac{1}{2\rmM} \left \{ \frac{\partial^2
\Psi}{\partial\rho^2}+\frac{5}{\rho}\frac{\partial
\Psi}{\partial\rho} +\frac{4}{\rho^2}A^{(0)}\Psi \right\}
+V\Psi=E\Psi.
\end{equation}
\item{ The case of $l=1$ and of even parity:}
{
\begin{equation}
-\frac{1}{2\rmM} \left \{ \frac{\partial^2
\Psi}{\partial\rho^2}+\frac{9}{\rho}\frac{\partial
\Psi}{\partial\rho} +\frac{4}{\rho^2} A^{(2)}\Psi\right\}
+V\Psi=E\Psi.
\end{equation}
}
\item{The case of $l=1$ and of odd parity:}
\begin{eqnarray}
\fl \qquad &-\frac{1}{2\rmM} \left \{ \frac{\partial^2
\psi_0}{\partial\rho^2}+\frac{7}{\rho} \frac{\partial
\psi_0}{\partial\rho} +\frac{4}{\rho^2} \left[A^{(1)}\psi_0+B_1
\psi_0+B_2 \psi_1 \right ] \right\} +V\psi_0=E\psi_0,
\nonumber \\[0.5em]
\fl &-\frac{1}{\rm 2M} \left \{ \frac{\partial^2
\psi_1}{\partial\rho^2}+\frac{7}{\rho} \frac{\partial
\psi_1}{\partial\rho} +\frac{4}{\rho^2} \left[ A^{(1)}\psi_1-B_1
\psi_1+B_2 \psi_0 \right ] \right\} +V\psi_1=E\psi_1.
\end{eqnarray}
\end{enumerate}

\subsubsection[Eigenfunctions of the operator $A^{(l)}$]
{Eigenfunctions of the angular differential operator $A^{(l)}$}

Here we give the eigenfunctions for the following type of angular
differential operator:
\begin{equation}
A^{(l)}\psi=(1-\xi^2)\frac{\partial^2 \psi}{\partial\xi^2}
+\frac{1-(l+3)\xi^2}{\xi}\frac{\partial\psi}{\partial\xi}
+\frac1{\xi^2}\frac{\partial^2\psi}{\partial\beta^2}.
\end{equation}
Let $P_n^{(a,b)}(x)$ be the Jacobi polynomials, then the following
set of functions constitutes a complete family of eigenfunctions of
$A^{(l)}$:
\begin{equation}
\left\{ J_{\pm m,n}^{(l)}(\xi,\beta)= \rme^{\pm \rmi\, m\beta}\xi^m
P_{n}^{(\frac{l}{2},m)}(2\xi^2-1), \,\,m,n\geq 0 \right\},
\end{equation}
 with their respective eigenvalues given by
\begin{equation}
\lambda_{m,n}^{(l)}=- \left\{ 4n\left(1+\frac{l}{2}+m+n\right
)+m(l+m+2) \right\}.
\end{equation}
The details and some  properties of $J_{\pm m,n}^{(l)}$ are given in
Appendix A.

\subsubsection[Matrices $B_1$ and $B_2$]{Matrices $B_1$ and $B_2$ applied to the
eigenfunctions  of $A^{(l)}$} Applying  $B_1$ and $B_2$ on the
eigenfunctions of $A^{(l)}$, \jab, one obtains the following three
cases.
\begin{enumerate}
\item The case $m=0$:
\begin{eqnarray}
B_1 J_{0,n}^{(l)}&=&2\sum_{k=1}^{n}b_{0,n,k}^{(l)}
\left\{J_{1,n-k}^{(l)}+J_{-1,n-k}^{(l)}\right\},
\nonumber \\
\rmi\,B_2 J_{0,n}^{(l)}&=&2\sum_{k=1}^{n}b_{0,n,k}^{(l)}
\left\{J_{1,n-k}^{(l)}-J_{-1,n-k}^{(l)}\right\}.
\end{eqnarray}
\item The case of positive indices:
\begin{eqnarray}
\fl \qquad &\hspace{0.5em}B_1
J_{m,n}^{(l)}=2\left(\sum_{k=1}^{n}b_{m,n,k}^{(l)}J_{m+1,n-k}^{(l)}
+\sum_{k=0}^{n}\widetilde{b}_{m,n,k}^{(l)}J_{m-1,n-k}^{(l)}\right),
\nonumber \\[0.5em]
\fl &\rmi \, B_2
J_{m,n}^{(l)}=2\left(\sum_{k=1}^{n}b_{m,n,k}^{(l)}J_{m+1,n-k}^{(l)}
-\sum_{k=0}^{n}\widetilde{b}_{m,n,k}^{(l)}J_{m-1,n-k}^{(l)}\right).
\end{eqnarray}
\item The case of negative indices:
\begin{eqnarray}
\fl \qquad &\hspace{1.8em} B_1
J_{-m,n}^{(l)}=2\left(\sum_{k=1}^{n}b_{m,n,k}^{(l)}J_{-(m+1),n-k}^{(l)
}
+\sum_{k=0}^{n}\widetilde{b}_{m,n,k}^{(l)}J_{-(m-1),n-k}^{(l)}\right),&
\nonumber \\[0.5em]
\fl &-\rmi \,B_2
J_{-m,n}^{(l)}=2\left(\sum_{k=1}^{n}b_{m,n,k}^{(l)}J_{-(m+1),n-k}^{(l)
}
-\sum_{k=0}^{n}\widetilde{b}_{m,n,k}^{(l)}J_{-(m-1),n-k}^{(l)}\right).
\end{eqnarray}
\end{enumerate}
In the above,  the coefficients $b_{m,n,k}^{(l)}$ and
$\widetilde{b}_{m,n,k}^{(l)}$ are given by {
\begin{eqnarray}
\fl\hspace{4em} &b_{m,n,1}^{(l)}=\frac1{2}(a+m)+n,
\nonumber \\
\fl
&b_{m,n,2}^{(l)}=\frac{m+n}{a+m+n}\left(\frac1{2}(a+m)+n-1\right),
\nonumber \\
\fl & b_{m,n,3}^{(l)}=\frac{m+n}{a+m+n}\frac{m+n-1}{a+m+n-1}
\left(\frac1{2}(a+m)+n-2\right), etc.
\nonumber \\
\fl &\widetilde{b}_{m,n,0}^{(l)}=\frac{m+n}{a+m+n}
\left(\frac1{2}(a+m)+n\right),
\nonumber \\
\fl
&\widetilde{b}_{m,n,1}^{(l)}=\frac{m+n}{a+m+n}\frac{m+n-1}{a+m+n-1}
\left(\frac1{2}(a+m)+n-1\right),
\nonumber \\
\fl
&\widetilde{b}_{m,n,2}^{(l)}=\frac{m+n}{a+m+n}\frac{m+n-1}{a+m+n-1}
\frac{m+n-2}{a+m+n-2}\left(\frac1{2}(a+m)+n-2\right),etc. \nonumber
\\
\fl&
\end{eqnarray}}
where $a=\frac{l}{2}$.  The details of calculation can be found in
Appendix A. Denote
\begin{equation}
\left|m,n;l\right>=C_{m,n}^{(l)}J_{m,n}^{(l)},\quad (m\in Z,n \in
Z^+),
\end{equation}
where $C_{m,n}^{(l)}$ is the normalization constant of
$J_{m,n}^{(l)}$, $\cB_1^{(l)}$ and $\cB_2^{(l)}$ being the  matrices
of operators $B_1$ and $B_2$ on the function
space~$\left\{\left|m,n;l\right>,m\in Z,n \in Z^+\right\}$.
$\cB_1^{(l)}$ and $\cB_2^{(l)}$ have two essential properties:
\begin{enumerate}
\item For the  index $m$, both $\cB_1^{(l)}$ and $\cB_2^{(l)}$ are
tridiagonal block matrices with zero diagonal blocks.
\item For a fixed $m$, each none-zero sub-block matrix of $\cB_1^{(l)}$
and $\cB_2^{(l)}$ for the index $n$ is an upper triangular matrix.
\end{enumerate}
The above two properties will  be used to advantage in designing the
numerical scheme.

\subsubsection{Coulomb potential expressed in Jacobi-spherical coordinates \rxb}
From \eref{eq:potential0} and \eref{eq:r123}, it is easy to get the
potential in the Jacobi-spherical coordinates
\begin{equation}\label{eq:V_in_sph}
V=\frac{U(\xi,\beta)}{\rho} ,
\end{equation}
where
\begin{equation} \label{eq:V_xibeta}
\fl \qquad
 U(\xi,\beta)=\frac{Z_2Z_3{\left
(\frac{2m_2m_3}{1-m_1}\right)}^{1/2}} {\sqrt{{1+\xi \cos\beta}}}
+\frac{Z_3Z_1{\left (\frac{2m_3m_1} {1-m_2}\right)}^{1/2}}
{\sqrt{{1+\xi \cos(\beta+2\beta_{2})}}} +\frac{Z_1Z_2{\left
(\frac{2m_1m_2}{1-m_3}\right)}^{1/2}} {\sqrt{{1+\xi
\cos(\beta+2\beta_{3})}}}.
\end{equation}
From \eref{eq:V_in_sph} and \eref{eq:V_xibeta}, one can see that
\begin{enumerate}
\item The Coulomb potential is separable into  $\rho$ and $(\xi,\beta)$ components, which is crucial
for further reduction.
\item The potential has three singularities  in the $(\xi,\beta)$ space:
$(1,\pi)$, $(1,\pi-2\beta_2)$, and $(1,\pi-2\beta_3)$.  These
singularities  affect the convergence of the numerical calculations.
\end{enumerate}

\subsubsection[Matrix elements of $U(\xi,\beta)$]{Matrix elements of $U(\xi,\beta)$ with respect to the
eigenfunctions of $A^{(l)}$}

The matrix element of  $U(\xi,\beta)$ in the space $\left
\{\left|m,n;l\right>,\,m\in Z,n\in Z^+\right\}$ is defined as
\begin{eqnarray}
\fl\qquad &&{\cU^{(l)}}(m n,m^\prime n^\prime)= \int_{-\pi}^{\pi}
\int_{0}^{1} \exp{(-\rmi(m-m^\prime)\beta)}U(\xi,\beta) T_{m
n,m^\prime n^\prime}^{(l)}(\xi)\, \rmd\xi \rmd\beta,
\end{eqnarray}
where
\begin{equation}
\fl\quad T_{mn,m^\prime n^\prime}^{(l)}(\xi)
=C_{m,n}^{(l)}C_{m^\prime,n^\prime}^{(l)}
\xi^{|m|+|m^\prime|+1}(1-\xi^2)^{l/2} P_n^{(l/2,|m|)}(2\xi^2-1)
P_{n^\prime}^{(l/2,|m^\prime|)}(2\xi^2-1).
\end{equation}
Setting \beq \small U_0(\xi,\beta)=(1+\xi\cos\beta)^{-1/2}, \eeq and
denoting its matrix $\cU_0^{(l)}(mn,m^\prime n^\prime)$  in the
space  $\left \{\left|m,n;l\right>,m\in Z,n\in Z^+\right\}$  the
``universal matrix", one has
\begin{equation}\label{eq:u_matrix}
\cU^{(l)}(mn,m^\prime n^\prime)
=C(m-m^\prime)\cU_0^{(l)}(mn,m^\prime n^\prime),
\end{equation}
where
\begin{equation}
\scriptsize \fl C(n)=\sqrt{\frac{2m_2m_3}{1-m_1}}Z_2Z_3
+\sqrt{\frac{2m_3m_1}{1-m_2}}Z_3Z_1\exp(\rmi 2n\beta_{2})
+\sqrt{\frac{2m_1m_2}{1-m_3}}Z_1Z_2\exp(\rmi 2n\beta_{3})
\end{equation}
is denoted the ``modulation factor".  A few properties of the matrix
is noted below.
\begin{enumerate}
\item The physical quantities, i.e., masses and charges, appear only in the
modulation factor.  For the special case of particle 2 and particle
3 being identical~($m_2=m_3=m_0,\,Z_2=Z_3=Z$ and
$\beta_{2}=-\beta_{3}=\beta_0$), the modulation factor is  a real
number, given by
\begin{equation}
C(n)=\sqrt{m_0}Z^2+ 2\sqrt{\frac{2m_0m_1}{1-m_0}}ZZ_1\cos(2n
\beta_0).
\end{equation}
\item The universal matrix  is independent of  any specific system,
which is the nature of three-body Coulomb system.  In the numerical
scheme, this property will be used to advantage.
\end{enumerate}
In the particular case of helium-like ions with infinite nuclear
mass, we would like to note that the Hamiltonian is
\begin{equation}
H=-\frac1{2}(\Delta_1+\Delta_2)+\left(-\frac{Z}{r_1}
-\frac{Z}{r_2}+\frac1{r_{12}} \right),
\end{equation}
where $Z$ is the nuclear charge. In this case the modulation factor
becomes
\begin{equation}
C(n)=1-2\sqrt{2}Z\cos(n\pi/2).
\end{equation}

\subsection{Further reduction to a system of 1D ODEs}
\label{sec:ODEs}

The ``space-rotation reduction" enables us to completely remove the
three rotational  degrees of freedom, and reduce the \Sch~equation
to a system of PDEs solely in terms of the rotationally invariant
parameters \fff. In terms of the Jacobi-spherical coordinates \rxb,
the above reduced system of PDEs becomes well-organized, in the
sense that there can again be a separation of the angular component
with the radial component. The angular component consists of the sum
of $A^{(l+\lambda)}$ and the integral multiples $B_1$ and $B_2$. The
complete solutions of the eigen-functions of $A^{(l)}$  in terms of
the Jacobi polynomials, namely $\left \{J_{\pm
m,n}^{(l)}(\xi,\beta)\right\}$, can be explicitly obtained, as well
as analytic formulas of the matrices of the linear differential
operators $B_1$ and $B_2$ with respect to the above eigen-basis of
$A^{(l)}$. Furthermore, we provide a way of computing the matrix of
$U\left(\xi,\beta\right)$, which constitutes a separable part of the
Coulomb potential.  By combining all the above results, it is
straightforward to further reduce the system of PDEs in \rxb~to an
infinite system of ODEs in terms of $\rho$.  By writing
\begin{equation}
\psi_q^{(\lambda)}=\sum_{m=-\infty}^{\infty}\sum_{n=0}^{\infty}
f_{q,m,n}^{(\lambda)}(\rho) \left|m,n;l+\lambda\right>,\,
(\lambda=0,1;\lambda \leq q \leq l)
\end{equation}
and substituting the above into the system of PDEs \eref{eq:pde2},
we obtain a set of ODEs on the expansion coefficients.  That is, let
$\cF_q^{(\lambda)}(\rho)$ be the column vector with
$\left\{f_{q,m,n}^{(\lambda)}(\rho)\right\}$  as the components, and
let $\cB_1^{(l+\lambda)}$, $\cB_2^{(l+\lambda)}$,
$\cU^{(l+\lambda)}$ be respectively the matrices of the linear
operators  $B_1$, $B_2$ and $U(\xi,\beta)$  with respect to the
eigenfunctions of $A^{(l+\lambda)}$. Then the set
$\left\{\cF_q^{(\lambda)}(\rho),\,\lambda \leq q \leq l\right\}$
satisfies the following system of coupled ODEs:
\begin{equation}
\fl \qquad  -\frac{1}{\rm 2M} \left\{
\begin{array}{l}
\left( \frac{\partial^2}{\partial \rho^2}+\frac{5+2l+2\lambda}{\rho}
\frac{\partial}{\partial\rho} \right)\cF_q^{(\lambda)}
\\
+\frac{4}{\rho^2} \left[
\begin{array}{l}
\cA^{(l+\lambda)} \cF_q^{(\lambda)}
\\
+(l-2q+\lambda)\cB_1^{(l+\lambda)}\cF_q^{(\lambda)}
\\
+(q-\lambda)\cB_2^{(l+\lambda)}\mathcal{F}_{q-1}^{(\lambda)}
\\
+(l-q)\cB_2^{(l+\lambda)}\cF_{q+1}^{(\lambda)}
\end{array}
\right]
\end{array}
\right\}
+\frac{1}{\rho}\cU^{(l+\lambda)}\cF_q^{(\lambda)}=E\cF_q^{(\lambda)},
\end{equation}
where $\lambda=0$ or 1, $\lambda \leq q \leq l$, and
$\mathcal{A}^{(l+\lambda)}$ is the diagonal matrix of the
eigenvalues of  $A^{(l+\lambda)}$.  In particular, we note the
following special cases.
\begin{enumerate}
\item The case of $l=0$:
\begin{eqnarray}
-\frac{1}{\rm 2M} \left\{ \left( \frac{\partial^2}{\partial
\rho^2}+\frac{5}{\rho} \frac{\partial}{\partial\rho} \right) \cF
+\frac{4}{\rho^2}\cA^{(0)}\cF \right\}
+\frac1{\rho}\cU^{(0)}\cF=E\cF.
\end{eqnarray}
\item The case of $l=1$ and  even parity:
\begin{eqnarray}
-\frac{1}{\rm 2M} \left\{ \left( \frac{\partial^2}{\partial
\rho^2}+\frac{9}{\rho} \frac{\partial}{\partial\rho} \right)
\mathcal{F} +\frac{4}{\rho^2}\cA^{(2)}\cF \right\}
+\frac1{\rho}\cU^{(2)}\cF=E\cF.
\end{eqnarray}
\item The case of $l=1$ and odd parity:
\begin{equation}\label{eq:temp10}
\fl \qquad
\begin{array}{c}
-\frac{1}{\rm 2M} \left\{
\begin{array}{l}
\left( \frac{\partial^2}{\partial \rho^2}+\frac{7}{\rho}
\frac{\partial}{\partial\rho} \right)\cF_0
\\
+\frac{4}{\rho^2} \left[\cA^{(1)}\cF_0 +\cB_1^{(1)}\cF_0
+\cB_2^{(1)}\cF_1 \right]
\end{array}
\right\} +\frac1{\rho}\cU^{(1)}\cF_0=E\cF_0,
\\[1em]
-\frac{1}{\rm 2M} \left\{
\begin{array}{l}
\left( \frac{\partial^2}{\partial \rho^2}+\frac{7}{\rho}
\frac{\partial}{\partial\rho} \right)\cF_1
\\
+\frac{4}{\rho^2} \left[\cA^{(1)}\cF_1 -\cB_1^{(1)}\cF_1
+\cB_2^{(1)}\cF_0 \right]
\end{array}
\right\} +\frac1{\rho}\cU^{(1)}\cF_1=E\cF_1.
\end{array}
\end{equation}
By using the following notations
\begin{equation}
\begin{array}{lr}
\cF= \left(
\begin{array}{c}
\cF_0 \\
\cF_1
\end{array}
\right), &
 \cA= \left(
\begin{array}{cc}
\cA^{(1)} &\cO \\
\cO &\cA^{(1)}
\end{array}
\right),
\\[1em]
\cB= \left(
\begin{array}{cc}
\cB_1^{(1)} &\cB_2^{(1)} \\
\cB_2^{(1)}&-\cB_1^{(1)}
\end{array}
\right), & \cU= \left(
\begin{array}{cc}
\cU^{(1)} &\mathcal{O} \\
\mathcal{O} &\cU^{(1)}
\end{array}
\right),
\end{array}
\end{equation}
where $\mathcal{O}$ represents the zero matrix, and $\mathcal{A}$,
$\mathcal{B}$ and $\mathcal{U}$ are all block matrices, the  two
coupled equations~\eref{eq:temp10} can be abbreviated as
\begin{equation}
\fl\qquad -\frac{1}{\rm 2M} \left\{ \left(
\frac{\partial^2}{\partial \rho^2}+\frac{7}{\rho}
\frac{\partial}{\partial\rho} \right)\mathcal{F}
+\frac{4}{\rho^2}[\mathcal{A}\mathcal{F} +\mathcal{B}\mathcal{F} ]
\right\} +\frac1{\rho}\mathcal{U}\mathcal{F}=E\cF.
\end{equation}
\end{enumerate}
In the general case of $l\geq 2$,  we  can use the block-matrix
notations to write the system of ODEs in a more compact  form:
\begin{equation} \label{eq:ode0}
-\frac{1}{\rm 2M} \left\{
\begin{array}{l}
 \left( \frac{\partial^2}{\partial
\rho^2}+\frac{5+2l+2\lambda}{\rho} \frac{\partial}{\partial\rho}
\right)\cF^{[\lambda]}
\\
+\frac{4}{\rho^2} \left[\cA^{[\lambda]}\cF^{[\lambda]}
+\cB^{[\lambda]}\cF^{[\lambda]} \right]
\end{array}
\right\}
+\frac1{\rho}\cU^{[\lambda]}\cF^{[\lambda]}=E\cF^{[\lambda]},
\end{equation}
where $\lambda=0$ or 1, $\cF^{[\lambda]}$  is the collection of
$\left\{\cF_q^{(\lambda)}(\rho),\,\lambda\leq q \leq l\right\}$,
$\cA^{[\lambda]}$ is the collection of the matrices
$\cA^{(l+\lambda)}$, $\cB^{[\lambda]}$ is the collection of the
matrices $\cB_1^{(l+\lambda)}$ and  $\cB_2^{(l+\lambda)}$,  and
$\cU^{[\lambda]}$  is the collection of the matrices
$\cU^{(l+\lambda)}$. All of  $\cA^{[\lambda]}$, $\cB^{[\lambda]}$
and $\cU^{[\lambda]}$ are $(l+1-\lambda)\times(l+1-\lambda)$ block
matrices. As it turns out, these block-matrix notations are not only
convenient for analyzing the dependence of  $\cF^{[\lambda]}(\rho)$
on $\rho$,  but they also provide more insight into the structure of
the three-body system,  which will greatly benefit the numerical
calculations.

    At this point it should be noted that while the starting point of
the three-body problem is a nine-dimensional problem, in
\eref{eq:ode0} the problem has been reduced to a one-dimensional
problem, since $\cF^{[\lambda]}(\rho)$  depends only on $\rho$.
Below we provide a way to solve this one-dimensional coupled ODEs by
converting them into a linear eigenvalue problem.

\subsection{Conversion to a linear algebraic eigenvalue problem}
\label{sec:Linear}

In \eref{eq:ode0}, there is a common differential operator,
$\frac{\partial^2}{\partial
\rho^2}+\frac{5+2l+2\lambda}{\rho}\frac{\partial} {\partial \rho}$,
for each component of the vector $\cF^{[\lambda]}(\rho)$.  The
energy level $E$  of a stationary state should be negative, and if
we analyze the ``asymptotic behavior" of \eref{eq:ode0}, the
limiting equations are all reduced to  the following single ODE:
\begin{equation}\label{eq:limit}
\left(-\frac{1}{\rm 2M} \frac{\partial^2}{\partial \rho^2}-E
\right)\cF^{[\lambda]}=0\quad (E<0).
\end{equation}
\Eref{eq:limit} clearly implies the exponential asymptotic decay of
the wave functions of stationary states and, moreover, the order of
such exponential decay is given by
\begin{equation}\label{eq:k}
k=\sqrt{-{\rm 2M}E},\,\,E<0.
\end{equation}
In other words,
\begin{equation}
\cF^{[\lambda]}\sim \exp(-k\rho),\,\,\rm{as} \,\, \rho\rightarrow
\infty.
\end{equation}
Setting $x=2k\rho$,
$\cF^{[\lambda]}(\rho)=e^{-k\rho}\cY^{[\lambda]}(x)$ and
\begin{equation}
L^{(l)}=\left(\frac{\partial^2}{\partial \rho^2}
+\frac{2l+5}{\rho}\frac{\partial}{\partial \rho}-k^2\right),
\end{equation}
then
\begin{equation}
\fl \qquad \rho e^{k\rho} L^{(l+\lambda)}(\rho) =2k\left\{
x\frac{\rmd^2}{\rmd \,x^2} +(2l+2\lambda+5-x)\frac{\rmd}{\rmd\, x}
-(l+\lambda+\frac{5}{2}) \right\}\cY^{[\lambda]}.
\end{equation}

\subsubsection{Expansion of the ODE system in terms of the Laguerre polynomials.}
If we substitute the above results into \eref{eq:ode0}, the system
is transformed into a system of ODEs with respect to the new
independent variable $x$:
\begin{equation} \label{eq:ode1}
\fl \qquad \widetilde{L}^{(l+\lambda)}\cY^{[\lambda]}
-\left(l+\lambda+\frac{5}{2}\right) \cY^{[\lambda]}+\frac{4}{x}
\left(\cA^{[\lambda]}\cY^{[\lambda]}+\cB^{[\lambda]}\cY^{[\lambda]}
\right) =\frac{\rm M}{k}\cU^{[\lambda]}\cY^{[\lambda]},
\end{equation}
where $\lambda=0$ or 1, and
\begin{equation}
\widetilde{L}^{(l)}=x\frac{\rmd^2}{\rmd x^2} +(2l+5-x)\frac{\rmd}
{\rmd x}.
\end{equation}
These Laguerre polynomials \lpx~are the eigenfunctions of
$\widetilde{L}^{(l)}$, i.e.,
\begin{equation}
\label{eq:lag} \left(x\frac{\rmd^2}{\rmd
x^2}+(2l+5-x)\frac{\rmd}{\rmd x}
\right)L_{p}^{(2l+4)}=-pL_{p}^{(2l+4)}.
\end{equation}
\Eref{eq:lag} shows that it is advantageous to use the family of
Laguerre polynomials \lpxlam~as the basis functions to solve
\eref{eq:ode1}.

\subsubsection{Infinite linear equations for the coefficients.}

Let $v_p^{(\lambda)}$ be the undetermined coefficients in the
expression of $\cY^{[\lambda]}(x)$  as the linear combination of the
Laguerre polynomials \lpxlam:
\begin{equation}
\cY^{[\lambda]}(x)=\sum_{p=0}^{\infty}\,v_p^{(\lambda)}
\,L_{p}^{(2l+2\lambda+4)}(x).
\end{equation}
Here  $v_p^{(\lambda)}$ is more precisely defined as the set
$\left\{v_{q,p,m,n}^{(\lambda)},\,\lambda\leq q\leq l\right\}$. By
omitting the sub-indices $q$, $m$ and $n$ in the block-matrix
notations, we obtain
\begin{equation}
\scriptsize \fl \qquad \left\{
\begin{array}{l}
-\sum_{p=0}^{\infty}\, \left(p+l+\lambda+\frac{5}{2}\right)
v_p^{(\lambda)}\,L_p^{(2l+2\lambda+4)}\cI_{\cA^{[\lambda]}}
\\
+
\sum_{p=0}^{\infty}\,\frac{4}{x}\,v_p^{(\lambda)}\,L_p^{(2l+2\lambda+4)}
\,\left[\cA^{[\lambda]}+\cB^{[\lambda]}\right]
\end{array}
\right\} = \frac{\rm
M}{k}\sum_{p=0}^{\infty}\,v_p^{(\lambda)}\,L_p^{(2l+2\lambda+4)}
\,\cU^{[\lambda]},
\end{equation}
where $\lambda=0$ or 1, and $\cI_{\cA^{[\lambda]}}$ is the identity
matrix, with the same rank as $\cA^{[\lambda]}$.    Multiplying the
above equations by $kx$ and then making use of the recurrence
relations of the Laguerre polynomials, we get the systems of linear
relations in terms of the Laguerre polynomials.  By applying the
orthogonality relations of these polynomials, a system of algebraic
linear equations for the coefficients
$\left\{v_{q,p,m,n}^{(\lambda)}\right\}$ is obtained. In the
block-matrix notations, they may be expressed as
\begin{equation} \label{eq:linear}
\scriptsize{
 \fl  \quad k \left\{
\begin{array}{l}
\left[
\begin{array}{l}
p \left(p+l+\lambda+\frac{3}{2}\right)v_{p-1}^{(\lambda)}
\\
-2\left(p+l+\lambda+\frac{3}{2}\right)v_p^{(\lambda)}
\\
+\left(p+l+\lambda+\frac{7}{2}\right)\left(p+2l+2\lambda+5\right)
v_{p+1}^{(\lambda)}
\end{array}
\right] \cI_{\cA^{[\lambda]}}
\\
+4v_p^{(\lambda)}\left(\cA^{[\lambda]} +\cB^{[\lambda]}\right)
\end{array}
\right\} ={\rm M} \left\{
\begin{array}{l}
-pv_{p-1}^{(\lambda)}
\\
+\left(2p+2l+2\lambda+5\right) v_p^{(\lambda)}
\\
-\left(p+2l+2\lambda+5\right)v_{p+1}^{(\lambda)}
\end{array}
\right\} \cU^{[\lambda]} }
\end{equation}
where $\lambda=0$ or 1. Let  $\cD_P^{(l)}$ and $\cG_P^{(l)}$ be the
tridiagonal matrices related to the index $p$, and $\cI_P^{(l)}$ be
the identity matrix with the same size of $\cD_P^{(l)}$ and
$\cG_P^{(l)}$, where
\begin{eqnarray}
\cD_P^{(l)}(p,p-1)&=&p\left(p+l+\frac{3}{2}\right),
\nonumber \\
\cD_P^{(l)}(p,p)&=&-2\left(p+l+\frac{5}{2}\right)^2,
\nonumber \\
\cD_P^{(l)}(p,p+1)&=&
\left(p+l+\frac{7}{2}\right)\left(p+2l+5\right).
\nonumber \\[.5em]
\cG_P^{(l)}(p,p-1)&=&-p,
\nonumber \\
\cG_P^{(l)}(p,p)&=&(2p+2l+5),
\nonumber \\
\cG_P^{(l)}(p,p+1)&=&-(p+2l+5).
\end{eqnarray}
Equation~(\ref{eq:linear}) can be re-written into a compact form:
\begin{equation}\label{eq:linear20}
k\left\{
\begin{array}{l}
\cD_P^{(l+\lambda)}\otimes\cI_{\cA^{[\lambda]}}
\\
+4\cI_P^{(l+\lambda)} \otimes(\cA^{[\lambda]}+\cB^{[\lambda]})
\end{array}
\right\}v^{(\lambda)} ={\rm M} \left\{\cG_P^{(l+\lambda)} \otimes
\cU^{[\lambda]}\right\}v^{(\lambda)},
\end{equation}
where the symbol $\otimes$ denotes outer-product between two
matrices. Three points should be noted.
\begin{enumerate}
\item Equation~(\ref{eq:linear20}) has a robust structure,
mainly due to the fact that $\rho$ and $(\xi,\beta)$ are separable
in the potential. $\cD_P^{(l+\lambda)}$ and $\cG_P^{(l+\lambda)}$
are unchanged when there are identical particles in the system.
\item The left-hand  of  (\ref{eq:linear20}) can be partitioned
into two parts, namely
\begin{equation}
\fl \qquad \cM_0=\cD_P^{(l+\lambda)} \otimes
\cI_{\cA^{[\lambda]}}+4\cI_P^{(l+\lambda)} \otimes
\cA^{[\lambda]},\quad \cM_1=4\cI_P^{(l+\lambda)} \otimes
\cB^{[\lambda]},
\end{equation}
where $\cM_0$ is a super-tridiagonal matrix that dominates the left.
Consequently, we can solve the linear equation, $\cM_0x=y$, just
like solving a  tridiagonal equation. Furthermore, the equation
$(\cM_0+\cM_1)x=y$ can also be solved based on an iteration process,
 detailed in the next section.
\item The outer-product form  provides a way to  easily and efficiently
get  the matrix-on-vector product.
\end{enumerate}

\section{Numerical scheme}
\label{sec:scheme}

\Sch~equation for the Coulomb three-body system has been reduced
from a nine dimensional problem to a one dimensional problem and
finally to a linear algebraic problem in \sref{sec:formulation}.
Here we present a numerical scheme designed to solve the relevant
linear equations. The core of the scheme is the iterative solution
of the matrix-eigenvalue problem. The advantage of the iteration
procedure is that it only requires the matrix-vector product, which
can significantly facilitate the numerical calculations due to the
sparse-block structure of the matrices in our problem.  The general
linear eigenvalue problem is written into a standard form, and by
applying a special integration rule, presented in Appendix B, the
calculation of the potential energy~(Coulomb interaction) matrix
elements and the product of that matrix with a vector are combined
together, to  achieve a high degree of numerical accuracy with
minimal computational resources. In present numerical calculations,
the aim is to solve a system of sparse linear equations, and an
iteration solver is adopted for that purpose. Since truncation is
necessary, an extrapolation procedure is applied to accelerate the
convergence.

\subsection{Solution of the linear eigenvalue problem}

After introducing the Jacobi polynomials, the system of PDEs is
reduced to a system of ODEs, and the ODE system is further reduced
to a linear eigenvalue problem through expansion in terms of the
Laguerre polynomials. In the block-matrix notations, the linear
algebraic equation can be expressed as
\begin{equation}\label{eq:linear2}
\fl \hspace{3em}
k\left\{\cD_P^{(l+\lambda)}\otimes\cI_{\cA^{[\lambda]}}
+4\cI_P^{(l+\lambda)} \otimes
\left(\cA^{[\lambda]}+\cB^{[\lambda]}\right) \right\}v^{(\lambda)}
={\rm M} \left\{\cG_P^{(l+\lambda)} \otimes
\cU^{[\lambda]}\right\}v^{(\lambda)},
\end{equation}
where the symbols' definitions can be found in
\sref{sec:formulation}, and $k$ is related to the energy $E$ by
\begin{equation}
k=\sqrt{-2{\rmM}E},
\end{equation}
where $\rmM$ is the total mass of three particles. \Eref{eq:linear2}
may be compactly expressed as
\begin{equation}\label{eq:general}
k_M\cM_Lv=\cM_Rv,
\end{equation}
where $k_M=k/\rmM$, and
\begin{eqnarray}
\cM_L&=&\cD_P^{(l+\lambda)}\otimes\cI_{\cA^{[\lambda]}}
+4\cI_P^{(l+\lambda)} \otimes
\left(\cA^{[\lambda]}+\cB^{[\lambda]}\right),
\nonumber \\[0.5em]
\cM_R&=&\cG_P^{(l+\lambda)} \otimes \cU^{[\lambda]}.
\end{eqnarray}
In general, only the low-lying states are of interest, which implies
only a few of the largest eigenvalues of \eref{eq:general} are
needed. One can write this general eigenvalue problem in a standard
form:
\begin{equation}\label{eq:std}
\cM_L^{-1}\cM_Rv=k_Mv,
\end{equation}
where $\cM_L^{-1}$ means the inverse of $\cM_L$.  Hence, the task
becomes to design a scheme to find a few of the largest eigenvalues
of \eref{eq:std}.

We use one of the software packages, such as ARPACK~\cite{ARPACK},
designed to compute a few eigenvalues and their corresponding
eigenvectors of a general n-by-n matrix  $A$.  It is most
appropriate for large sparse or structured matrices where structured
means that a matrix-vector product $w\leftarrow Av$ requires order
$n$,  rather than the usual order  $n^2$,  floating point
operations.  This particular software routine is based on an
algorithmic variant of the Arnoldi process called the Implicitly
Restarted Arnoldi Method.  By using this package, only the operation
of the matrix-vector product is needed.  In our problem, the
following two steps are used to calculate the matrix-vector product
$\cM_L^{-1}\cM_Rv$.
\begin{enumerate}
\item First calculate  $y\leftarrow \cM_Rv$.  This step is the most
time-consuming part, and the block-matrix structure of $\cM_R$ is
found to be suitable for distributed computing systems, such as  PC
clusters.
\item  Calculate $y\leftarrow \cM_L^{-1}y$.  The difficult part of this
step is to
solve a large sparse equation, and a sparse solver is adopted.
\end{enumerate}
Below we detail each of the two steps.

\subsubsection{Matrix-vector product ($\cM_Rv$) evaluation}

The matrix $\cM_R$, derived from the Coulomb potential, is the
tensor formed from the outer product of $\cG_P^{(l+\lambda)}$ and
$\cU^{[\lambda]}$:
\begin{equation}
\cM_R=\cG_P^{(l+\lambda)}\otimes\cU^{[\lambda]},
\end{equation}
where $\cG_P^{(l+\lambda)}$ is a tridiagonal matrix, and
$\cU^{[\lambda]}$ is a block-diagonal matrix in which every
sub-block is the same, denoted $\cU^{(l+\lambda)}$.

The Kronecker tensor product, $X \otimes Y $, of two matrices is
a larger matrix formed from all possible products of the elements of
$X$ with those of  $Y$.  If  $X$ is m-by-n and  $Y$ is p-by-q, then
$X \otimes Y $ is  an mp-by-nq matrix.  The elements are arranged in
the following order:
\begin{equation}
X\otimes Y=\left (
\begin{array}{lccr}
X_{1,1}*Y&X_{1,2}*Y& \ldots&X_{1,n}*Y \\
X_{2,1}*Y&X_{2,2}*Y& \ldots&X_{2,n}*Y\\
\vdots & \vdots &\ddots &\vdots \\
X_{m,1}*Y&X_{m,2}*Y& \ldots&X_{m,n}*Y
\end{array}
\right ).
\end{equation}
For the matrix-vector product $\{X \otimes Y\}\mathbf{v}$, where
$\mathbf{v}$ is a vector, it is advantageous to reshape $\mathbf{v}$
to a q-by-n matrix $V$, and to calculate $R=(Y*V)*X^{T}$.  The
vector reshaped from the p-by-m matrix $R$ is the final result.

The particular structure of $\cM_R$ and $\cU^{[\lambda]}$ makes it
easy to calculate the matrix-vector product.  The tensor form of
$\cM_R$ provides a straightforward way to calculate the
matrix-vector product, and the block structure of $\cU^{[\lambda]}$
lets one to focus on only the sub-block matrix  $\cU^{(l+\lambda)}$.
The elements of matrix $\cU^{(l)}$ are expressible as
\begin{equation}\label{eq:u_matrix}
\cU^{(l)}(mn,m^\prime n^\prime)
=C(m-m^\prime)\cU^{(l)}_0(mn,m^\prime n^\prime),
\end{equation}
where both the ``modulation factor" $C(n)$ and $\cU^{(l)}_0$ the
``universal matrix" are given previously.  From the definition \beqa
\fl \hspace{3em} &\cU^{(l)}_0(mn,m^{\prime}n^{\prime})
=2\int_{\beta=0}^{\pi}\!\!\int_{\xi=0}^1
\frac{\cos{(m-m^\prime)\beta}}{\sqrt{1+\xi \cos\beta}}
w_l(\xi)\,P_{mn}^{(l)}(\xi)\,P_{m^\prime n^\prime}^{(l)}(\xi)\,
\rmd\xi \rmd\beta, \eeqa where
\begin{equation} w_l(\xi)=\xi(1-\xi^2)^{l/2},\quad
P_{mn}^{(l)}(\xi)=C_{m,n}^{(l)}\xi^{|m|} P_n^{(l/2,|m|)}(2\xi^2-1) ,
\end{equation}
with $\left\{P_n^{(l/2,|m|)}(x)\right\}$ being the set of Jacobi
polynomials.  Set
\begin{equation}
u_m(\xi)=\int_0^\pi {\cos m\beta}
\sqrt{\frac{1-\xi}{1+\xi\cos\beta}}\,\rmd\beta,
\end{equation}
and
\begin{equation}
U_m^{(l)}(\xi)=\frac{2C(m)w_l(\xi)u_m(\xi)}{\sqrt{1-\xi}},
\end{equation}
then \eref{eq:u_matrix} becomes
\begin{equation}
\label{eq:ummnn} \cU^{(l)}(mn,m^\prime n^\prime)=\int_0^1
U_{m-m^\prime}^{(l)}(\xi)\,P_{mn}^{(l)}(\xi)\,P_{m^\prime
n^\prime}^{(l)}(\xi)\,\rmd\xi.
\end{equation}
The integration implied by \eref{eq:ummnn} is performed using the
IMT integration scheme~(see Appendix B).

\subsubsection{Calculation of $\cM_L^{-1} \mathbf{v} $}

One needs to solve a linear algebraic equation to get the vector
$\cM_L^{-1} \mathbf{v} $.  From the previous discussion, it is known
that $\cM_L$ is composed of two parts:
\begin{equation}
\cM_L=\cM_0+\cM_1,
\end{equation}
where
$\cM_0=\cD_P^{(l+\lambda)}\otimes \cI_{\cA^{[\lambda]}}
+4\cI_P^{(l+\lambda)} \otimes \cA^{[\lambda]}$ is a
super-tridiagonal matrix that dominates the matrix $\cM_L$, and
$\cM_1=4\cI_P^{(l+\lambda)}\otimes\cB^{[\lambda]}$ is relatively
small. Because of the super-tridiagonal structure of $\cM_0$, one
can obtain the vector $\cM_0^{-1} \mathbf{v}$ exactly like solving a
tridiagonal equation.  Obtaining $\cM_L^{-1} \mathbf{v}$ may be
implemented as
\begin{equation}
\label{eq:Ml_inv}
\cM_L^{-1}\mathbf{v}=(\cI+\cM_0^{-1}\cM_1)^{-1}\cM_0^{-1}\mathbf{v},
\end{equation}
where $\cI$ represents the identity matrix, through the following
two steps.
\begin{enumerate}
\item First solve for
\begin{equation}
\mathbf{y} \longleftarrow \cM_0^{-1} \mathbf{v}.
\end{equation}
As described above, this step can be easily implemented.
\item  Solve for
\begin{equation}
\mathbf{y} \longleftarrow (\cI+\cM_0^{-1}\cM_1)^{-1} \mathbf{y}.
\end{equation}
Since the matrix $\cM_0$ dominates,  thus $\cM_0^{-1}\cM_1$ is small
in some sense, so one can consider  the matrix $\cI+\cM_0^{-1}\cM_1$
to be close to the identity matrix, i.e.,
\begin{equation}
\cI+\cM_0^{-1}\cM_1\sim \cI+\epsilon.
\end{equation}
Thus an iteration solver should be efficient. In our calculations,
the Conjugate-Gradients-Squared~(CGS)~\cite{Sonneveld:1989} method
is taken as the solver. The CGS method just requires the user to
provide the matrix-vector product, i.e., one  works out the vector
\begin{equation} (\cI+\cM_0^{-1}\cM_1)
\mathbf{v}
\end{equation}
during the intermediate iteration steps by first calculating
$\mathbf{t}\leftarrow\cM_1 \mathbf{v}$, and then calculating
$\mathbf{t}\leftarrow\cM_0^{-1} \mathbf{t}$.  The final result is
given by  $\mathbf{v}+\mathbf{t}$. In practice, after about 12
iterations, the relative error of the solution is reduced to
$10^{-14}$.
\end{enumerate}

Since the infinite matrices in the problem are truncated into finite
matrices, the truncation is done by following a rule, specified
below. Results obtained from smaller scale calculations, i.e., the
eigenvalues and their eigenvectors, are taken as the initial guesses
for the subsequent, increasingly larger cases.  This process fully
takes the advantage of the previous results, and significantly
speeds up the calculation.  It also provides a sequence of data from
which one can apply the extrapolation procedure~(see below) to
obtain a few more digits of accuracy.

\subsection{Truncation and extrapolation procedures}
In our numerical calculations, the expansion of the partial wave
function $\psi_{q}^{(\lambda)}$ has to be truncated, i.e.,
\begin{equation}
\psi_{q}^{(\lambda)}=\rme^{-k\rho}\sum_{m=-M}^{M}\sum_{n=0}^{N}\sum_{p=0}^{
P}v_{q,m,n,p}^{(\lambda)}J^{(l+\lambda)}_{m,n}(\xi,\beta)L_p^{(2l+2\lambda
+4)}(2k\rho),
\end{equation}
where $M$, $N$ and $P$ represent the three truncation indices for
the $m$, $n$ and $p$, respectively. One solves the finite
dimensional algebraic eigenvalue equation to obtain the lowest order
eigenvalues and their corresponding eigenvectors.  These eigenvalues
are functions of $M$, $N$ and $P$, i.e.,
\begin{equation}
E=E(M,N,P),
\end{equation}
where the desired answer is the limiting value
\begin{equation}
E_{\infty}=E(\infty,\infty,\infty).
\end{equation}

\subsubsection{Truncation path}
We have carefully checked the dependence of $E$ on the three numbers
$M$, $N$ and $P$.  In our procedure, we chose a ``path" in the
$\{M,N,P\}$  space to approach the limit $E_{\infty}$.  First, it
was found that for sufficiently large $P$~(typically $P\sim 70$),
increase in $P$ does not provide any significant improvement on $E$.
This behavior owes to the fact that $P$ controls how many terms in
the Laguerre polynomials are included in the expansion, but there is
a direct relation between the spatial extend of the system and the
spherical variable $\rho$, so for large enough $P$ the terms of
Laguerre polynomials in the expansion are sufficient to delineate
the spatial domain.  Different states have different ``appropriate"
$P$  values, and after that value is determined, denoted $P_0$, the
following ``path" in $\{M,N\}$ space is taken:
\begin{equation}
M=M_0+20K,\quad
N=N_0+10K,\,\quad K=0,1,\cdots ,
\end{equation}
where $M_0$ and $N_0$ are the
initial truncations for $m$ and $n$, generally in the range of 70
and 50, respectively. The three-dimensional sequence $E(M,N,P)$ is
thus mapped to a one-dimensional sequence:
\begin{equation}
E(K)=E(M_0+20K,N_0+10K,P_0).
\end{equation}
There is a reason why we increase
$M$ twice as fast as $N$. It is known that the eigenvalues of
operator $A^{(l)}$ have the form
\begin{equation}
\lambda_{m,n}^{(l)}=- \left\{
4n\left(1+\frac{l}{2}+m+n\right )+m(l+m+2) \right\}.
\end{equation}
As $m$ and
$n$ approach infinity, $\lambda_{m,n}^{(l)}$ has the asymptotic form
\begin{equation}
\lambda_{m,n}^{(l)} \propto -(m+2n)^2,
\end{equation}
so that the
stipulated sequence is designed to coincide with this asymptotic
behavior.

\subsubsection{Extrapolation procedure}
For extrapolation, an auxiliary variable $x$, defined as \beq
x=\frac{100}{M+2N}=\frac{100}{M_0+2N_0+40K}, \eeq is used to
determine the form of $E(x)$  as provided by the data sequence. A
power law form is assumed, such that \beq
E(x)=E_{\infty}-c_0x^a(1+c_1x+c_2x^2+\cdots). \eeq The task is to
identify the leading term $x^a$.  Sequences of $E(k)\,(E(x))$  are
collected in the numerical calculations.  The leading term is
identified by studying the relation of $E(x)$ versus $x^{a_t}$,
where $a_t$ is an estimated  value of $a$.  If  the  plot is a
straight line, the leading term can be immediately identified. The
process of Richardson extrapolation~\cite{Brezinski:1991}  is
then performed on the sequence $\{E(x_i),i=0,1,\cdots\}$ to estimate
the limiting value $E_{\infty}$.  The Richardson extrapolation
utilizes the following sequences:
\begin{eqnarray}
\label{eq:extrap}
&&E(x_0)=E_{\infty}-c_0x_0^a-c_1x_0^{a+1}-c_2x_0^{a+2}-\cdots
\nonumber
\\
&&E(x_1)=E_{\infty}-c_0x_1^a-c_1x_1^{a+1}-c_2x_1^{a+2}-\cdots
\nonumber
\\
&&E(x_2)=E_{\infty}-c_0x_2^a-c_1x_2^{a+1}-c_2x_2^{a+2}-\cdots
\nonumber
\\
&&\cdots \nonumber \\
&&E(x_n)=E_{\infty}-c_0x_n^a-c_1x_n^{a+1}-c_2x_n^{a+2}-\cdots,
\end{eqnarray}
to approach $E_{\infty}$.  By defining
\begin{equation}
g_1(n)=x_n^a,\,g_2(n)=x_n^{a+1},\,g_3(n)=x_n^{a+2},\cdots,
\end{equation}
and denoting $E(x_n)$ as $E_n$, \eref{eq:extrap} becomes
\begin{equation}
E_n=E_{\infty}-c_0g_1(n)-c_1g_2(n)-c_2g_3(n)-\cdots,\,\,n=0,1,2,\cdots.
\end{equation}
Then the recursive  $E-algorithm$ declares that the limiting value
$E_{\infty}$ can be approached  by the following rule. Let
\begin{eqnarray}
&&E_0^{(n)}=E_n,\quad n=0,1,\cdots \nonumber \\
&&g_{0,i}^{(n)}=g_i(n),\quad n=0,1,\cdots\, \textrm{and}
\,i=1,2,\cdots.
\end{eqnarray}
For $k=1,2,\cdots$ and $n=0,1,\cdots$ one has
\begin{equation}
E_k^{(n)}= E_{k-1}^{(n)}-\frac{E_{k-1}^{(n+1)}-E_{k-1}^{(n)}}
{g_{k-1,k}^{(n+1)}-g_{k-1,k}^{(n)}}\cdot g_{k-1,k}^{(n)},
\end{equation}
where the $g_{k-1,k}^{(n)}$'s are auxiliary quantities recursively
computed by
\begin{equation}
g_{k,i}^{(n)}=g_{k-1,i}^{(n)}-\frac{g_{k-1,i}^{(n+1)}-g_{k-1,k}^{(n)}}
{g_{k-1,k}^{(n+1)}-g_{k-1,k}^{(n)}}\cdot
g_{k-1,k}^{(n)},\,i=k+1,k+2,\cdots.
\end{equation}
The sequences $\{E_k^{(n)},n=0,1,\cdots,\,\, \textrm{and}\,\, k\geq
1\}$ are more convergent than the initial sequence
$\{E_0^{(n)},n=0,1,\cdots\}$. From the convergence pattern of
$\{E_k^{(n)}\}$,  one may  determine the limiting value with high
accuracy.

\section{Results on three-body systems}
\label{sec:results}

In this section we present results on some three-body systems, and
discuss the properties of their relevant wave functions. In order to
display the data in physical space, we first give a description of
the relevant angles, as well as the procedure by which the wave
functions are re-constituted from numerical data.

\subsection{Euler angles and wave functions}

\begin{figure}
\centering
\includegraphics [width=0.8\textwidth]{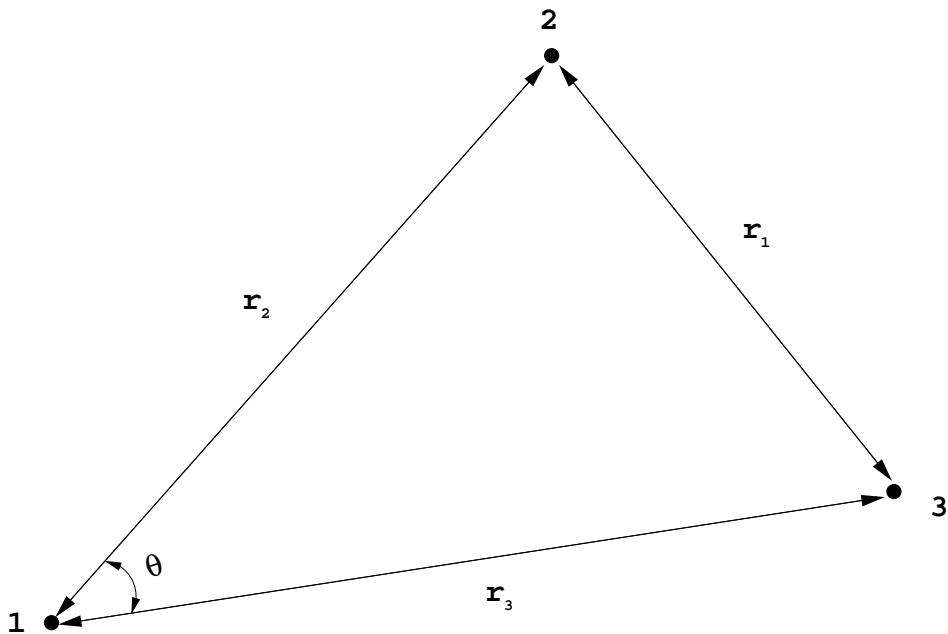}
\caption {\label{fig:r123} The triangle formed by three particles.}
\end{figure}

\subsubsection{Euler angles and the distribution function}

The three-body  wave function depends upon the shape of the triangle
formed by three particles and the three Euler
angles~\cite{Zare:1998}, such that
\begin{equation}
\Psi=\Psi(\bigtriangleup;\Omega),
\end{equation}
where $\bigtriangleup$ represents  the triangle which can be
described by the three spherical variables \rxb, or its three edges
$(r_1,r_2,r_{3})$, or $(r_2,r_3,\cos \theta)$ as shown in
\fref{fig:r123}, and $\Omega$ denotes the Euler angles that depict
the rotational orientation of the triangle in space. Here $r_2$ and
$r_3$  denote the distances of particles 2 and 3 from particle 1,
respectively, and $r_1$ denotes the distance between the two
identical particles 2 and 3.  The $\cos \theta$   is between
particles 2 and 3~(the identical pair). For a six-dimensional
function, visualization is an issue.   The Euler angles are noted to
take part in the wave function through the angular momentum
eigenfunctions, but do not enter in the rotation-invariant
functions, which indicates that in order to deal with the Euler
angles, we just need to focus on the angular momentum eigenfunctions
\hlqsxy.  After the Euler angles are integrated out from the wave
function, a three-variable probability distribution function is
obtained, i.e.,
\begin{equation}
{\rm{P}}(\bigtriangleup) \equiv \int_{\Omega}
|\Psi(\bigtriangleup;\Omega)|^2 \rmd\Omega,
\end{equation}
which can be easier to visualize. Let
${\rm{P}}(\bigtriangleup)=g(r_1,r_2,r_{3})$. In order to display the
information contained in $g(r_1,r_2,r_{3})$, we propose to use the
following probability distribution functions~(PDF).
\begin{enumerate}
\item One-variable PDF, e.g.,  the distribution of $r_{1}$:
\begin{equation}
{\rm{P}}(r_{1})dr_{1}\equiv\int_{r_2}\int_{r_3}g(r_1,r_2,r_{3})
\rmd\tau,
\end{equation}
where $d\tau$ is the volume element.
\item Two-variable PDF, e.g.,  the radial correlation function:
\begin{equation}
{\rm{P}}(r_2,r_{3})\rmd r_2\rmd
r_{3}\equiv\int_{r_1}g(r_1,r_2,r_{3})\rmd\tau,
\end{equation}
or the conditional distribution
\begin{equation}
{\rm{P}}(r_2,r_{3};r_1=a)\rmd r_2 \rmd r_{3}\equiv
g(r_1=a,r_2,r_{3})\rmd\tau/\rmd r_1,
\end{equation}
where $a$ is a given value, and a PDF is obtained for  each given
$a$.
\end{enumerate}
We noted that  the volume element $\rmd \mathbf{x} \rmd \mathbf{y}$
in different coordinate systems has the following explicit form:
\begin{equation}
\fl \hspace{3em} \rmd\mathbf{x} \rmd\mathbf{y}\propto \rho^5\xi
\rmd\rho \rmd\xi \rmd\beta \rmd\Omega \propto r_2^2r_3^2 \rmd r_2
\rmd r_3 \rmd\cos \theta \rmd\Omega=r_1r_2r_{3} \rmd r_1 \rmd r_2
\rmd r_{3} \rmd\Omega.
\end{equation}
In calculations, we always normalize  the wave function, i.e.,
\begin{equation}
\int|\Psi|^2\rho^5\xi \, \rmd\rho \rmd\xi \rmd\beta \rmd\Omega=1.
\end{equation}
If there are two identical particles in the three-body system, we
identify them as particles 2 and 3.

\subsection{Negative hydrogen ion}

The negative hydrogen ion $\textrm{H}^-$  is an interesting special
case of helium-like ions~(Z=1). It is marginally  stable against
dissociation into a neutral hydrogen atom plus a free electron. The
dissociation energy $J$ of $\textrm{H}^-$ ground state is only about
$0.75 eV$ and this ion possesses no other bound
state~\cite{Hill:1977a,Hill:1977b}.

The negative hydrogen ion has been found to be of great importance
for the opacity of sun's atmosphere. The ionization potential of
$\textrm{H}^-$, $J\approx 0.75eV$, corresponding to about
$8700^\circ $~K, is only slightly higher than the temperature for
the solar atmosphere. As free electrons are released by the
ionization of the metal elements present in the gas, and since
neutral hydrogen is by far the main constituent of the solar
atmosphere, many of these electrons will be captured to form  an
abundant source of $\textrm{H}^-$.  The radiation  flux coming from
the sun's interior would be absorbed by the $\textrm{H}^-$ ions,
accompanied by their dissociation.  The  electrons released can
again be captured by $\textrm{H}^-$ atoms with the emission of
radiation, and so on. The process $\textrm{H}^-\rightleftharpoons
\textrm{H}+e^-$ is the main source of observed opacity in the solar
atmosphere.  The absorption coefficient of $\textrm{H}^-$ has been
studied extensively and used in the theory of the solar atmosphere.
In fact, discrepancies between early calculations and observational
evidence on the sun's radiation have pointed out the inaccuracies of
the calculated $\textrm{H}^-$ wave functions  available then.
Calculation of the wave function and energy of $\textrm{H}^-$ is
also of purely methodological interest, since this most
loosely-bound of all Helium-like ions provides a severe test for the
various approximation schemes.

\subsubsection{The ground state}
\begin{figure}
\centering
\includegraphics [width=0.8\textwidth]{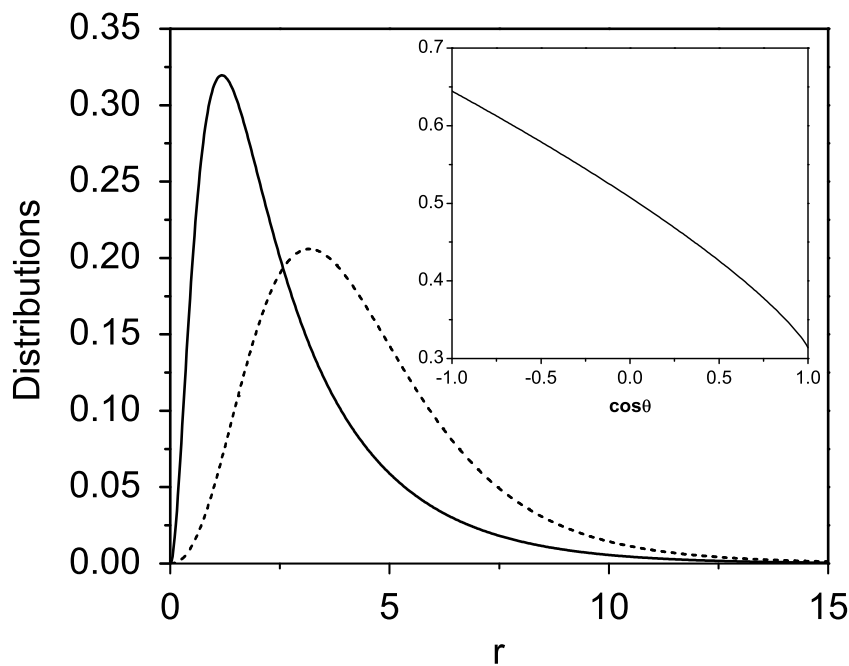}
\caption {\label{fig:HSr} Distributions of $r_1$, $r_2$~( or $r_3$)
and $\cos \theta$ of the ground state of ${}^\infty \textrm{H}^-$.
The solid line is the distribution of $r_2$ or $r_3$~(same), and the
dashed line is the distribution of $r_1$.  Inset is the distribution
of $\cos \theta$.}
\end{figure}
Bethe was the first to give an unambiguous proof of this ion as a
bound system~\cite{Bethe:1929}. Using the Hylleraas variational wave
functions, Bethe concluded that the resulting Rayleigh-Ritz upper
bound on the energy lies below -0.5~a.u..  In 1944, Chandrasekhar
introduced a two-parameter trial wave function,
\begin{equation}
\label{eq:chand}
\Psi=\exp(-\alpha r_2-\beta r_3)+\exp(-\alpha
r_2-\beta r_3),
\end{equation}
and showed that the energy minimum at  $\alpha=1.03925$ and
$\beta=0.28309$ is sufficient to provide binding for $\textrm{H}^-$.
The function shown in \eref{eq:chand} exhibits the specific nature
of electron-electron correlation in the ground state. The two
electrons are on very different  footings, one bound much closer to
the nucleus than the other, which is weakly held at a distance
 $\simeq 4-5$ from the nucleus, and this electron can be regarded as
weakly bound in a short-range attractive potential well. In modern
variational calculations, many $r_{1}^j$ terms are necessary  in the
trial wave function in order to fully account for  the
electron-electron correlation. While  many-parameter variational
calculations can give great accuracy for the ground state energy,
the best experimental values come from a high resolution~($0.03\rm
cm^{-1}$) laboratory photodetachment laser experiment. The binding
energy has been determined to be $6082.99\pm 0.15 \rm cm^{-1}$ for
$\textrm{H}^-$ and $6086.2\pm 0.6 \rm cm^{-1}$ for the similar ${\rm
D}$ states~\cite{Lykke:1991}. In \tref{table:H1S_comp}, our results
on the ground states of $^\infty \textrm{H}^-$ and $\textrm{H}^-$
are compared with other calculations. In general, all our results
have $10\sim 12$ significant figures when double precision
programming is used. The accuracy in this case is less than those
obtained variationally.  It is well known that the precision of the
operator expectation values in the variational calculations usually
has  two less significant figures than that for the energy. Thus
much more efforts is required to obtain the wave function
expectation values to the same accuracy. In our case the wave
function is calculated at the same time as the eigenvalues.  No
extra effort is required.

\begin{figure}
\centering
\includegraphics [width=0.8\textwidth]{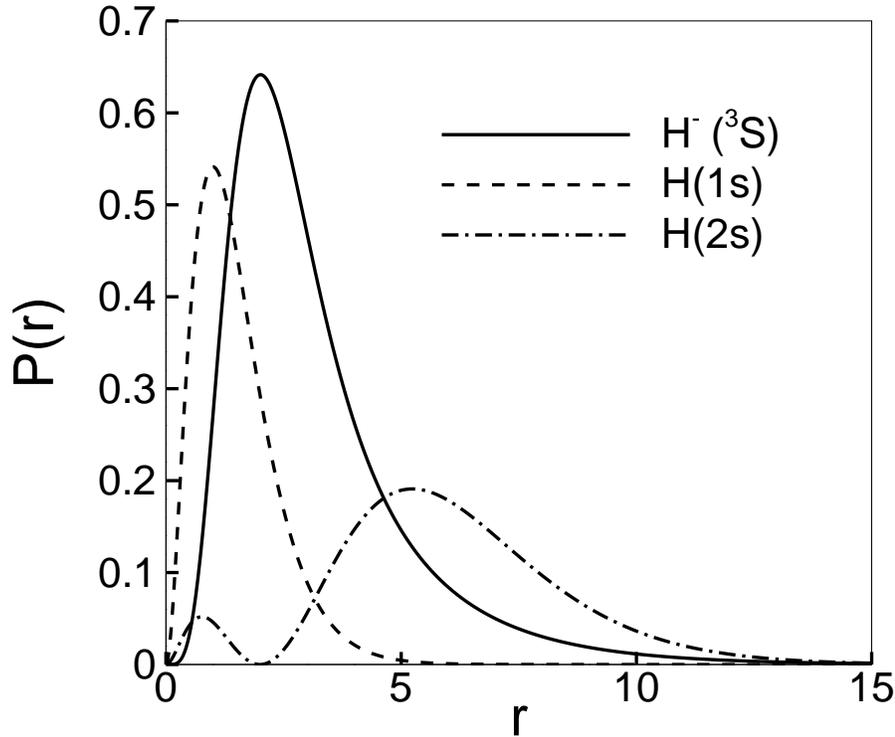}
\caption {\label{fig:HSr_comp} Radial density probability
distribution ${\rm P}(r)$ for
${}^{\infty}\textrm{H}^-$(${}^3\rm{S}$), $\textrm{H}(1s)$ and
$\textrm{H}(2s)$.}
\end{figure}

\Fref{fig:HSr} shows the distributions of  $r_1$ , $r_2$, $r_3$ and
$\cos \theta$ of the ground state of ${}^\infty \textrm{H}^-$. The
distribution of $\cos \theta$ is a curve that shows a maximum at
$\theta=\pi$.  If the shell model is valid,  the curve of the  $\cos
\theta$ distribution should be close to a horizontal line at 0.5.
For comparison, we also plot the radial density distribution ${\rm
P}(r)$, for $^{\infty}\textrm{H}^-$(${}^3\rm{S}$), $\textrm{H}(1s)$
and $\textrm{H}(2s)$ in \fref{fig:HSr_comp}.   A long tail is seen
in the distribution of $r_1$, $r_{2}$ (or $r_3$). In
\fref{fig:HSr_comp}, the electron distribution of ${}^\infty
\textrm{H}^-$ is seen to be very different from the neutral $1s$ or
$2s$ state. Thus the electron-electron correlation  plays an
important role in $\textrm{H}^-$, and independent-electron model can
not give an accurate picture.

\begin{figure}
\centering
\includegraphics [width=0.8\textwidth]{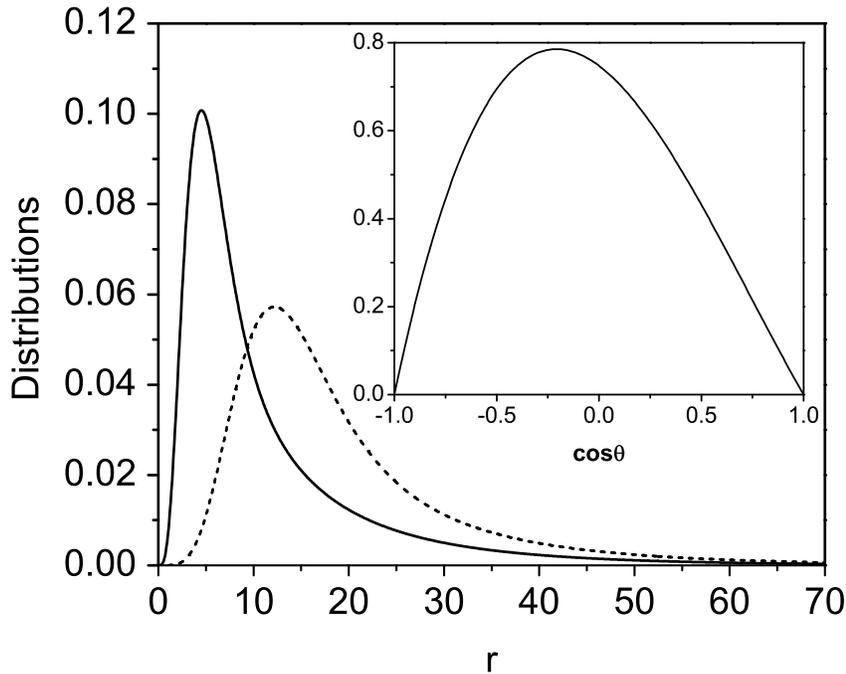}
\caption{\label{fig:HPr} Distributions of $r_1$, $r_2$~(or $r_3$)
and $\cos \theta$ of the ${}^3P^e$ state of $^\infty \textrm{H}^-$.
The solid line is the distribution of $r_2$ or $r_3$~(same),  and
the dashed line  is the distribution of $r_1$. Inset is the
distribution of $\cos \theta$.}
\end{figure}
\subsubsection{The ${}^3P^e$ state}

The even-parity  ${}^3P^e$ state of ${}^\infty \textrm{H}^-$ and
$\textrm{H}^-$ is quasi-stable, and its energy is just below the n=2
threshold (-0.125 a.u.) of the hydrogen atom.  For  the
${}^\infty\textrm{H}^-$ ion,  the existence of the even-parity
${}^3P^e$ state   was predicted computationally nearly 40 years ago,
followed by many variational calculations. J\'{a}uregui and
Bunge~\cite{Jauregui:1979} used a
configuration-interaction(CI)-expansion of 108 configuration terms
and obtained the energy -0.125 354 716 6~a.u.. They analyzed the
convergence pattern of their computations and extrapolated to -0.125
355 08(10)~a.u.. The largest Hylleraas-type computation for this
state, done by Drake and then repeated by~\cite{Banyyard:1992} gave
only -0.125 335 6~a.u., though they used about 50 thousand terms
containing powers of $r_{1}$ up to 82. The best result, as we know,
was obtained by Bylicki and Bednarz~\cite{Bylicki:2003} who applied
the Hyllerass configuration-interaction correlated expansions and
used 1442 terms to give -0.125 355 451 24~a.u.. More efforts are
needed than that for the ground state. In our approach, this problem
is reconsidered from a new viewpoint, and the finite mass of proton
is naturally taken into account.

We compare our results with other calculations in
\tref{table:H3P_comp}.  The convergence is just as good in this case
as for the ground state, and the leading term of all the sequences
is estimated to be $x^5$. Our energy eigenvalue is 0.125 355 451 242
a.u. for ${}^{\infty}\rm{H}^-$, better than most of previous
calculations.

\begin{figure}
\centering
\includegraphics [width=0.8\textwidth]{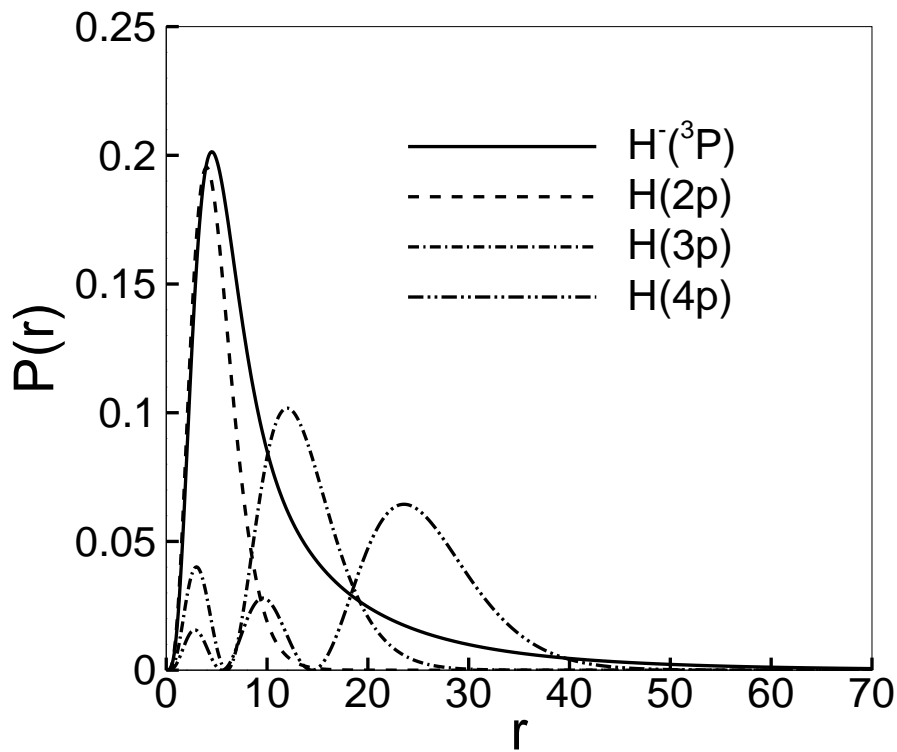}
\caption{\label{fig:HPr_comp} Radial density probability
distribution ${\rm P}(r)$ for
${}^{\infty}\textrm{H}^-$(${}^3\rm{P}$), $\textrm{H}(2p)$,
$\textrm{H}(3p)$ and $\textrm{H}(4p)$.}
\end{figure}

As a negative ion,  the extra electron is significantly affected by
the other electron. \Fref{fig:HPr} shows the distribution of $r_1$,
$r_2$, $r_3$ and $\cos \theta$ of the ${}^\infty\textrm{H}^-$. The
radial distribution has a very long tail;  even at $r=60$ the
probability does not vanish.  If the electron-electron interaction
is ignored so as to approximate this state by two $2p$-state
electrons, the charge distribution of the $2p$ state would to vanish
at about $r=15$ (see \fref{fig:HPr_comp}), very far from $r=60$.
This long tail exponentially decays with a decay length $l\sim18.3$,
with a form $r^a \exp\left(-{r}/{l}\right)$ with $a\approx 0.5$. In
this case, the independent-electron  picture is very far removed
from the present excited state.  The variational method is observed
to be less accurate for this state due to the very strong
electron-electron correlation.  For some of the expectation values,
such as $<r_i>$, comparable results can not be found in the
literature.  To our knowledge, our calculated result on this state
is probably the most accurate so far.

It is interesting to observe that in this state , the two electrons
form nearly an angle of $\pi/2$ relative to the nucleus~(the value
of $<\cos(\mathbf{r_{12}},\mathbf{r_{13}})>\approx 0$). Also, the
value of $<r_1>$ is nearly twice that of the ground state.

\subsection{Helium and helium-like ions}
\begin{figure}
\centering
\includegraphics [width=0.8\textwidth]{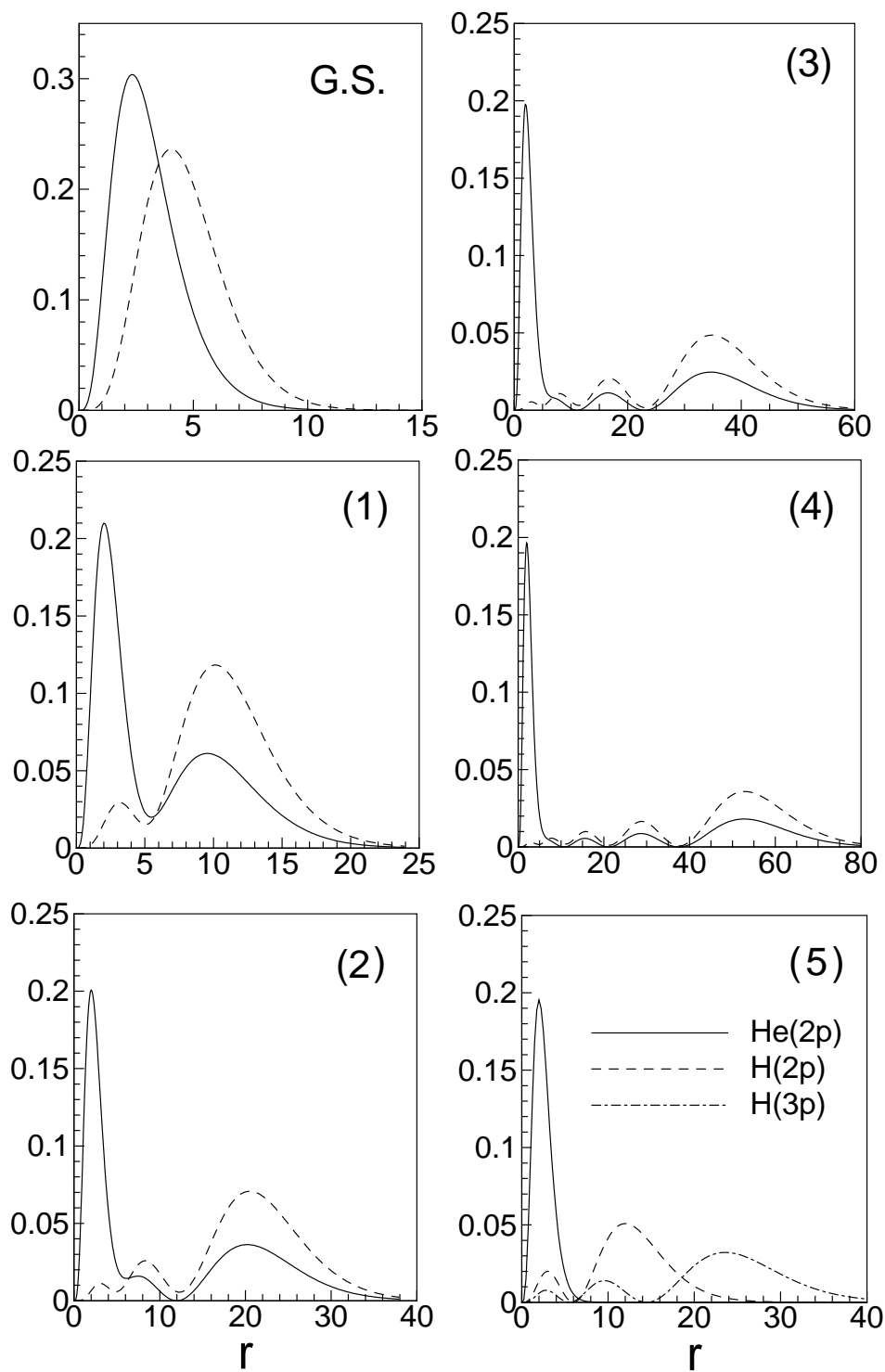}
\caption{\label{fig:He_Pr_Pr12_3Pe}Distributions of $r_1$, $r_2$~(or
$r_3$) of the five low-lying ${}^3P^e$ states of
${}^{\infty}\textrm{He}$~(ground state plus four excited states).
The solid line is the distribution of $r_2$ or $r_3$~(same), and the
dashed line is the distribution of $r_1$. For comparison, we also
plot the radial density distribution of the neutral ${\textrm
He}(2p)$, ${\textrm H}(2p)$ and ${\textrm H}(3p)$ states in (5)
(normalized to 1/2).}
\end{figure}

Except for the hydrogen atom, helium atom~(or helium-like ions) is
perhaps the simplest system in quantum mechanics. However, the
calculation of its properties is not trivial and represents a real
challenge.  It has been 80 years since the beginning of quantum
mechanics, and there are still efforts trying to understand the
system from some new perspectives.

In atomic theory, the shell model is the starting point to
understand complex atoms.  This model is an independent-electron
picture, and the interaction between  electrons are averaged by a
mean field.   For helium, we denote one of the electrons as the
inner-electron, and the other as the outer-electron.   But when the
correlation of two electrons is strong, this picture fails.  For
example, in the shell model language the two electrons in the ground
state are in the same  orbit.  Interaction would thus become
important, which would make the independent-electron picture not
suitable.  For the doubly excited states in which both electrons are
not in the ground  state, interaction can affect the slow-moving
electrons very significantly.

\subsubsection{The $S$ states of helium}
These states have been  extensively studied, and there are many
methods used to treat the $S$ states.   For extrapolation, we
estimate the leading term of  $E(x)$~(for all of the $S$ states) to
be $x^3$, and a list of the limiting values are given in
\tref{table:He13S_comp} after performing the Richardson
extrapolation.  Except for the ground state ${}^1S^e$, the other
states are close to the independent-electron picture.  In
tables~\ref{table:He1Se_properties}
and~\ref{table:He3Se_properties}, some of expectation values are
given for the ${}^{1,3}S^{e}$ states of ${}^{\infty}\textrm{He}$.

\subsubsection{The odd parity $P$ states}
The leading term of $E(x)$ of all of these states is estimated to be
$x^2$. \Tref{table:He13Po} shows the convergence for the three
low-lying  ${}^{1,3}P^{o}$ states of ${}^\infty\textrm{He}$.

\subsubsection{The parity-unfavorable ${}^{1,3}P^e$, ${}^{1,3}D^o$,
${}^{1,3}F^{e}$ states}

The doubly excited states ${}^{1,3}P^{e}$ and ${}^{1,3}D^o$  have
been observed in experiments long ago. Dolye \etal~\cite{Dolye:1971}
calculated the positions of the ${}^{1,3}P^{e}$ and ${}^{1,3}D^o$
states of the helium isoelectronic sequence using the $1/Z$
expansion method. By using a variational scheme
Bhatia~\cite{Bhatia:1972} had calculated the ${}^{1,3}D^o$ states of
helium again and obtained a few higher-precision results. The best
results for the states of helium were obtained by Goodson
\etal~\cite{Goodson:1990}, who took the advantage of the
interdimensional degeneracies of the problem and used the
variational method in the five-dimensional space to obtain the
results.  The observation of the ${}^{1,3}F^{e}$ states is rarely
reported, and its energies had been computed by Gal{$\acute{\rm
a}$}n and Bunge~\cite{Galan:1981}.  In this subsection, we report
the energies re-calculated by our approach.

For the ${}^{1,3}P^{e}$ states, we compare our results with other
calculations in \tref{table:He13Pe_comp}. As a non-variational
method, the performance of our approach is excellent.   The
distributions of  $r_1$, $r_2$, $r_3$ and $\cos\theta$ are shown in
\fref{fig:He_Pr_Pr12_3Pe} for ${}^\infty\textrm{He}$.  Some
properties of these states are given in
tables~\ref{table:He1Pe_properties}
and~\ref{table:He3Pe_properties}. It should be mentioned that only
in the case of $E_2$ of the ${}^{1}P^{e}$  state is our result
higher in energy than the variational approach, but has more
significant figures.   We think that in this case our result is
correct, because our excited states' eigen-energies  and their
related wavefunctions were all obtained simultaneously with the
ground state properties. Hence the accuracy of our excited states'
properties are insured by the demonstrated accuracy of the ground
state eigenvalue. For helium-like ions ($Z=3 \sim 6$), the
eigen-energies are summarized in \tref{table:Helike13Pe}.

We have also calculated four low-lying ${}^{1,3}D^{o}$ states of
${}^\infty\textrm{He}$.   We compare our results with Bathia's in
\tref{table:He13Do_comp}.   The three low-lying states ${}^{1}F^{e}$
states of ${\rm{Li}}^+$ ion are given in \tref{table:Li13Fe}. It is
seen that our results are better than those obtained by other
methods.

\subsection{$\textrm{H}_2^+$ and $\textrm{Ps}^-$}
\begin{figure}
\centering
\includegraphics [width=0.8\textwidth]{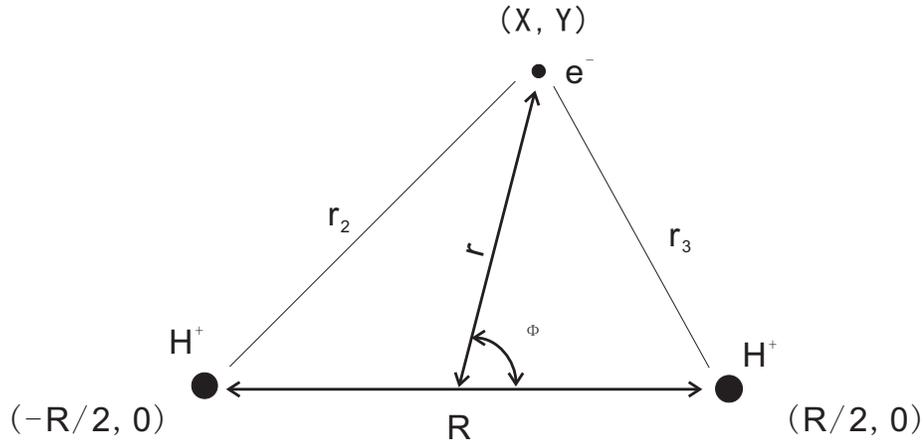}
\caption{\label{fig:Rr_H2} The coordinate system of
$\textrm{H}_2^+$.}
\end{figure}

\begin{figure}
\centering
\includegraphics [height=.9\textheight,width=0.8\textwidth]{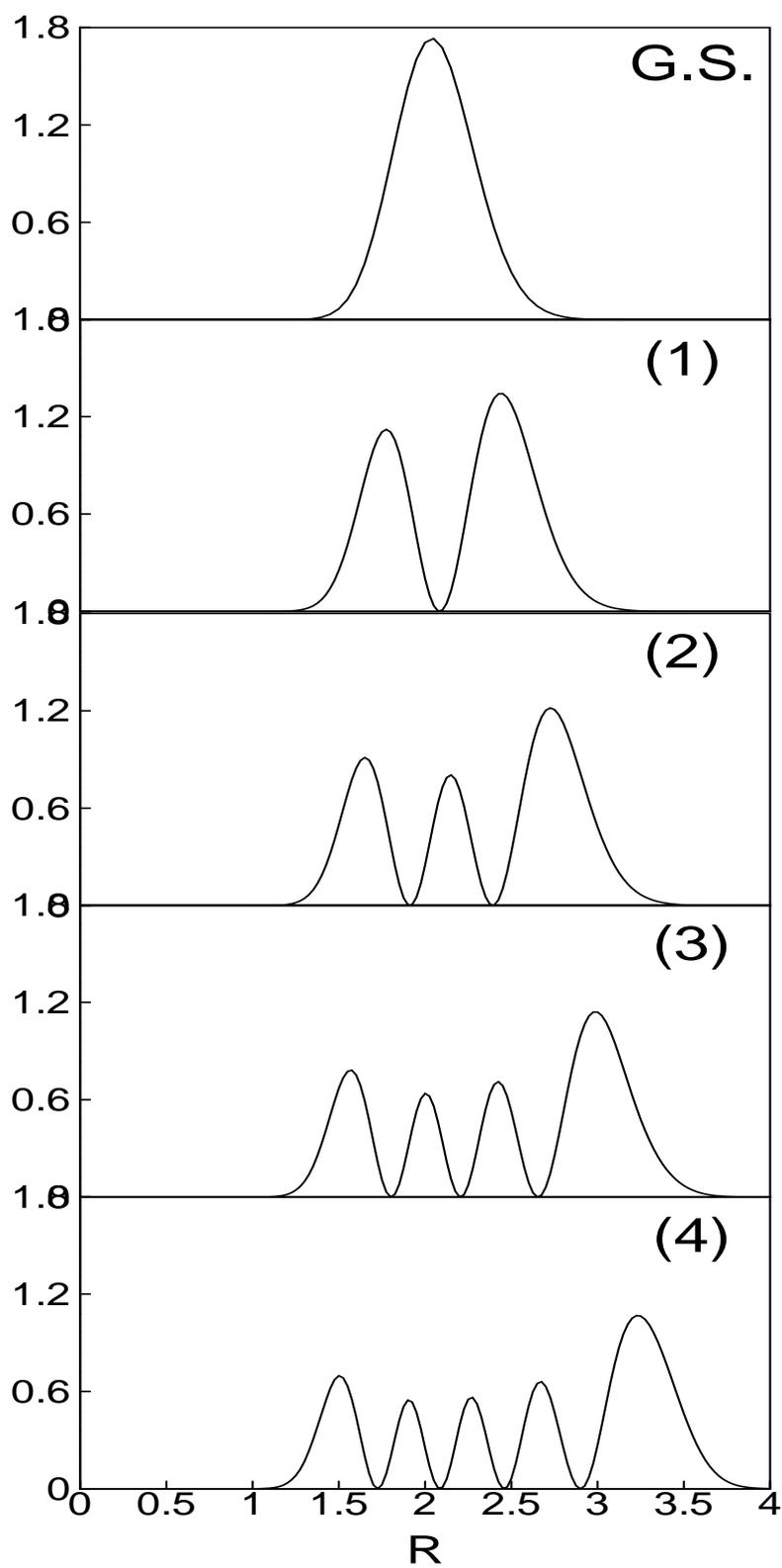}
\caption{\label{fig:HH_PR} Distributions of $R$ for the five
low-lying $S$ states of $\textrm{H}_2^+$. Similarity to oscillator
is noted}
\end{figure}
In order to obtain a better understanding of  the underlying
dynamical property of the Coulomb  three-body system with two
identical particles,  three typical systems are the most
interesting:  (1) where the mass ratio $m_1$ is large compared to
 $m_2$,  $m_3$  in the identical particles pair, e.g., for a helium atom or a
helium-like ion (such as the $\textrm{H}^-$ ion),  (2) where
$m_1=m_2=m_3$ for a positronium negative ion $\textrm{Ps}^-$, and
(3) where $m_1$ is very small, e.g., for a hydrogen molecular ion
$\textrm{H}_2^+$. It should be noted that the case of hydrogen
molecular  ion is very close to a one-body system in which the
single electron moves in the field of two heavy protons.

\subsubsection{The hydrogen molecular ion  $\textrm{H}_2^+$ }
\begin{figure}
\centering
\includegraphics [width=0.8\textwidth,bb=0 0 751 345]{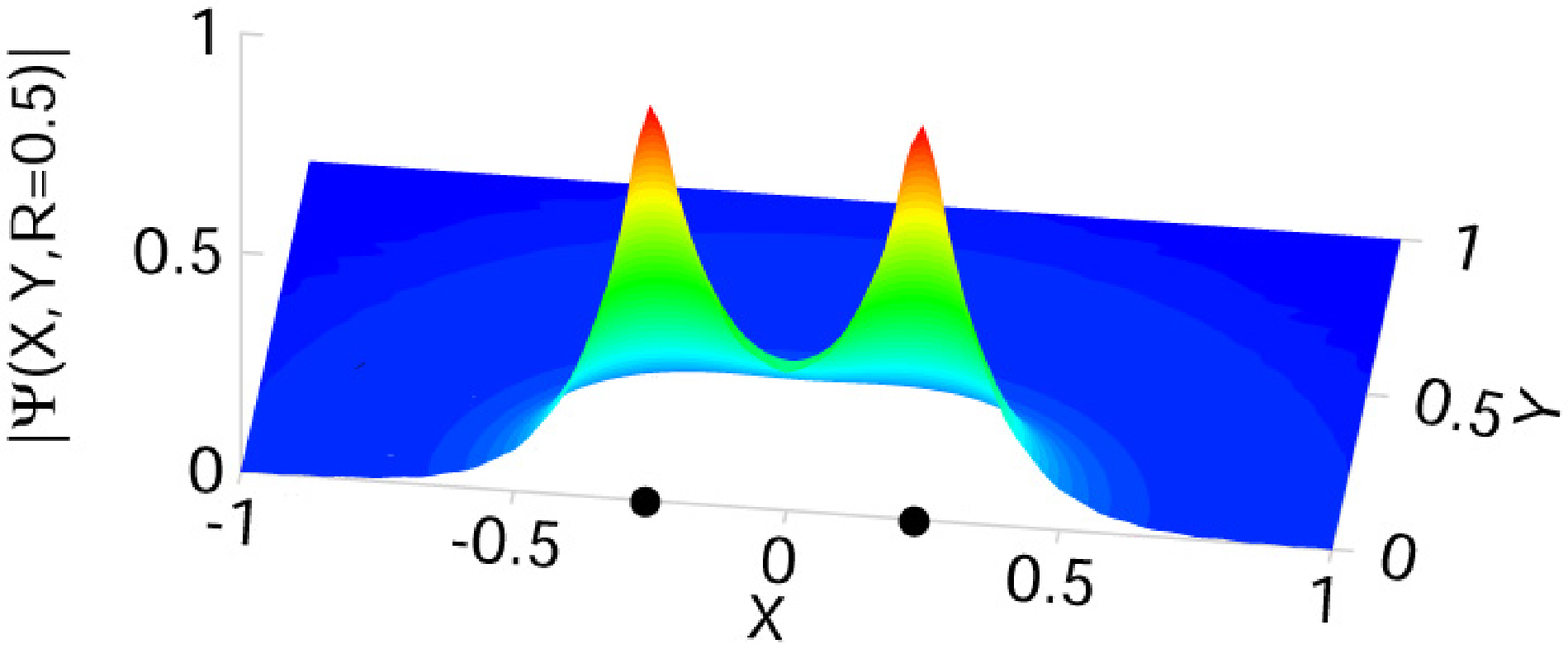}
\includegraphics [width=0.8\textwidth,bb=0 0 730 391]{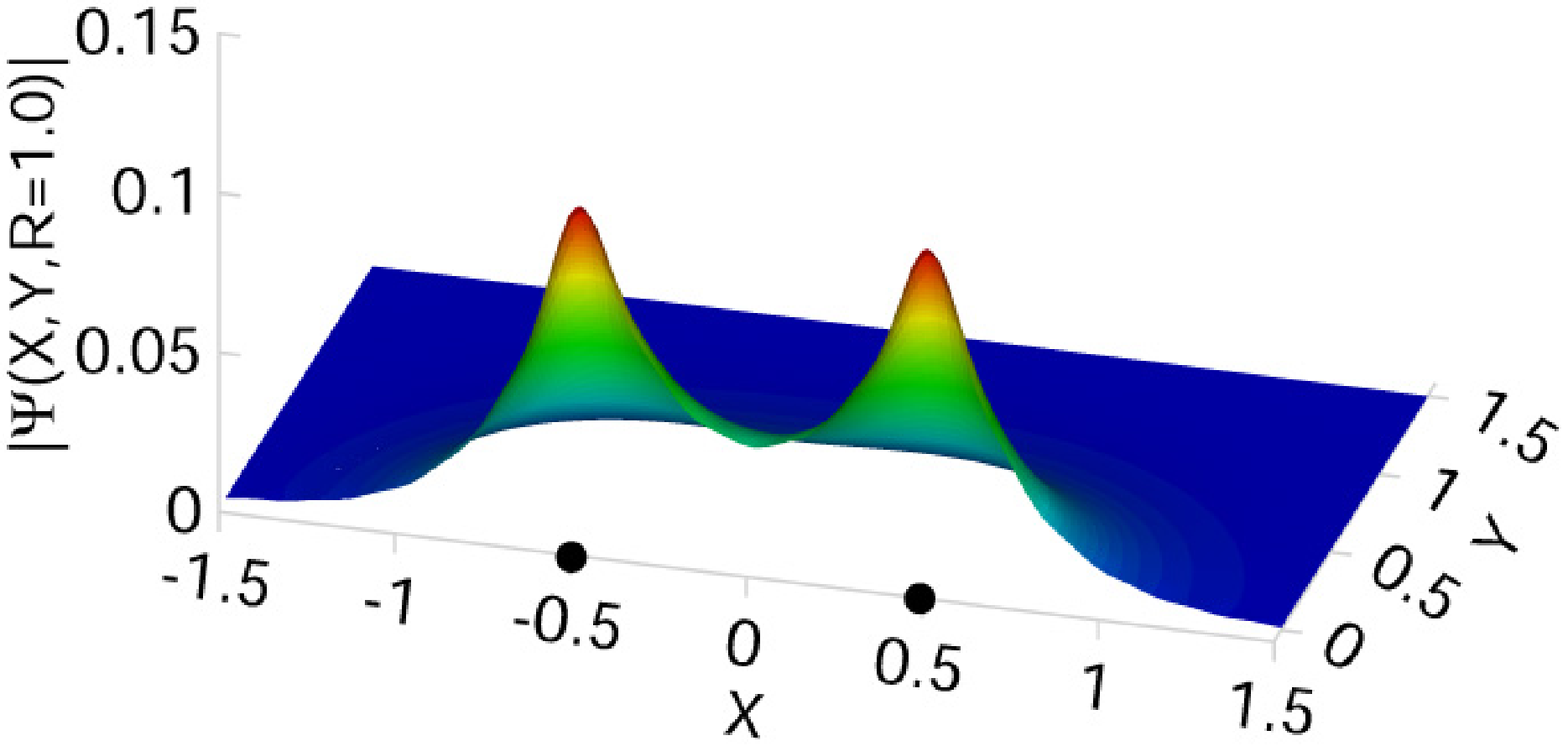}
\caption[The  function $|\Psi(X,Y,R=R_0)|$ of the ground state of
$\textrm{H}_2^+$]{\label{fig:HH_w0xy_0} The  function
$|\Psi(X,Y,R=R_0)|$ for the ground state of  $\textrm{H}_2^+$, where
$R_0=0.5,\,\, 1.0$. The electron density is noted to peak directly
above the positions of the protons, with a sharp cusp.}
\end{figure}

\begin{figure}
\centering
\includegraphics [width=0.8\textwidth,bb=0 0 723
357]{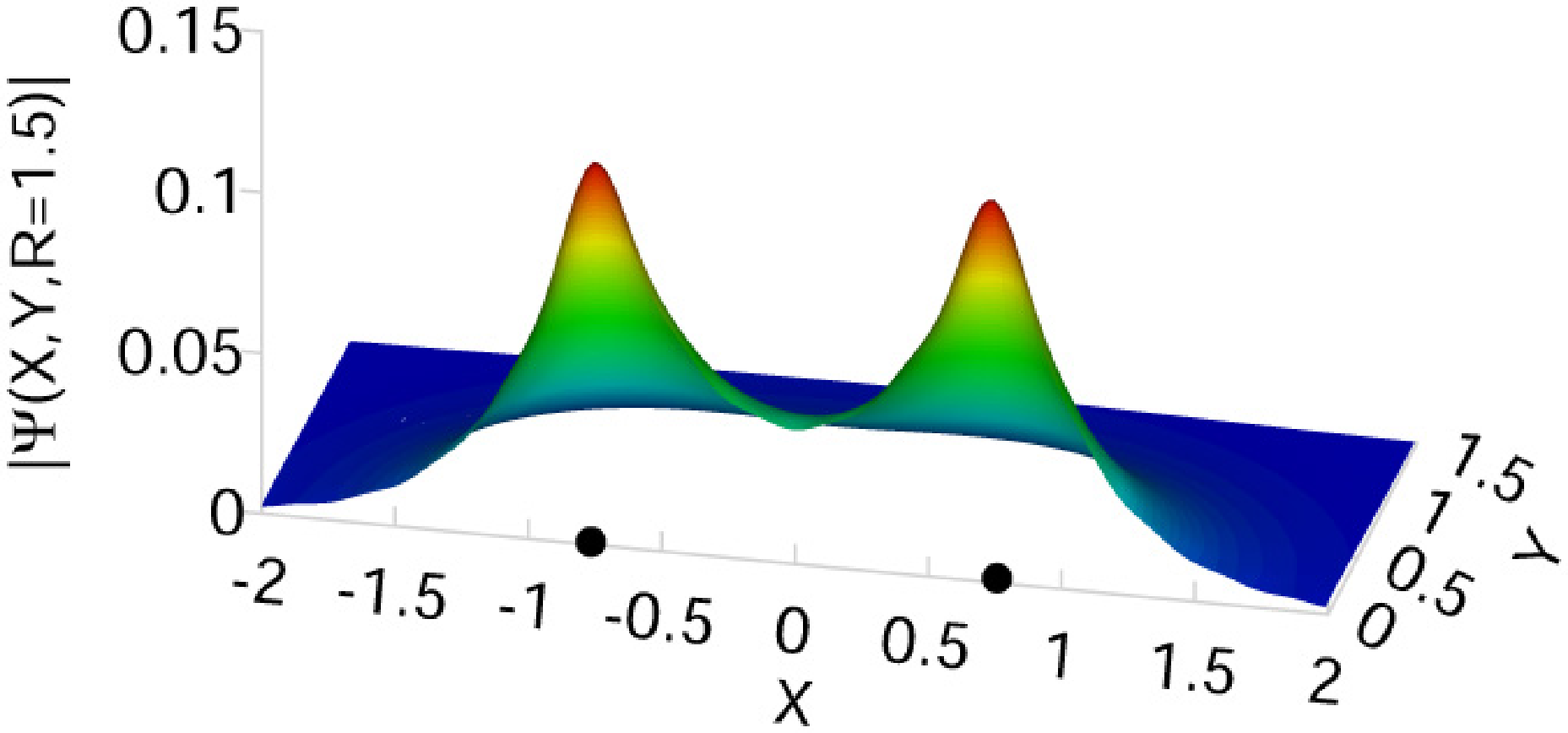}\\
\includegraphics [width=0.8\textwidth,bb=0 0 650 341]{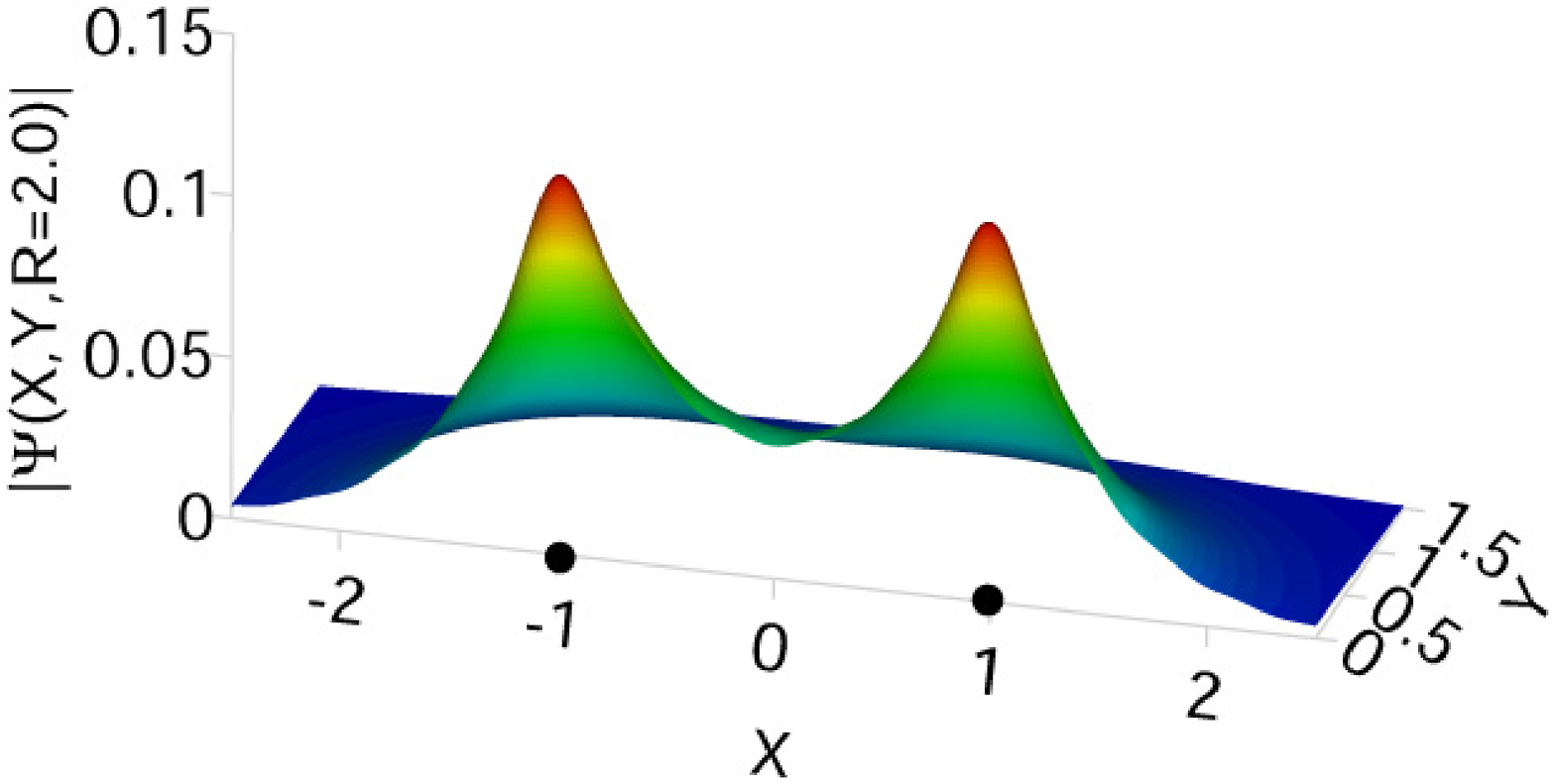}
\caption{\label{fig:HH_w0xy_1} Same as in~\fref{fig:HH_w0xy_0}, but
$R_0=1.5,\,\, 2.0$.}
\end{figure}

We have calculated 15 low-lying $S$ states for this ion, the most
achieved so far, and the numerical results are summarized in
\tref{table:HH_s}. For exploring the properties of $\textrm{H}_2^+$,
the coordinate system ($X,Y,R$) as shown in~\fref{fig:Rr_H2} is
used,  and the wave function has the form:
\begin{equation}
\Psi=\Psi(X,Y,R).
\end{equation}
Here $R$ is the distance between the two protons, and  $X$, $Y$
denote the coordinate of the electron relative to the two protons.
Figure~\ref{fig:HH_PR} shows the distribution of $R$ for the
calculated states.  It is obvious that there is vibrational motion
between the two protons, similar to an anharmonic oscillator.  In
this case the attraction between the two protons is clearly mediated
by the electron.  Figures~\ref{fig:HH_w0xy_0}
and~\ref{fig:HH_w0xy_1} show the function $\left|\Psi(X,Y,R)\right|$
of the ground state when $R=0.5,\,\, 1.0, \,\, 1.5 , \,\, 2.0$.  It
is clear from these pictures that the electron probability density
is the highest directly above each of the protons, with a sharp cusp
at the proton locations.

\subsubsection{The positronium negative ion $\textrm{Ps}^-$}
\begin{figure}
\centering
\includegraphics [width=0.8\textwidth]{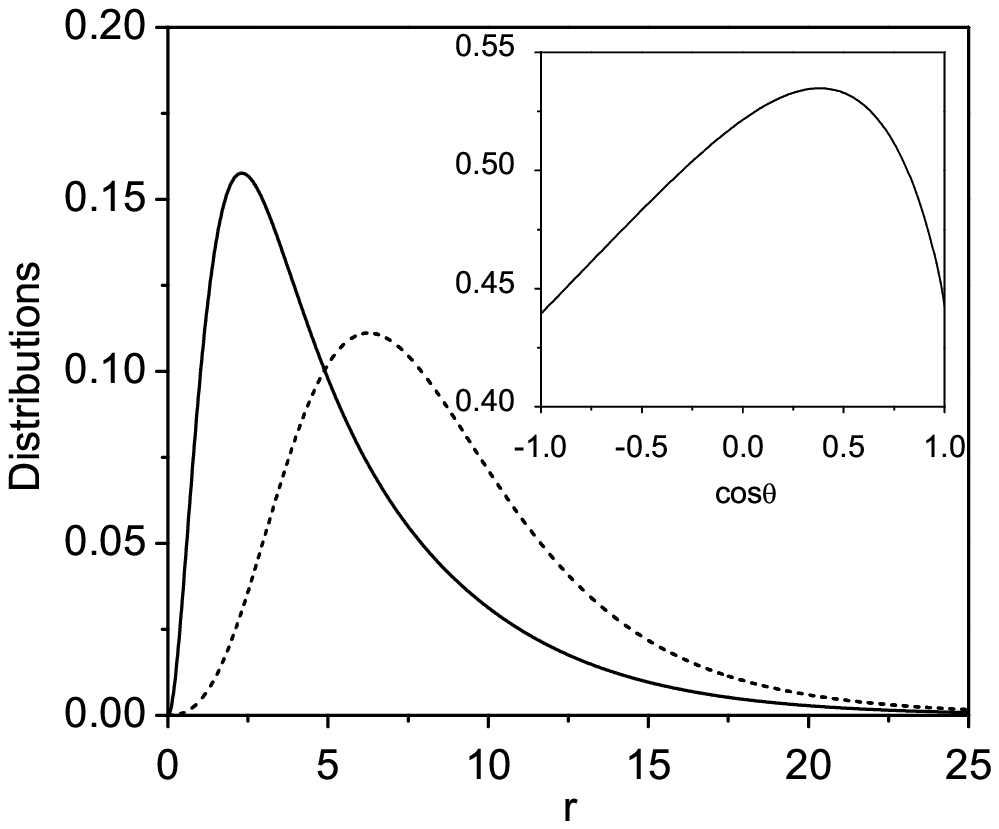}
\caption{\label{fig:Ps_Pr} Distributions of $r_1$, $r_2$~(or $r_3$)
and $\cos\theta$ of the ground state of $\textrm{Ps}^-$. The solid
line is the distribution of $r_2$ or $r_3$~(same), and the dashed
line is the distribution of $r_1$. Inset is the distribution of
$\cos\theta$.}
\end{figure}
The experimental and theoretical studies of the positronium negative
ion $\rm{Ps}^-$, consisting of two electrons and one positron, have
attracted considerable interest since the work of
Wheeler~\cite{Wheeler:1946}, who proved $\rm{Ps}^-$ to be stable
against dissociation into a free electron and  a positronium atom.
The formation of $\textrm{Ps}^-$ has been discussed in the
$e^+-\textrm{He}$~\cite{Ferrante:1969} and
$\textrm{Ps}-\textrm{H}$~\cite{Stancanelli:1970} scattering
calculations. The binding energy has been calculated variationally
by several authors. Recent interests include the calculations of
autoionizing doubly-excited states and the investigations of the
possible existence of the so-called second bound state ${}^3P^e$.
These positronium negative ions have been observed in the laboratory
by Mills~\cite{Mills:1981, Mills:1983}.  A large number of the
doubly-excited states of a positronium negative ion were calculated
by the method of complex-coordinate rotations~\cite{Lin:1995}.

We compare our results with other calculations in
\tref{table:Ps_Comp}.  \Fref{fig:Ps_Pr} shows the distributions for
$r_1$, $r_2$~(or $r_3$) and $\cos\theta$ between the two electrons.
It is seen that the two electrons plus the positron form a triangle
in which the angle formed by  $r_2$ and $r_3$ is greater than 60
degrees, consistent with the fact that the separation between the
two electrons is larger than their respective distances to the
positron.  It should be noted that the curve of the $\cos\theta$
distribution  is very different from that of the
$^{\infty}\textrm{H}^-$~(see \fref{fig:HSr}). Since the only
difference between the two systems lies in the mass of the positive
charge, the comparison is useful to delineate the importance of mass
ratio.  In the present case, the mass of the positive charge is the
same as that of the electron, so the positron is envisioned to be
much more mobile, whereas in the case of $\textrm{H}^-$ the proton
is more localized. This comparison is perhaps helpful in answering
the question~\cite{Rost:1992}: Is the positronium system a molecule
or an atom?

\section{Further developments}
The kinetic energy operator approach is noted to have applications
potential for a number of interesting problems. The approach is
obviously applicable to the three-body problem in two dimensions.
The addition of a perpendicular magnetic field plus a circularly
symmetric potential is noted to be possible in this case, without
breaking the symmetry that is essential to the solution approach.
Thus it would be particularly interesting to examine the interaction
of three electrons under a strong magnetic field, a configuration
which might be relevant to a fractional quantum Hall state.

The fact that we can reduce the dimensionality of the three-body
problem also promises numerical efficiency in the case of three-body
scattering, which requires very extensive computational resources at
present~\cite{Rescigno:1999}.   There is also the possibility of
carrying out time-dependent calculations, with implications to the
enumeration of doubly excited states in the continuum.

\begin{appendix}
\section{Eigenfunctions of $A^{(l)}$}
\subsection{Eigenfunctions of $A^{(l)}$}
In this appendix, we give the details for computing the
eigenfunctions of the operator
\begin{equation}
A^{(l)}=(1-\xi^2)\frac{\partial^2}{\partial\xi^2}
+\frac{1-(l+3)\xi^2}{\xi}\frac{\partial}{\partial\xi}
+\frac1{\xi^2}\frac{\partial^2}{\partial\beta^2},
\end{equation}
and list some useful properties of the Jacobi polynomials.

Let $x=2\xi^2-1$ and $\psi=\rme^{\pm \rmi m\beta}\xi^mF(x)$, where
$m=0,\,1,\,2,\cdots$. Straightforward computation shows that
\begin{equation}
 A^{(l)}\psi=\rme^{\pm \rmi m\beta}\xi^m \left\{
L_m^{(l)}F(x)-m (l+m+2)F(x) \right\},
\end{equation}
where
\begin{equation}\label{eq:auxf_l}
\fl \qquad L_m^{(l)}F(x)=4 \left \{ (1-x^2)F^{''}(x)+ \left[
\left(m-\frac{l}{2}\right)-\left(2+\frac{l}{2}+m\right)x
\right]F^{'}(x) \right\}.
\end{equation}
The Jacobi polynomial $P^{(a,b)}_n(x)$ satisfies the following
differential equation~$\left(D:=\frac{\rmd}{\rmd x}\right)$:
\begin{eqnarray}\label{eq:jacobi}
(1-x^2)D^2P^{(a,b)}_n(x)+\left[ b-a-(2+a+b)x \right]DP^{(a,b)}_n(x)
\nonumber \\
\qquad \qquad +n(1+a+b+n)P^{(a,b)}_n(x)=0.
\end{eqnarray}
The orthogonal relations of Jacobi polynomials
$\left\{P^{(a,b)}_n(x),\,n=0,\,1,\,2,\,\cdots\right\}$ are given by
\begin{equation}
\fl
\int_{-1}^{1}(1-x)^{a}(1+x)^{b}P^{(a,b)}_n(x)P^{(a,b)}_{n^{\prime}}(x)\rmd
x = \frac{2^{1+a+b}\Gamma(1+a+n)\Gamma(1+b+n)}
{n!(1+a+b+2n)\Gamma(1+a+b+n)}\delta_{n,n^{\prime}}.
\end{equation}
From \eref{eq:auxf_l} and \eref{eq:jacobi}, we can see that the
following set of functions constitute a complete family of
eigenfunctions of $A^{(l)}$, i.e.,
\begin{equation}
\left \{ J^{(l)}_{\pm m,n}(\xi,\beta)=\rme^{\pm \rmi m\beta}\xi^m
P^{(\frac{l}{2},m)}_n(2\xi^2-1),\, m,n\geq 0 \right \},
\end{equation}
with their eigenvalues given by
\begin{equation}
\lambda^{(l)}_{m,n}=- \left\{ 4n
\left(1+\frac{l}{2}+m+n\right)+m(l+m+2) \right\}.
\end{equation}
The orthogonal relations become
\begin{eqnarray}
\fl \int_{-\pi}^{\pi}\int_{0}^{1} J^{(l)}_{m_1,n_1}
J^{(l)}_{m_2,n_2} \xi (1-\xi)^{l/2} \rmd \xi \rmd \beta=\frac{\pi
\Gamma(1+\frac{l}{2}+n)\Gamma(1+m+n)}
{n!(1+\frac{l}{2}+m+2n)\Gamma(1+\frac{l}{2}+m+n)}\delta_{m_1,m_2}\delta_{n_1,n_2}.
\nonumber \\
\end{eqnarray}

\subsection[Properties of the Jacobi polynomials]{Recurrence relations of the Jacobi polynomials and the matrices of $B_1$ and
$B_2$}
\subsubsection{Matrices of $B_1$ and $B_2$}

For the two linear differential operators
$$
B_1=\cos\beta \frac{\partial
}{\partial\xi}-\frac{\sin\beta}{\xi}\frac{\partial}{\partial \beta},
\quad {\rm and }\quad B_2=\sin\beta \frac{\partial }{\partial
\xi}+\frac{\cos\beta}{\xi}\frac{\partial}{\partial \beta}, \quad
$$
we have
\begin{enumerate}
\item The case of $m=0$:
\begin{eqnarray*}
B_1J^{(l)}_{0,n}&=&(2\rmi \rme^{\rmi\beta}\xi+2\rmi
\rme^{-\rmi\beta}\xi)
DP^{(\frac{l}{2},0)}_n(x),\\
B_2J^{(l)}_{0,n}&=&(-2\rmi \rme^{\rmi\beta}\xi+2\rmi
\rme^{-\rmi\beta}\xi) DP^{(\frac{l}{2},0)}_n(x).
\end{eqnarray*}
\item The case of positive indices:
\begin{eqnarray*}
B_1J^{(l)}_{m,n}&=&2\rme^{\rmi(m+1)\beta}\xi^{m+1}DP^{(\frac{l}{2},m)}_n(x)\\
&{ }&+\rme^{\rmi(m-1)\beta}\xi^{m-1}
\left[mP^{(\frac{l}{2},m)}_n(x)+(1+x)DP^{(\frac{l}{2},m)}_n(x)\right],
\\
B_2J^{(l)}_{m,n}&=&-2\rmi\rme^{\rmi(m+1)\beta}\xi^{m+1}DP^{(\frac{l}{2},m)}_n(x)\\
&{ }&+\rmi \rme^{\rmi(m-1)\beta}\xi^{m-1}
\left[mP^{(\frac{l}{2},m)}_n(x)+(1+x)DP^{(\frac{l}{2},m)}_n(x)\right].
\end{eqnarray*}
\item The case of negative indices:
\begin{eqnarray*}
B_1J^{(l)}_{-m,n}&=&2\rme^{-\rmi(m+1)\beta}\xi^{m+1}DP^{(\frac{l}{2},m)}_n(x)\\
&{ }&+\rme^{-\rmi(m-1)\beta}\xi^{m-1}
\left[mP^{(\frac{l}{2},m)}_n(x)+(1+x)DP^{(\frac{l}{2},m)}_n(x)\right],
\\
B_2J^{(l)}_{-m,n}&=&2\rmi \rme^{-\rmi(m+1)\beta}\xi^{m+1}DP^{(\frac{l}{2},m)}_n(x)\\
&{ }&-\rmi \rme^{-\rmi(m-1)\beta}\xi^{m-1}
\left[mP^{(\frac{l}{2},m)}_n(x)+(1+x)DP^{(\frac{l}{2},m)}_n(x)\right].
\end{eqnarray*}
\end{enumerate}
Therefore, we need explicit formulas which express
$DP^{(\frac{l}{2},m)}_n(x)$
($mP^{(\frac{l}{2},m)}_n(x)+(1+x)DP^{(\frac{l}{2},m)}_n(x)$) as
linear combination of
$P^{(\frac{l}{2},m+1)}_k(x)$($P^{(\frac{l}{2},m-1)}_k(x)$). Here,
the existence of a collection of recurrence relations among Jacobi
polynomials provides a handy tool for such a purpose.

\subsubsection{Recurrence relations of the Jacobi polynomials}
We recall some of the mixed-type ones from \S 138 of
Rainville~\cite{Rainville:1960}:
\begin{eqnarray*}
\fl \qquad
DP^{(a,b)}_n(x)&=&\frac{1}{2}(1+a+b+n)P^{(a+1,b+1)}_{n-1}(x) \\
\fl \qquad         &=&\frac{1}{2}(b+n)P^{(a+1,b)}_{n-1}(x)+\frac{1}{2}(a+n)P^{(a,b+1)}_{n-1}(x)\\
\fl \qquad (1+x)DP^{(a,b)}_n(x)&=&nP^{(a,b)}_n(x)+(b+n)P^{(a+1,b)}_{n-1}(x)\\[1.5em]
\fl \qquad (a+b+2n)P^{(a,b-1)}_n(x)
&=&(a+b+n)P^{(a,b)}_n(x)+(a+n)P^{(a,b)}_{n-1}(x)\\
\fl \qquad (a+b+2n)P^{(a-1,b)}_n(x)
&=&(a+b+n)P^{(a,b)}_n(x)-(b+n)P^{(a,b)}_{n-1}(x)\\
\fl \qquad P^{(a,b-1)}_n(x)-P^{(a-1,b)}_n(x) &=&P^{(a,b)}_{n-1}(x).
\end{eqnarray*}
We have the following specific recurrence relations:
\begin{eqnarray*}
\fl \qquad DP^{(a,b)}_n(x)&=&
\left[\frac{1}{2}(a+b)+n\right]P^{(a,b+1)}_{n-1}(x)+\frac{b+n}{a+b+n}DP^{(a,b)}_{n-1}(x)
, \\[.5em]
\fl \qquad \left[ bP^{(a,b)}_n(x)+(1+x)DP^{(a,b)}_n(x) \right]
&=&(b+n)P^{(a+1,b-1)}_n(x)\\
\fl \qquad &=&\frac{b+n}{a+b+n} \left[
bP^{(a,b)}_{n-1}(x)+(1+x)DP^{(a,b)}_{n-1}(x) \right] \\
\fl \qquad &&+\frac{(b+n)(a+b+2n)}{a+b+n}P^{(a,b-1)}_{n}(x).
\end{eqnarray*}
By using the above formulas, we get the explicit expressions for the
matrices of $B_1$ and $B_2$. The results are given in section 3.

\section{IMT Integration Scheme}

Consider the real value function $f(x)$ defined on the interval
$[\,0,1]$, which is continuous and differentiable sufficiently many
times on $[\,0,1]$. For the N-point trapezoidal rule, there is the
Euler-Maclaurin formula
\begin{eqnarray}\label{eq:Euler}
\int_{0}^{1}f(x)\,\rmd x&=&\frac{1}{N}\left[\frac{1}{2}(f(0)+f(1))+\sum_{n=1}^{N-1}f(n/N)\right] 
\nonumber \\
&&+\sum_{r=1}^{m-1}(-1)^{r}\frac{B_r}{(2r)!}\frac{1}{N^{2r}}
\left[f^{(2r-1)}(1)-f^{(2r-1)}(0)\right]+R_m,
\end{eqnarray}
where $B_r$ is the $r$-th Bernoulli number and $R_m$ is the reminder
term. Noting that  the expression for the truncation error depends
only on values of  the functional derivatives at the integration end
points, we design a change of  variable, so that
\begin{equation}
f^{(2r-1)}(1)=f^{(2r-1)}(0)\qquad \mbox{for} \quad r=1,2,\cdots.
\end{equation}
It follows from \eref{eq:Euler} that  all  error terms will vanish,
and high precision can be achieved. The
IMT-rule~\cite{Ooura:1974,Davis:1984}, also denoted the
``double-exponential formula", is one type of integration technique.
Consider
\begin{eqnarray}
Q&=&\int_{0}^{1}\exp{\left(-\frac{1}{t}-\frac{1}{1-t}\right)}\, \rmd
t
\nonumber \\
&=&0.00702\, 98584\, 06609\, 65623\, 92412\,70530\,\cdots.
\end{eqnarray}
Define the function $\varphi(t)$ by
\begin{equation}
\varphi{(t)}=\int_{0}^{t}\varphi^\prime(\tau)\,\rmd\tau,\quad
\varphi^\prime(t)=\frac{1}{Q}\exp\left(-\frac1{t}-\frac1{1-t}\right).
\end{equation}
The transformation $x=\varphi(t)$ maps the variable $x\in[\,0,1]$
into the variable $t\in[\,0,1]$, and we have a new expression for
the integration of $f(x)$ over $[\,0,1]$:
\begin{equation}
\label{eq:int} \int_{0}^{1}f(x)\,\rmd x=\int_{0}^{1}g(t)\,\rmd
t,\quad \mbox{with} \quad g(t)=f(\varphi(t))\varphi^\prime(t).
\end{equation}
If the function $f(x)$ is differentiable infinitely many times on
$(0,1)$ and has an algebraic singularity such as $x^\alpha$ or
$(1-x)^\beta$ $(\alpha,\beta>-1)$, then the function $g(t)$ also can
be differentiable many times and all its derivatives vanish at the
two ends of the interval $[\,0,1]$ due to the strong singularity of
$\varphi^\prime(t)$, so that
\begin{equation}
g^{(m)}(0)=g^{(m)}(1)=0,\quad \mbox{for}\quad m=0,1,2,\cdots.
\end{equation}
Applying the trapezoidal rule on \eref{eq:int}, we get the
approximation for the  integration:
\begin{equation}
\fl \qquad
S_N=\frac{1}{N}\sum_{n=1}^{N-1}w_n^{(N)}f(x_n^{(N)}),\quad
\mbox{and}\quad
x_n^{(N)}=\varphi(n/N),\,w_n^{(N)}=\varphi^\prime(n/N).
\end{equation}
If we define
$$c_k=\int_0^1g(t)\exp(\rmi 2\pi kt)\,\rmd t,$$
then obviously
$$g(t)=\sum_{k=-\infty}^{\infty}c_k\exp(-\rmi 2\pi kt)$$
and
$$\int_0^1f(x)\,\rmd x=\int_0^1g(t)\,\rmd t=c_0.$$
The approximation $S_N$ is given by
\begin{equation}
S_N=\frac{1}{N}\sum_{n=0}^{N-1}g(n/N)=c_0+\sum_{p=1}^{\infty}(c_{pN}+c_{-pN}).
\end{equation}
The truncation error $\varepsilon_N$ is estinated as:
\begin{eqnarray}
\varepsilon_N&=& S_N-\int_0^1g(t)\,\rmd t
\nonumber \\
&=&\sum_{p=1}^{\infty}(c_{pN}+c_{-pN})
\nonumber \\
&=&2\sum_{p=1}^{\infty} \mbox{Re}\,c_{pN}
\nonumber \\
&\sim& 2\mbox{Re}\,c_N.
\end{eqnarray}
For $f(x)$ with singularity $x^\alpha$ or $(1-x)^\alpha$, an
estimate of $\varepsilon(N,\alpha)$ is given by
\begin{eqnarray}
\fl \qquad \varepsilon(N,\alpha)&=&\frac{\sqrt{4\pi}}{(\rme
Q)^{\alpha+1}}\cdot \frac{(\alpha+1)^{1/4+\alpha}}{(2\pi
N)^{3/4+\alpha}}
\nonumber \\
\fl \qquad &&\cdot \exp{\left[-\sqrt{4\pi (\alpha+1)N}\right]}
\cdot\cos{\left[\sqrt{4\pi
(\alpha+1)N}+\frac{3+4\alpha}{8}\pi\right]}.
\end{eqnarray}
\end{appendix}

\section*{References}

\Tables
%
%

\Table{\label{table:H1S_comp}Comparison of the ground state of
${}^{\infty}\textrm{H}^-$ and $\textrm{H}^-$~($m_p=1836.152701$)
with other theoretical calculations.} \br
&\multicolumn{1}{c}{${}^{\infty}\textrm{H}^-$}&\multicolumn{1}{c}{$\textrm{H}^-$}\\
\mr
-$E$&\00.527 751 016 54&\00.527 445 881 1\\
  &\tblue{\00.527 751 016 544 302}$\,^{\rm a}$&\tblue{\00.527 445 881 114 104}$\,^{\rm a}$\\
  &\tblue{\00.527 751 016 532}$\,^{\rm b}$&\tblue{\00.527 445 881 110}$\,^{\rm c}$\\
$<{1}/{r_2}>$=$<{1}/{r_3}>$&\00.683 261 767&\00.682 853 385\\
&\tblue{\00.683 261 767 654 0}$\,^{\rm a}$&\tblue{\00.682 853 384 854}$\,^{\rm a}$\\
&\tblue{\00.683 261 768}$\,^{\rm b}$&\tblue{\00.682 853 384 96}$\,^{\rm c}$\\
$<{1}/{r_1}>$&\00.311 021 502 2&\00.310 815 007\\
&\tblue{\00.311 021 502 219 1}$\,^{\rm a}$&\tblue{\00.310 815 007 479}$\,^{\rm a}$\\
&\tblue{\00.311 021 503}$\,^{\rm b}$&\tblue{\00.310 815 007 66}$\,^{\rm c}$\\
$<{r_2}>$=$<{r_3}>$&\02.710 178 27&\02.712 095 6\\
&\tblue{\02.710 178 278 34}$\,^{\rm a}$&\tblue{\02.712 095 626 51}$\,^{\rm a}$\\
&\tblue{\02.710 178 263}$\,^{\rm b}$&\tblue{\02.712 095 621 4}$\,^{\rm c}$\\
$<{r_1}>$&\04.412 694 50&\04.415 692 6\\
&\tblue{\04.412 694 497 79}$\,^{\rm a}$&\tblue{\04.415 692 603 31}$\,^{\rm a}$\\
&&\tblue{\04.415 692 593 4}$\,^{\rm c}$\\
$<{r_2^2}>$=$<{r_3^2}>$
&11.913 699 6&11.931 747 7\\
&\tblue{11.913 699 681 6}$\,^{\rm a}$&\tblue{11.931 747 760}$\,^{\rm a}$\\
&\tblue{11.913 699 235}$\,^{\rm b}$&\tblue{11.931 747 62}$\,^{\rm c}$\\
$<{r_1^2}>$
&25.202 025 2&25.237 175\\
&\tblue{25.202 025 298 2}$\,^{\rm a}$&\tblue{25.237 175 614 1}$\,^{\rm a}$\\
&&\tblue{25.237 175 34}$\,^{\rm c}$\\
$<\cos(\mathbf{r_{12}},\mathbf{r_{13}})>$
&\0\-0.105 147 693 7&\0\-0.104 996 606\\
&\tblue{\0\-0.105 147 693 566 0}$\,^{\rm a}$&\tblue{\0\-0.104 996 606 303 1}$\,^{\rm a}$\\
$<\cos(\mathbf{r_{12}},\mathbf{r_{32}})>$
&\00.649 871 581&\00.694 795 647\\
&\tblue{\00.649 871 581 193 9}$\,^{\rm a}$&\tblue{\00.694 795 646 586}$\,^{\rm a}$\\
\br
\end{tabular}
\item[]$^{\rm a}$Ref.~\cite{Frolov:1998}    $^{\rm b}$Ref.~\cite{Ackermann:1995}   $^{\rm c}$Ref.~\cite{Ackermann:1996}
\end{indented}
\end{table}

%
%

\Table{\label{table:H3P_comp}Comparison of the ${}^3P^e$ state of
${}^{\infty}\textrm{H}^-$ and $\textrm{H}^-$~($m_p=1836.152701$)
with other theoretical calculations.} \br
&\multicolumn{1}{c}{${}^{\infty}\textrm{H}^-$}&\multicolumn{1}{c}{$\textrm{H}^-$}\\
\mr
-$E$&\0\00.125 355 451 242&\0\00.125 283 157 034\\
&\tblue{\0\00.125 354 7}$\,^{\rm a}$&\\
&\tblue{\0\00.125 355 451 24}$\,^{\rm b}$&\\
&\tblue{\0\00.125 354 705}$\,^{\rm c}$&\tblue{\0\00.125 282 391 9}$\,^{\rm c}$\\
&\tblue{\0\00.125 351 3}$\,^{\rm d}$&\tblue{\0\00.125 279 0}$\,^{\rm d}$\\
&\tblue{\0\00.125 355 08}$\,^{\rm e}$&\\
$<{1}/{r_2}>$=$<{1}/{r_3}>$&\0\00.160 520 878 4&\0\00.160 397 165\\
$<{1}/{r_1}>$&\0\00.0703 308 548&\0\00.0702 280 17\\
             &\tblue{\0\00.0706 30}$\,^{\rm a}$&\\
$<{r_2}>$=$<{r_3}>$&\011.657 657 7&\011.683 161 8\\
$<{r_1}>$&\019.585 091&\019.632 096 8\\
         &\tblue{\019.237}$\,^{\rm a}$&\\
$<{r_2^2}>$=$<{r_3^2}>$&271.263 4&273.015 9\\
$<{r_1^2}>$&557.259 2&560.744\\
           &\tblue{517.09}$\,^{\rm a}$&\\
$<\cos(\mathbf{r_{12}},\mathbf{r_{13}})>$&\0\0\-0.093 867 209 4&\0\0\-0.093 625 370\\
$<\cos(\mathbf{r_{12}},\mathbf{r_{32}})>$&\0\00.651 280 257&\0\00.651 086 687\\
\br
\end{tabular}
\item[] $^{\rm a}$Ref.~\cite{Banyyard:1992}, $^{\rm b}$Ref.~\cite{Bylicki:2003},
$^{\rm c}$Ref.~\cite{Bhatia:1970}, $^{\rm d}$Ref.~\cite{Hesse:2001},
$^{\rm e}$Ref.~\cite{Jauregui:1979}.
\end{indented}
\end{table}

\Table{\label{table:He13S_comp} Comparison  for the energies of the
five low-lying  ${}^{1,3}S^{e}$ states of ${}^\infty \textrm{He}$
and $\textrm{He}$ with others calculations.}
\br
\multicolumn{2}{c}{$-E({}^1S^e)$}&\multicolumn{2}{c}{$-E({}^3S^e)$}\\
\mr This work&Ref.~\cite{Burgers:1995}
&This work&Ref.~\cite{Burgers:1995}\\
\mr
\multicolumn{4}{c}{${}^\infty \textrm{He}$}\\
2.903 724 377 03&2.903 724 377 034 119&2.175 229 378 2&2.175 229 378 236\\
2.145 974 046&2.145 974 046 054&2.068 689 067&2.068 689 067 47\\
2.061 271 98&2.061 271 989 7&2.036 512 0&2.036 512 083\\
2.033 586&2.033 586 7&2.022 6&2.022 618\\
2.021 17&2.021 17&2.015&2.015 377\\[0.5em]
\multicolumn{4}{c}{$\textrm{He}\,(m_{\alpha}=7294.299507)$}\\
2.903 304 557 7&&2.174 930 190 6&\\
2.145 678 587&&2.068 405 243&\\
2.060 989 07&&2.036 232 73&\\
2.033 307 6&&2.022 280&\\
2.020 77&&2.013 34&\\
\br
\end{tabular}
\end{indented}
\end{table}

\Table{\label{table:He13Po} Convergence study for  the  three
low-lying  ${}^{1,3}P^{o}$ states of ${}^\infty\textrm{He}$.} \br
($M$, $N$,$P$)& \multicolumn{1}{c}{$-E_0$}&
\multicolumn{1}{c}{$-E_1$}&
\multicolumn{1}{c}{$-E_2$}\\
\mr
\multicolumn{4}{c}{${}^{1}P^{o}$}\\[.2em]
(290,160,60)&2.121 781 471 934&2.045 206 888 028&2.007 282 642 688\\
(310,170,60)&2.122 001 335 701&2.046 159 516 026&2.009 274 234 036\\
(330,180,60)&2.122 187 837 151&2.046 981 326 187&2.011 018 748 522\\
(350,190,60)&2.122 347 396 423&2.047 695 333 577&2.012 556 633 845\\
(370,200,60)&2.122 484 960 882&2.048 319 671 610&2.013 920 175 194\\
(390,210,60)&2.122 604 392 689&2.048 868 793 171&2.015 135 435 817\\
(410,220,60)&2.122 708 742 385&2.049 354 337 962&2.016 223 670 874\\
(430,230,60)&2.122 800 445 240&2.049 785 768 826&2.017 202 374 649\\[1.0em]
{Extrap.}&2.123 842 8 &2.055 149&2.031 0\\[.5em]
Ref.~\cite{Schiff:1965} &2.123 843 085 800 &2.055 146 355
4&2.031 069 591\\[.5em]
\multicolumn{4}{c}{${}^{3}P^{o}$}\\[.5em]
(290,160,60)&2.131 433 366 926&2.049 022 358 625&2.009 815 747 062\\
(310,170,60)&2.131 618 811 233&2.049 898 916 341&2.011 718 260 682\\
(330,180,60)&2.131 775 987 950&2.050 654 119 626&2.013 383 329 203\\
(350,190,60)&2.131 910 360 569&2.051 309 458 014&2.014 849 959 935\\
(370,200,60)&2.132 026 134 097&2.051 881 840 210&2.016 149 272 454\\
(390,210,60)&2.132 126 587 511&2.052 384 725 339&2.017 306 376 122\\
(410,220,60)&2.132 214 308 072&2.052 828 939 042&2.018 341 737 893\\
(430,230,60)&2.132 291 359 280&2.053 223 271 537&2.019 272 194 566\\[1.0em]
{Extrap.}&2.133 164 06 &2.058 081&2.032 36\\[.5em]
Ref.~\cite{Schiff:1965} &2.133 164 190 534 &2.058 081 081 6
&2.032 324 325\\
\br
\end{tabular}
\end{indented}
\end{table}

\begin{table}
\caption{\label{table:He1Se_properties} Properties  of the five
low-lying ${}^{1}S^{e}$ states of ${}^\infty \textrm{He}$.}
\footnotesize
\begin{tabular}{clllll}
\br \multicolumn{1}{c}{State} &\multicolumn{1}{c}{1 ${}^{1}S^{e}$}
&\multicolumn{1}{c}{2 ${}^{1}S^{e}$} &\multicolumn{1}{c}{3
${}^{1}S^{e}$} &\multicolumn{1}{c}{4 ${}^{1}S^{e}$}
&\multicolumn{1}{c}{5
${}^{1}S^{e}$}\\
\mr
$<{1}/{r_2}>$ &1.688 316 800 71 &1.135 407 686 &1.058 514 75
&1.032 484 8
&1.021 298\\
$<{1}/{r_1}>$ &0.945 818 448 80 &0.249 682 652 &0.111 514 95 &0.062
760 2
&0.041 429\\
$<{r_2}>$ &0.929 472 294 87 &2.973 061 12 &6.511 676 &11.550 6
&17.573\\
$<{r_1}>$ &1.422 070 255 5 &5.269 696 20 &12.304 521 &22.368 1
&34.407\\
$<{r_2^2}>$ &1.193 482 995 0 &16.089 233 2 &85.890 18 &281.33
&660.7\\
$<{r_1^2}>$ &2.516 439 312 9 &32.302 380 3 &171.838 66 &562.71
&132 1.4\\
$<\cos(\mathbf{r_{12}},\mathbf{r_{13}})>$ &-0.064 202 614 217
&-0.014 657 043 3 &-0.004 317 036 7 &-0.001 795 63
&-0.000 958 99\\
$<\cos(\mathbf{r_{12}},\mathbf{r_{32}})>$ &0.648 017 667 47 &0.557
144 578 &0.526 466 53 &0.515 120 &0.510 057\\
\br
\end{tabular}
\end{table}

\begin{table}
\caption{\label{table:He3Se_properties} Properties  of the  five
low-lying ${}^{3}S^{e}$ states of ${}^\infty \textrm{He}$.}
\footnotesize
\begin{tabular}{clllll}
\br State &\multicolumn{1}{c}{2${}^{3}S^{e}$} &\multicolumn{1}{c}{3
${}^{3}S^{e}$} &\multicolumn{1}{c}{4 ${}^{3}S^{e}$}
&\multicolumn{1}{c}{5 ${}^{3}S^{e}$} &\multicolumn{1}{c}{6
${}^{3}S^{e}$}\\
\mr
$<{1}/{r_2}>$ &1.154 663 198 4 &1.063 661 050 &1.034 491 5
&1.021 740
&1.019 08\\
$<{1}/{r_1}>$ &0.268 197 633 6 &0.117 316 73 &0.065 253 6 &0.042 086
&0.037 343\\
$<{r_2}>$ &2.550 464 78 &5.856 041 6 &10.661 70 &16.702 8
&19.945\\
$<{r_1}>$ &4.447 538 89 &10.998 910 1 &20.592 6 &32.666 8
&39.150\\
$<{r_2^2}>$ &11.464 340 5 &68.710 08 &238.596 &596.06
&870.56\\
$<{r_1^2}>$ &23.046 235 5 &137.478 44 &477.225 &119 2.20
&174 1.3\\
$<\cos(\mathbf{r_{12}},\mathbf{r_{13}})>$ &-0.015 839 217 1 &-0.004
245 085 8 &-0.001 686 96 &-0.000 853 68
&\-0.000 753 55\\
$<\cos(\mathbf{r_{12}},\mathbf{r_{32}})>$ &0.562 788 947 4 &0.528
301 29 &0.515 917 &0.510 32 &0.509 16\\
\br
\end{tabular}
\end{table}

\begin{table}
\caption{\label{table:He1Pe_properties} Properties  of the  five
low-lying ${}^{1}P^{e}$  states of ${}^\infty \textrm{He}$.}
\footnotesize
\begin{tabular}{clllll}
\br State &\multicolumn{1}{c}{3 ${}^{1}P^{e}$}
&\multicolumn{1}{c}{4${}^{1}P^{e}$}
&\multicolumn{1}{c}{5${}^{1}P^{e}$}
&\multicolumn{1}{c}{6${}^{1}P^{e}$}
&\multicolumn{1}{c}{7${}^{1}P^{e}$}\\
\mr $<{1}/{r_2}>$&.320101054072&.28662333040&.27259779205
&.2653459435
&.261119621\\
$<{1}/{r_1}>$ &.119911271506&.066410151506&.0455874718&.0320333447&.021194314\\
$<{r_2}>$ &5.68748370235&10.289515048&16.39163259&23.9930126&33.043005\\
$<{r_1}>$ &9.3831001400&18.33321725&30.43563233&45.5874716&63.658 480\\
$<{r_2^2}>$ &48.197972382&187.5444301&517.8278870&1160.99752&2260.9388\\
$<{r_1^2}>$ &97.935626136&376.133414&1036.36343&2322.49992&4522.255\\
$<\cos(\mathbf{r_{12}},\mathbf{r_{13}})>$&-0.031985135411&-0.01274515627&-0.00621248298&-0.0034696118
&-0.002133103\\
$ <\cos(\mathbf{r_{12}},\mathbf{r_{32}})>$
&.608746828961&.56272353076&.54033973838 &.5280463517&.520635797\\
\br
\end{tabular}
\end{table}

\begin{table}
\caption{\label{table:He3Pe_properties} Properties  of the five
low-lying ${}^{3}P^{e}$  states of ${}^\infty \textrm{He}$.}
\footnotesize
\begin{tabular}{clllll}
\br State &\multicolumn{1}{c}{2 ${}^{3}P^{e}$} &\multicolumn{1}{c}{3
${}^{3}P^{e}$} &\multicolumn{1}{c}{4 ${}^{3}P^{e}$}
&\multicolumn{1}{c}{5 ${}^{3}P^{e}$} &\multicolumn{1}{c}{6
${}^{3}P^{e}$}\\
\mr $<{1}/{r_2}>$ &.41809816667770 &.312484187824 &.28403411128
&.27140054988
&.2646919161\\
$<{1}/{r_1}>$&.2513923553540&.114310953848&.06440206759&.04109304812
&.028447092\\
$<{r_2}>$ &3.089879033146&6.460157023&11.30744532&17.67087220&25.5392910\\
$<{r_1}>$ &4.676371886388&10.831719973&20.33743461&32.98000076&48.6725310\\
$<{r_2^2}>$ &11.79098787761&65.49966426&232.2699200&609.9032260&1325.7638\\
$<{r_1^2}>$ &25.06816973162&132.73779825&465.6256393&1220.528804&2652.0392\\
$<\cos(\mathbf{r_{12}},\mathbf{r_{13}})>$&-0.0714037175082&-0.031229607240
&-0.01339606321&-0.00680633501&-0.0038988799\\
$<\cos(\mathbf{r_{12}},\mathbf{r_{32}})>$
&.6800429403893&.603424804436&.56048920421&.53923906130&.527421792\\
\br
\end{tabular}
\end{table}

\Table{\label{table:He13Pe_comp}Comparison for the energies of the
five low-lying  ${}^{1,3}P^{e}$ states of ${}^\infty \textrm{He}$
and $\textrm{He}$ with other calculations.}
\br
\multicolumn{2}{c}{$-E({}^1P^e)$}&\multicolumn{2}{c}{$-E({}^3P^e)$}\\
\mr
\multicolumn{1}{c}{This work}&\multicolumn{1}{c}{other results}
&\multicolumn{1}{c}{This work}&\multicolumn{1}{c}{other results}\\
\mr
\multicolumn{4}{c}{${}^\infty \textrm{He}$}\\
\mr 0.580 246 472 594&\tblue{0.580 246 472 594 392}$\,^{\rm a}$
&0.710 500 155 678 3&\tblue{0.710 500 155 678 334 3}$\,^{\rm a}$\\
&\tblue{0.580 246 472 594 388}$\,^{\rm b}$&&\tblue{0.710 500 155 678 23}$\,^{\rm b}$\\
&&&\tblue{0.710 500 155 656 78}$\,^{\rm c}$\\
0.540 041 590 93&\tblue{0.540 041 590 938 1}$\,^{\rm a}$
&0.567 812 898 725 1&\tblue{0.567 812 898 725 31}$\,^{\rm a}$\\
&\tblue{0.540 041 590 938 52}$\,^{\rm b}$&&\tblue{0.567 812 898 725 16}$\,^{\rm b}$\\
0.524 178 981 8&\tblue{0.524 179 01}$\,^{\rm a}$
&0.535 867 188 767&\tblue{0.535 867 188 71}$\,^{\rm a}$\\
0.516 208 609 4&\tblue{0.516 03}$\,^{\rm a}$
&0.522 254 575 706&\tblue{0.522 253}$\,^{\rm a}$\\
0.511 624 834&&0.515 160 203 83&\\[1em]
\multicolumn{4}{c}{$\textrm{He}\,(m_{\alpha}=7294.299507)$}\\
\mr 0.580 165 768 725&\tblue{0.580 165 768 308 4}$\,^{\rm b}$&0.710
396 457
557&\tblue{0.710 396 457 021 81}$\,^{\rm b}$\\
&\tblue{0.580 165 768}$\,^{\rm d}$&&\tblue{0.710 396 457}$\,^{\rm d}$\\
0.539 967 178 01&\tblue{0.539 967 177 633}$\,^{\rm b}$&0.567 733
870 122&\tblue{0.567 733 869 714 03}$\,^{\rm b}$\\
&\tblue{0.539 967 2}$\,^{\rm d}$&&\tblue{0.567 733 87}$\,^{\rm d}$\\
0.524 106 954 1&&0.535 793 284 74&\\
0.516 137 755 4&&0.522 182 770 7&\\
0.511 554 646 4&&0.515 089 462&\\
\br
\end{tabular}
\item[] $^{\rm a}$Ref.~\cite{Goodson:1990}, $^{\rm b}$Ref.~\cite{Hesse:2001},
$^{\rm c}$Ref.~\cite{Mukherjee:2004}, $^{\rm
d}$Ref.~\cite{Duan:2001}.
\end{indented}
\end{table}

\Table{\label{table:Helike13Pe}The eigen-energies for the ${}^{1,3}P^{e}$ states of  helium-like 
ions $(Z=3 \sim 6)$.} \br
 \multicolumn{1}{c}{$-E({}^{1}P^{e})$}
&\multicolumn{1}{c}{$-E({}^{3}P^{e})$}
&\multicolumn{1}{c}{$-E({}^{1}P^{e})$}
&\multicolumn{1}{c}{$-E({}^{3}P^{e})$}\\
\mr
\multicolumn{2}{c}{Z=3}&\multicolumn{2}{c}{Z=4}\\
1.401 410 927 020&1.796 648 099 720&2.583 994 187 432&3.382 712 420 777\\
1.269 787 972 287&1.373 589 535 176&2.312 232 549 880&2.540 768 798 391\\
1.214 520 958 34&1.260 545 265 63&2.194 982 661 13&2.298 106 658 34\\
1.185 881 076 3&1.210 287 247 90&2.133 407 126 0&2.188 554 817 30\\
1.169 099 559&1.183 580 072 8&2.097 031 380& 2.129 923 592
6\\[0.5em]
\multicolumn{2}{c}{Z=5}&\multicolumn{2}{c}{Z=6}\\
4.127 776 355 386&5.468 730 984 923
&6.032 706 408 00&8.054 724 273 276\\
3.667 237 481 32&4.069 185 038 025
&5.334 768 645 76&5.958 774 148 469\\
3.465 482 470 00&3.648 310 490 768
&5.025 999 284 37&5.311 076 536 74\\
3.358 735 604 6&3.456 902 285 67
&4.861 852 996 5&5.015 282 040 9\\
3.295 388 021 8&3.354 092 319 0
&4.764 160 404&4.856 057 141\\
\br
\end{tabular}
\end{indented}
\end{table}

\Table{\label{table:He13Do_comp} Comparison for the energies of the
five low-lying  ${}^{1,3}D^{o}$ states of ${}^\infty \textrm{He}$
with Bhatia's~\cite{Bhatia:1972}} \br
\multicolumn{2}{c}{$-E({}^1D^o)$}&\multicolumn{2}{c}{$-E({}^3D^o)$}\\
\mr \mbox{\hspace{2em}}This work&\mbox{\hspace{1em}}Bhatia
&\mbox{\hspace{1em}}This work&\mbox{\hspace{1em}}Bhatia\\
\mr
\mbox{\hspace{1em}}0.563 800 420 4&0.563 800 405&0.559 328 263 0&0.559 328 25\\
\mbox{\hspace{1em}}0.534 576 384&0.534 576 015&0.532 678 600&0.532 678 075\\
\mbox{\hspace{1em}}0.521 659 00&0.521 642 77&0.520 703 44&0.520 693 865\\
\mbox{\hspace{1em}}0.514 833 4&0.514 269 06&0.514 288 2&0.514 235 78\\
\br
\end{tabular}
\end{indented}
\end{table}

\Table{\label{table:Li13Fe}Convergence study for the three low-lying
${}^{1}F^{e}$ states of  ${\textrm{Li}}^+$ ion.} \br ($M$,$N$,$P$)
&\multicolumn{1}{c}{$-E_0$} &\multicolumn{1}{c}{$-E_1$}
&\multicolumn{1}{c}{$-E_2$}\\
\mr
(310,170,40)&1.252 443 716 475 54&1.206 193 104 449&1.181 010 621 655\\
(330,180,40)&1.252 445 059 532 88&1.206 204 195 186&1.181 059 878 109\\
(350,190,40)&1.252 446 091 667 05&1.206 212 775 544&1.181 098 378 639\\
(370,200,40)&1.252 446 895 466 84&1.206 219 497 607&1.181 128 811 337\\
(390,210,40)&1.252 447 529 007 27&1.206 224 824 200&1.181 153 115
955\\[0.5em]
\multicolumn{1}{c}{Extrap.}
&1.252 450 636 &1.206 251 57&1.181 279 1\\[0.5em]
&\multicolumn{3}{l}{\tblue{1.252 445 1}\hspace{1.2em}Ref.~\cite{Galan:1981}}\\
&\multicolumn{3}{l}{\tblue{1.252 258}\hspace{2em}Ref.~\cite{Lipsky:1977}}\\
\br
\end{tabular}
\end{indented}
\end{table}

\Table{\label{table:HH_s} 15 low-lying $S$ states of
$\textrm{H}_2^+$} \br
\mbox{\hspace{1em}}n&\multicolumn{1}{c}{$-E_n$}&\mbox{\hspace{1em}}n&\multicolumn{1}{c}{$-E_n$}&\mbox{\hspace{1em}}n&\multicolumn{1}{c}{$-E_n$}\\
\mr
\mbox{\hspace{1em}}0\mbox{\hspace{1em}}&0.597 139&\mbox{\hspace{1em}}5\mbox{\hspace{1em}}&0.552 841&\mbox{\hspace{1em}}10\mbox{\hspace{1em}}&0.521 699\\
\mbox{\hspace{1em}}1&0.587 156&\mbox{\hspace{1em}}6&0.545 593&\mbox{\hspace{1em}}11&0.517 002\\
\mbox{\hspace{1em}}2&0.577 752&\mbox{\hspace{1em}}7&0.538 858&\mbox{\hspace{1em}}12&0.512 827\\
\mbox{\hspace{1em}}3&0.568 909&\mbox{\hspace{1em}}8&0.532 631&\mbox{\hspace{1em}}13&0.509 189\\
\mbox{\hspace{1em}}4&0.560 609&\mbox{\hspace{1em}}9&0.526 911&\mbox{\hspace{1em}}14&0.506 11 \\
\br
\end{tabular}
\end{indented}
\end{table}

\Table{\label{table:Ps_Comp} Comparison of the ground state of
$\textrm{Ps}^-$  with other theoretical calculations.} \br
\multicolumn{1}{c}{$-E$}&\multicolumn{1}{c}{$<{1}/{r_2}>$=$<{1}/{r_3}>$}&\multicolumn{1}{c
}{$<{1}/{r_1}>$}
\\
\mr
\00.262 005 070 2&0.339 821 023&\00.155 631 905 7\\
\tblue{\00.262 004 857}$\,^{\rm a}$&\tblue{0.339 831 3}$\,^{\rm a}$&\tblue{\00.155 654 3}$\,^{\rm a}$\\
\tblue{\00.262 005 070 0}$\,^{\rm b}$&\tblue{0.339 821 02}$\,^{\rm b}$&\tblue{\00.155 631 90}$\,^{\rm b}$\\
\tblue{\00.262 005 070 232 94}$\,^{\rm c}$&\tblue{0.339 821 023 06}$\,^{\rm c}$&\tblue{\00.155 631 905 653}$\,^{\rm c}$\\
\tblue{\00.262 005 070 232 978}$\,^{\rm d}$&\tblue{0.339 821 023 059 27}$\,^{\rm d}$&\tblue{\00.155 631 905 652 66}$\,^{\rm d}$\\
\multicolumn{1}{c}{$<{r_2}>$=$<{r_3}>$}&\multicolumn{1}{c}{$<{r_1}>$}&\multicolumn{1}{c}{$
<{r_2^2}>$=$<{r_3^2}>$}
\\ \mr
\05.489 633 2&8.548 580 6&48.418 936\\
\tblue{\05.488 352}$\,^{\rm a}$&\tblue{8.546 111 29}$\,^{\rm a}$&\tblue{48.379 317}$\,^{\rm a}$\\
\tblue{\05.489 633 3}$\,^{\rm b}$&\tblue{8.548 580 8}$\,^{\rm b}$&\tblue{48.418 936}$\,^{\rm b}$\\
\tblue{\05.489 633 252}$\,^{\rm c}$&\tblue{8.548 580 655}$\,^{\rm c}$&\tblue{48.418 937 2}$\,^{\rm c}$\\
\tblue{\05.489 633 252 38}$\,^{\rm d}$&\tblue{8.548 580 655 16}$\,^{\rm d}$&\tblue{48.418 937 230}$\,^{\rm d}$\\

\multicolumn{1}{c}{$<{r_1^2}>$}&\multicolumn{1}{c}{$<\cos(\mathbf{r_{12}},\mathbf{r_{13}})
>$}
&\multicolumn{1}{c}{$<\cos(\mathbf{r_{12}},\mathbf{r_{32}})>$}
\\ \mr
93.178 63&0.019 769 632 8&\00.591 981 70\\
\tblue{93.100 697 0}$\,^{\rm a}$&&\\
\tblue{93.100 633} $\,^{\rm b}$&&\\
\tblue{93.178 633 80}$\,^{\rm c}$&&\\
\tblue{93.178 633 855}$\,^{\rm d}$&\tblue{0.019 769 632 816 7}$\,^{\rm d}$&\tblue{\00.591 981 701 149 2}$\,^{\rm d}$\\
\br
\end{tabular}
\item[]$^{\rm a}$Ref.~\cite{Haftel:1988b} $^{\rm b}$Ref.~\cite{Petelenz:1987}
$^{\rm c}$Ref.~\cite{Ho:1993} $^{\rm d}$Ref.~\cite{Frolov:1999}
\end{indented}
\end{table}


\end{document}